\documentclass[apj]{emulateapj}

\usepackage{times}
\usepackage{morefloats}
\usepackage{subfigure}
\usepackage{amsmath}
\usepackage[utf8]{inputenc}
\usepackage{hyperref}
\hypersetup{
  pdftitle={The HST Cluster SN Survey: Hosts},
  pdfauthor={Joshua Meyers},
  colorlinks=true, 
  linkcolor=blue, 
  citecolor=blue, 
  urlcolor=blue, 
  bookmarksnumbered
}

\newcommand{\OII}{[O~\textsc{ii}]}
\shorttitle{Dust-free sans spectroscopy: Hi-z SNe via HST}
\shortauthors{Meyers et al.}

\begin{document}

\title{The {\it Hubble Space Telescope}\footnotemark[*]
  Cluster Supernova Survey: \\ III. Correlated properties of Type Ia
  Supernovae and their hosts at $0.9 < $~\MakeLowercase{z}~$ <
  1.46$
}
\footnotetext[*]{Based on observations made with the NASA/ESA Hubble
  Space Telescope and obtained from the data archive at the Space
  Telescope Institute. STScI is operated by the Association of
  Universities for Research in Astronomy, Inc. under the NASA contract
  NAS 5-26555.  The observations are associated with program 10496.
}

\author{
J.~Meyers\altaffilmark{1,2},
G.~Aldering\altaffilmark{2},
K.~Barbary\altaffilmark{1,2},
L.~F.~Barrientos\altaffilmark{3},
M.~Brodwin\altaffilmark{4,21},
K.~S.~Dawson\altaffilmark{5},
S.~Deustua\altaffilmark{6},
M.~Doi\altaffilmark{7},
P.~Eisenhardt\altaffilmark{8},
L.~Faccioli\altaffilmark{2},
H.~K.~Fakhouri\altaffilmark{1,2},
A.~S.~Fruchter\altaffilmark{6},
D.~G.~Gilbank\altaffilmark{9},
M.~D.~Gladders\altaffilmark{10},
G.~Goldhaber\altaffilmark{1,2,23},
A.~H.~Gonzalez\altaffilmark{11},
T.~Hattori\altaffilmark{12},
E.~Hsiao\altaffilmark{2},
Y.~Ihara\altaffilmark{7,22},
N.~Kashikawa\altaffilmark{13},
B.~Koester\altaffilmark{10,14},
K.~Konishi\altaffilmark{15},
C.~Lidman\altaffilmark{16},
L.~Lubin\altaffilmark{17},
T.~Morokuma\altaffilmark{7,13,22},
T.~Oda\altaffilmark{18},
S.~Perlmutter\altaffilmark{1,2},
M.~Postman\altaffilmark{6},
P.~Ripoche\altaffilmark{2},
P.~Rosati\altaffilmark{19},
D.~Rubin\altaffilmark{1,2},
E.~Rykoff\altaffilmark{2},
A.~Spadafora\altaffilmark{2},
S.~A.~Stanford\altaffilmark{17,20},
N.~Suzuki\altaffilmark{2},
N.~Takanashi\altaffilmark{13},
K.~Tokita\altaffilmark{7},
N.~Yasuda\altaffilmark{15} \\
(The Supernova Cosmology Project)
}
\email{jmeyers314@berkeley.edu}

\altaffiltext{1}{Department of Physics, University of California
  Berkeley, Berkeley, CA 94720, USA}
\altaffiltext{2}{E.O. Lawrence Berkeley National Lab, 1 Cyclotron Rd.,
  Berkeley CA, 94720, USA}
\altaffiltext{3}{Departmento de Astronomía, Pontificia Universidad
  Católica de Chile, Santiago, Chile}
\altaffiltext{4}{Harvard-Smithsonian Center for Astrophysics, 60
  Garden Street, Cambridge, MA 02138, USA}
\altaffiltext{5}{Department of Physics and Astronomy, University of
  Utah, Salt Lake City, UT 84112, USA}
\altaffiltext{6}{Space Telescope Science Institute, 3700 San Martin
  Drive, Baltimore, MD 21218, USA}
\altaffiltext{7}{Institute of Astronomy, Graduate School of Science,
  University of Tokyo 2-21-1 Osawa, Mitaka, Tokyo 181-0015, Japan}
\altaffiltext{8}{Jet Propulsion Laboratory, California Institute of
  Technology, Pasadena, CA 91109, USA}
\altaffiltext{9}{Department of Physics and Astronomy,
  University Of Waterloo, Waterloo, Ontario, N2L 3G1 Canada}
\altaffiltext{10}{Department of Astronomy and Astrophysics, University
  of Chicago, Chicago, IL 60637, USA}
\altaffiltext{11}{Department of Astronomy, University of Florida,
  Gainesville, FL 32611, USA}
\altaffiltext{12}{Subaru Telescope, National Astronomical Observatory
  of Japan, 650 North Aóhaku Place, Hilo, HI 96720, USA}
\altaffiltext{13}{National Astronomical Observatory of Japan, 2-21-1
  Osawa, Mitaka, Tokyo,181-8588, Japan}
\altaffiltext{14}{Kavli Institute for Cosmological Physics, The University of
  Chicago, Chicago, IL 60637, USA}
\altaffiltext{15}{Institute for Cosmic Ray Research, University of
  Tokyo, 5-1-5, Kashiwanoha, Kashiwa, Chiba, 277-8582, Japan}
\altaffiltext{16}{Australian Astronomical Observatory, PO Box 296, Epping,
  NSW 1710, Australia}
\altaffiltext{17}{University of California, Davis, CA 95618, USA}
\altaffiltext{18}{Department of Astronomy, Kyoto University, Sakyo-ku,
  Kyoto 606-8502, Japan}
\altaffiltext{19}{ESO, Karl-Schwarzschild-Strasse 2, D-85748 Garching, Germany}
\altaffiltext{20}{Institute of Geophysics and Planetary Physics,
  Lawrence Livermore National Laboratory, Livermore, CA 94550, USA}
\altaffiltext{21}{W. M. Keck Postdoctoral Fellow at the
  Harvard-Smithsonian Center for Astrophysics}
\altaffiltext{22}{JSPS Fellow}
\altaffiltext{23}{Deceased}

\begin{abstract}
  Using the sample of Type Ia supernovae (SNe~Ia) discovered by the
  {\it Hubble Space Telescope (HST)} Cluster Supernova Survey and
  augmented with {\it HST}-observed SNe~Ia in the GOODS fields, we
  search for correlations between the properties of SNe and their host
  galaxies at high redshift.  We use galaxy color and quantitative
  morphology to determine the red sequence in 25 clusters and develop
  a model to distinguish passively evolving early-type galaxies from
  star-forming galaxies in both clusters and the field.  With this
  approach, we identify six SN~Ia hosts that are early-type cluster
  members and eleven SN~Ia hosts that are early-type field galaxies.
  We confirm for the first time at $z>0.9$ that SNe~Ia hosted by
  early-type galaxies brighten and fade more quickly than SNe~Ia
  hosted by late-type galaxies.  We also show that the two samples of
  hosts produce SNe~Ia with similar color distributions.  The
  relatively simple spectral energy distributions (SEDs) expected for
  passive galaxies enable us to measure stellar masses of early-type
  SN hosts. In combination with stellar mass estimates of late-type
  GOODS SN hosts from \citet{Thomson2011}, we investigate the
  correlation of host mass with Hubble residual observed at lower
  redshifts.  Although the sample is small and the uncertainties are
  large, a hint of this relation is found at $z>0.9$.  By
  simultaneously fitting the average cluster galaxy formation history
  and dust content to the red-sequence scatters, we show that the
  reddening of early-type cluster SN hosts is likely $E(B-V) \lesssim
  0.06$.  The similarity of the field and cluster early-type host
  samples suggests that field early-type galaxies that lie on the red
  sequence may also be minimally affected by dust.  Hence, the
  early-type hosted SNe~Ia studied here occupy a more favorable
  environment to use as well-characterized high-redshift standard
  candles than other SNe~Ia.
\end{abstract}


\section{Introduction}\label{sec:intro}
The use of Type Ia supernovae (SNe~Ia) as standard candles in
estimating astronomical distances has proven indispensable for modern
cosmology, leading to the remarkable discovery that the expansion of
the Universe is accelerating \citep{Riess1998,Perlmutter1999}.
Together with measurements of the cosmic microwave
background \citep{Dunkley2009, Komatsu2011} and baryon acoustic
oscillations \citep{Eisenstein2005, Percival2010}, our recent
compilation of the myriad SN surveys that led to and followed the
initial discovery has now constrained the dark energy equation of
state to be
$w=-0.974^{+0.054}_{-0.058}$(stat)$^{+0.075}_{-0.080}$(stat+sys)
\citep{Amanullah2010}.  The constraints from the 557 SNe~Ia in this
compilation are now limited by systematic uncertainties.  Additional
SN~Ia observations will not greatly improve dark energy constraints
until the systematic errors are reduced.

Improvements in SN~Ia distance estimates have historically followed
two paths.  The first path is to identify subsets of SNe with smaller
intrinsic brightness dispersion (e.g. the Type~Ia subset or
``Branch-normal'' SNe~Ia).  The second path is to make corrections
exploiting empirical correlations between SN brightness and other
observables (e.g. light curve shape and color).  The properties of
SN~Ia host galaxies offer additional variables with which to search
for smaller dispersion subsets or correlated observables.  In fact,
such a correlation has recently been reported: SNe~Ia in more massive
hosts are brighter after light curve shape and color
corrections \citep{Kelly2010, Sullivan2010, Lampeitl2010}.

The brightnesses of SNe~Ia are correlated with their colors at maximum
light: bluer SNe~Ia are brighter.  This trend is broadly consistent
with extinction due to dust as commonly described by $R_B = A(B) /
E(B-V)$ \citep{Cardelli1989}.  When relating SN~Ia colors to peak
magnitudes, the parameter $\beta = \Delta M_B / \Delta c$ is used,
where $c = \left(B-V\right)_{max} - \left<B-V\right>_{max}$; e.g. $c$
is the excess color at maximum brightness with respect to the average.
Direct fits of SN~Ia observations have found that $\beta$ is
considerably smaller than the value of $R_B$ observed in the Milky Way
diffuse interstellar medium, suggesting that an intrinsic component to
the bluer--brighter relation is also important \citep{Tripp1998,
Astier2006,Conley2007,Nobili2008}.  The systematic errors associated
with the interpretation of $\beta$, i.e., what fraction is
attributable to dust and how might that evolve with redshift, are now
comparable to the statistical errors of SN~Ia
surveys \citep{Kowalski2008,Wood-Vasey2007}.  However, if one were to
identify SN~Ia hosts with small extinction, then a single component of
the SN~Ia color-magnitude relation, the intrinsic component, could be
isolated.

In this paper we analyze the host galaxies of $z>0.9$ SNe~Ia drawn
from the {\it HST} Cluster SN Survey \citep[PI-Perlmutter:
GO-10496]{Dawson2009} and surveys of the GOODS
fields \citep{Blakeslee2003B,Riess2004,Riess2007}.  We investigate how
host properties correlate with SN~Ia properties and identify hosts
minimally affected by dust.  Particularly interesting host properties
such as age and metallicity are usually difficult to obtain, but are
correlated with the position of a galaxy on its cluster's red
sequence \citep{Gallazzi2006}.  The environment of the {\it HST}
Cluster SN Survey thus lends itself to a particular analysis strategy:
how does the location of a host galaxy on its cluster's red sequence
correlate with its SN, what does it imply about the SN type, and how
does it begin to parse the origin of the SN reddening law?

This paper is organized as follows: In \S \ref{sec:hosts} we describe
progress to date of SN host galaxy studies at lower redshifts.
In \S \ref{sec:data} we describe the {\it HST} Cluster SN Survey and
the {\it HST} SN surveys of the GOODS fields, the data taken, and the
various photometric, morphological, and spectroscopic quantities that
we derive from them.  In \S \ref{sec:analysis} we describe how, by
using the color-magnitude locations of spectroscopically confirmed
red-sequence cluster members, we predict the location of the red
sequence for each cluster, including those without extensive
spectroscopic coverage.  Additionally, we demonstrate how quantitative
morphology parameters can be used to enhance the contrast of the red
sequence to more easily fit a color-magnitude relation (CMR) for each
cluster.  In \S \ref{sec:classify} we describe criteria to classify
the observed SN hosts, and in \S \ref{sec:typing} we use expected
rates of SNe~Ia and core collapse SNe (SNe~CC) in their individual
galaxies to help type the SNe.  In \S \ref{sec:dust} we investigate
limits on extinction for SNe hosted by early-type galaxies.
In \S \ref{sec:correlations} we investigate correlations between SN
properties and properties of their host galaxies.  Finally,
in \S \ref{sec:conclusion} we summarize our conclusions and discuss
their implications.

This paper is one of a series of ten papers that report supernova
results from the {\it HST} Cluster Supernova Survey (PI: Perlmutter,
GO-10496), a survey to discover and follow SNe~Ia of very distant
clusters. Paper I \citep{Dawson2009} describes the survey strategy and
discoveries. Paper II \citep{Barbary2010} reports on the SN~Ia rate in
clusters.  The current work, Paper III, addresses the properties of
the galaxies that host SNe~Ia.  Paper IV \citet{Ripoche2011}
introduces a new technique to calibrate the ``zeropoint'' of the
NICMOS camera at low count rates, which is critical for placing
NICMOS-observed SNe~Ia on the Hubble diagram. Paper
V \citep{Suzuki2011} reports the SN~Ia light curves and cosmology from
the {\it HST} Cluster SN Survey. Paper VI \citep{Barbary2011} reports
on the volumetric field SN~Ia rate.  \citet{Melbourne2007}, one of
several unnumbered papers in this series, present a Keck Adaptive
Optics observation of a $z=1.31$ SN~Ia in {\it
H}-band.  \citet{Barbary2009} report the discovery of the
extraordinary luminous supernova, SN~SCP06F6.  \citet{Morokuma2010}
present the spectroscopic follow-up observations for SN candidates.
\citet{Hsiao2011} develop techniques to remove problematic artifacts
remaining after the standard STScI pipeline.  A separate series of
papers, ten to date, reports on cluster studies from the {\it HST}
Cluster SN Survey:
\citet[][]{Brodwin2011,Eisenhardt2008,Jee2009,Hilton2007,Hilton2009,
Huang2009,Santos2009,Strazzullo2010,Rosati2009}; and \citet{Jee2011}.

\section{SN~Ia Hosts}\label{sec:hosts}
In this section we briefly discuss the relationships between SNe and
their host galaxies with particular emphasis on early-type host
galaxies of SNe~Ia.  We also introduce the {\it HST} Cluster SN
Survey, which targets massive galaxy clusters to increase the yield of
SN~Ia discoveries, particularly those in low-dust early-type hosts.

\subsection{SN typing by host}\label{subsec:typing}

Early-type galaxies have almost never been observed to host Type~Ib/c
or Type~II SNe (hereafter SNe~CC), which are generally thought to
originate from the collapse of the cores of massive (and hence young)
stars.  In an extensive literature search of hundreds of
SNe~CC, \citet{Hakobyan2008} uncovered 22 examples reported as being
hosted by early-type galaxies, but ultimately reclassified 19 of these
galaxies as late-type.  The SN associated with one of the remaining
three galaxies has since been reclassified as a cataclysmic variable
star within the Milky Way \citep{Leonard2010}.  The SNe hosted by the
remaining two early-type hosts, SN2000ds and SN2005cz, are both
members of the SN2005E-like subset of faint ``Ca-rich'' SNe~Ib.  The
origins of SNe in this subset are still under debate, though it is
clear that they are readily distinguishable from SNe~Ia by their faint
peak magnitudes ($M_B \sim -15$ compared to $M_B \sim -18$ for even a
very faint SN~Ia) \citep{Perets2010, Kawabata2010}.  In a separate
analysis of the near ultraviolet and optical colors of early-type SN
hosts (including two which were reclassified by Hakobyan et
al.), \citet{Suh2011} find that the early-type hosts of SNe~CC have
had more recent star formation and occupy a different part of the
UV-optical color-magnitude diagram than identically selected
early-type hosts of SNe~Ia.

Despite these rare possible exceptions, SNe hosted by early-type
galaxies are overwhelmingly Type~Ia.  Typing high-redshift $(z>0.9)$
SNe spectroscopically can be done from space with {\it HST}
\citep{Riess2004} but requires a significant investment of orbits (4-8
per SN at $z\sim1.2$). Not only are SNe increasingly faint at higher
redshift, but the rest-frame features used to identify a Type Ia
become shifted out of the optical wavelengths.  Ground-based
spectroscopy faces the additional difficulty that the night sky
becomes increasingly bright at redder wavelengths.  With ideal
conditions and by targeting the SN near the peak of its light curve,
\citet{Morokuma2010} have shown it is possible to spectroscopically
type SNe~Ia as high as $z=1.34$ from the ground, but this is by no
means typical.  In fact, many of the lower redshift targets in this
paper (of which many also had higher signal-to-noise ratio spectra)
yielded inconclusive types.  Classifying SNe in ellipticals galaxies
as Type~Ia can therefore provide an efficient and robust alternative
to potentially expensive spectroscopic typing at high redshift.

\subsection{Dust in early-type SN~Ia hosts}\label{subsec:nearbyETGdust}

Early-type SN~Ia hosts offer an additional advantage to typing: they
generally contain only small amounts of dust.  Although {\it HST}
images of the cores of nearby early-type galaxies indicate that $\sim$
50\% exhibit optical absorption due to dust in small disks or
filaments, these are usually confined to the central few hundred
parsecs \citep{vanDokkum1995, Tomita2000, Rest2001, Tran2001,
Lauer2005, Ferrarese2006}.  In contrast, SNe~Ia hosted by early-type
galaxies are spatially distributed following the optical light of
their host galaxies \citep{Forster2008}.  Observations of emission
from dust at far-infrared (FIR) wavelengths suggest that additional
dust is diffusely distributed throughout at least some nearby
early-type galaxies \citep{Goudfrooij1995, Temi2004, Temi2007}.  The
inferred dust mass relative to the stellar mass of these galaxies
varies greatly; at fixed optical luminosity, \citet{Temi2007} measure
FIR luminosities that span two orders of magnitude.  At the low
dust-mass end, at least several early-type galaxies (3 elliptical and
3 S0) with $M_B \sim -18$ to $-20$, are undetected in the Herschel
Space Observatory survey of the Virgo cluster.  Upper limits constrain
the associate dust mass of these galaxies to be $<10^4$ M
$_\odot$ \citep{Clemens2010}.  For comparison, the Virgo cluster
spiral galaxies in the same $B$-band luminosity range also observed by
Herschel have dust masses of $\sim 10^7 - 10^8$ M $_\odot$)
\citep{Davies2011}.  Early-type galaxy dust masses derived
from Spitzer tend to fall in the range $10^4-10^6$
\citep{Kaneda2007}. To estimate the extinction associated
with a certain mass of dust, we use $\left<A_V\right> =
M_d\Gamma/\Sigma$, with $\Sigma$ the area of the dust feature,
$\left<A_V\right>$ the average $V$-band absorption in that area, and
$\Gamma \sim 6\times10^{-6}$ mag kpc$^2$ M$_\odot^{-1}$ the visual
mass absorption coefficient \citep{vanDokkum1995}.  For dust uniformly
distributed throughout a disc with radius $3$ kpc, which is
representative of the half-light radii of the galaxies considered in
this paper, the absorption ranges from $A_V < 0.002$ to $\sim 0.2$.
Since dust is not actually distributed in a uniform foreground screen,
but rather is embedded within the galaxy, these values only provide a
rough guide to the expected extinction along the lines of sight toward
early-type hosted SNe~Ia.

At high redshift, FIR observations of SN~Ia hosts are impractical.
However, optical observations of the integrated colors of galaxies can
also place constraints on the extinction expected from diffuse dust.
For example, consider the case where galaxy colors are the combined
result of stellar population age, metallicity, and dust.  The
contributions of age and metallicity to the integrated color can be
removed by using an auxilliary correlated variable.  At low redshift,
the spectral absorption feature ${\rm Mg}_2$ is such a variable.  The
equivalent width of ${\rm Mg}_2$ is expected to correlate with the
rest-frame intrinsic (i.e. unextinguished) $B-V$ color of elliptical
galaxies \citep{Faber1989}.  The line strength increases with
increasing galaxy age and metallicity, and consequently color, but
because the wavelength interval of the line is relatively narrow it
should not correlate significantly with dust.  \citet{Schlegel1998}
measured the scatter about the ${\rm Mg}_2$-color relation for nearby
elliptical galaxies to be just $\sigma_{BV} = 0.0257$ mag.  If at
least some elliptical galaxies are effectively dust-free ($A_V <
0.01$), which seems likely given the range of dust masses discussed
above, then this scatter measurement implies an absolute limit on dust
reddening at the $E(B-V) \sim 0.03$ level.  While spectroscopy of the
${\rm Mg}_2$ feature may not be feasible at high redshift, the red
sequence may provide a readily available alternative calibrator.

Two mechanisms have been proposed to explain the origin of the dust
found in elliptical galaxies.  Elliptical galaxies may accrete dust
and gas during (minor) mergers with dusty late-type galaxies.  This
scenario is supported by observations that the motions of dust and gas
in some nearby early-type galaxies seem unrelated to the motions of
their stars \citep{Goudfrooij1994,vanDokkum1995,Caon2000}.  As part of
the Galaxy-Zoo project, \citet{Kaviraj2011} find that early-type
galaxies with prominent dust lanes are likely associated with mergers
due to the frequency with which they present disturbed morphologies
compared to a control sample without dust lanes.  Prominent dust lane
early-type galaxies are relatively rare, however, constituting just
4\% of the early-type Galaxy-Zoo sample, and are even rarer in cluster
environments..

Noting that the dynamical freefall time for a merging galaxy (several
$\sim 10^8$\ yr) is comparable to the sputtering lifetime of dust
grains in the hot interstellar medium of early-type galaxies ($\sim
10^7 - 10^8$\ yr), \citet{Temi2007} argue that dust originating from
mergers must be resupplied at least every $\sim 10^8$\ yr.  Such a
high rate of early-type -- late-type mergers is not observed, however.
As an alternative, they propose an internal origin for diffuse dust in
early-type galaxies: dust is generated in the atmospheres of evolved
red giant stars \citep{Knapp1989,Athey2002} and may accumulate in the
concentrated disks of dust commonly observed in the centers of
early-type galaxies\citep{Mathews2003}.  Intermittent AGN activity can
then buoyantly transport this dust out to large radii where it will be
destroyed by sputtering in the hot interstellar medium.

One possible caveat to the expectation of low extinction SNe~Ia in
early-type galaxies is small amounts of recent star-formation seen in
some early-type galaxies, especially lenticular galaxies, and
associated dust \citep{Yi2005, Donas2007, Schawinski2007, Kaviraj2007,
Kaviraj2008, Temi2009a, Temi2009b}.  Together with the expected
increase in the SN~Ia rate with star-formation rate
(see \S \ref{subsec:SNprop} below), it is possible that an SN~Ia in a
mostly dust-free galaxy could, nevertheless, suffer from extinction.
Fortunately, recent star-formation in early-type galaxies is usually
confined to their inner regions \citep{Kuntschner2006, Sarzi2006},
whereas, as mentioned above, the spatial distribution of SNe~Ia
follows the optical light distribution.  Furthermore, we have directly
constrained star-formation using spectroscopy of our host galaxies and
its possible effect in increasing the galactic SN~Ia rate
(see \S \ref{sec:typing}).  We expect that most of our early-type
hosted SNe~Ia are not associated with recent star-formation.

\subsection{SN~Ia properties by host}\label{subsec:SNprop}

Several lines of evidence suggest that the demographics of SNe~Ia
hosted by early-type galaxies are different than those of SNe~Ia
hosted by late-type galaxies.  Using infrared magnitudes of
low-redshift SN~Ia host galaxies in the 2MASS survey,
\citet{Mannucci2005} measured the SN~Ia rate per unit stellar mass for
different host types.  They found a rate $\sim20$ times larger in
late-type galaxies than in E/S0 galaxies.  Splitting the galaxies by
color (as a proxy for star formation) they found a rate $\sim30$ times
larger in blue galaxies than in red galaxies.  \citet{Sullivan2006}
fit SED models to five-band galaxy photometry to directly constrain
the star formation rate and mass of intermediate redshift SN~Ia host
galaxies in the Supernova Legacy Survey (SNLS).  This analysis showed
that the SN~Ia rate is $\sim10$ times larger per unit stellar mass in
actively star-forming galaxies than in passive galaxies.  For a given
episode of star formation, the SN~Ia rate quickly peaks and declines
as the progenitor population ages.

Perhaps not surprisingly, the properties of SNe themselves are also
correlated with the properties of their hosts.  At low redshift,
SNe~Ia hosted by late-type, spiral galaxies tend to have broader
(slower) light curves on average \citep{Hamuy1996, Gallagher2005},
although interestingly, \citet{Hicken2009} find that the broadest SN
light curves are found in SNe hosted by Sb-Sc galaxies and not younger
Sd-Irr galaxies.  \citet{Sullivan2006} showed that this distinction
extends to intermediate redshift hosts with actively star-forming
galaxies (generally late-type) hosting SNe~Ia with broader light
curves.  Several analyses suggest that SNe~Ia hosted by early-type
galaxies may have a smaller intrinsic peak brightness dispersion after
light curve and color corrections are applied \citep{Sullivan2003,
  Jha2007, Sullivan2010, Lampeitl2010}, though at least one analysis
finds the opposite result \citep{Hicken2009}.

Since SN~Ia properties are correlated with the properties of their
host galaxies, and the demographics of galaxies change with redshift,
the demographics of SNe~Ia will also change with redshift.  This shift
can bias inferred cosmological parameters if not handled carefully.
Tests for bias in cosmology constructed by segregating SNe~Ia by
host-type were discussed in \citet{Perlmutter1997} and carried out
with a small sample of SNe~Ia (17) in \citet{Perlmutter1999}, and a
larger sample (39) in \citet{Sullivan2003} without any differences in
fitted cosmological parameters apparent within the statistical
uncertainties.  However, with a much larger sample from the SNLS
\citep{Sullivan2010}, it becomes possible to see subtle differences in
$\beta$ dependent on the host.  SNe~Ia hosted by lower specific
star-formation rate galaxies obey a shallower relation than SNe~Ia
hosted by higher specific star-formation rate galaxies ($\beta\sim2.8$
compared to $\beta\sim3.5$; the exact value depends on the choice of
specific star-formation rate delineating the low and high star-forming
samples).  Similarly, using SNe drawn from the Sloan Digital Sky
Survey (SDSS), \citet{Lampeitl2010} find $\beta \sim 2.5$ for passive
hosts and $\beta \sim 3$ for star-forming hosts.  Since $\beta$
reflects the combined effect of host galaxy dust extinction and an
intrinsic SN color-luminosity relation, it is not surprising that
$\beta$ should be closer to the Milky Way dust total-to-selective
extinction coefficient of $R_B = 4.1$ in dusty star-forming galaxies
than in passive galaxies.  This result also suggests that as the
fraction of low star-forming early-type galaxies decreases with
redshift and the fraction of high star-forming galaxies increases, the
average value of the relation $\left<\beta\left(z\right)\right>$ will
increase and introduce a bias on measurements of $w$ if the
subpopulations are not treated separately.

Perhaps a more important consideration, however, is the recently
reported evidence for a correlation between host galaxy mass and
Hubble residual after applying corrections for light curve shape and
color.  By fitting SEDs derived from the PEGASE2 \citep{Fioc1997,
  Fioc1999} stellar population synthesis models to SDSS photometry of
host galaxies, \citet{Kelly2010} measured the host masses of the low-z
SNe in the \citet{Hicken2009} analysis and compared these to their
Hubble residuals obtained with a variety of light curve fitters.  They
found that SNe~Ia in more massive galaxies are brighter (after stretch
and color corrections) by about $\sim0.1$ mag.  \citet{Sullivan2010}
found a similar relation in SNLS SNe and host galaxy photometry using
PEGASE2 SEDs and the SiFTO light curve fitter \citep{Conley2008}.
Finally, \citet{Lampeitl2010} found a similar relationship in the
SDSS-II SN survey by segregating SNe by host type, with passive
galaxies hosting brighter post-correction SNe.  However, because the
passive galaxies in this dataset are on average more massive than the
star-forming galaxies, this relation can also be framed in terms of
host galaxy mass.  These results are particularly important for
targeted surveys, such as those that target massive galaxies at low
redshift or the HST Cluster SN Survey which targets clusters of
massive galaxies at high redshift.  The demographics of hosts from
these surveys are substantially different than for untargeted surveys.
The SNe they discover should be on average brighter (after color and
light curve shape corrections) than those in other surveys, a property
which needs to be corrected for in cosmological analyses.

\subsection{The \textup{HST} Cluster SN Survey}\label{subsec:survey}

By targeting massive galaxy clusters, the {\it HST} Cluster SN Survey
was designed to efficiently discover well-characterized SNe~Ia at high
redshift.  Clusters are rich in elliptical galaxies, which constitute
a linear red sequence in a color-magnitude diagram.  The observed
evolution of the red-sequence slope indicates that ellipticals in the
cores of clusters have passively evolved since forming at
high-redshift ($z>2$) and that the redder colors of more massive
galaxies is due to increased metallicity \citep{Kodama1997,
Kodama1997b, Kauffmann1998, Gladders1998}.  With filters straddling
the $4000$\AA\ break, the red sequence can be readily isolated from
foreground and background galaxies, a technique also used to find
clusters \citep[e.g.][]{Gladders2000}.

Clusters also provide a convenient mechanism for probing the dust
contents of cluster early-type galaxies.  Just as the ${\rm Mg}_2$
absorption feature traces age and metallicity at low redshift, the
magnitude axis of the red sequence traces metallicity at high
redshift. A dispersion in the residuals from the color-magnitude
relation of roughly 3\% is found in clusters ranging from Coma to high
redshift (z $\sim$ 1.5) \citep{Bower1992, Ellis1997, Stanford1998,
Blakeslee2003, Mei2006a, Mei2006b, Lidman2008, Mei2009}.  Some of this
dispersion can be attributed to differences in galaxy ages, and what
remains sets a limit on dust.  Finally, we also note that the uniform
old stellar populations of elliptical galaxies imply simpler SEDs and
hence better mass estimates with which to address the trends
of \S \ref{subsec:SNprop}.

The completed {\it HST} Cluster SN Survey produced 19 $z>0.9$ SNe,
eight of which were found in the clusters themselves.  Deep two-color
images of the cluster galaxies (including SN hosts) were also produced
by stacking together SN search and follow-up epochs.  Similar data
exist from the SN surveys of the GOODS fields, which have discovered
26 $z>0.9$ SNe~Ia, all in the field.  We analyze the host galaxies of
these two sets of SNe in this paper.

\section{Data}\label{sec:data}
To identify a set of SNe~Ia hosted by a uniform stellar population and
minimally affected by dust, we compare the photometry of SN host
galaxies to the photometry of red-sequence galaxies and look for
star-formation indicators in SN host spectroscopy.  Identifying the
red sequence requires careful measurement of galaxy magnitudes and
colors, which we describe in this section.  We also measure
quantitative morphology parameters which we use to enhance the
contrast of the red sequence in color-magnitude diagrams by exploiting
the early-type galaxy population dominance along the red sequence.

\subsection{Image reduction and photometry}\label{subsec:image}

\begin{figure}
\begin{center}
\plotone{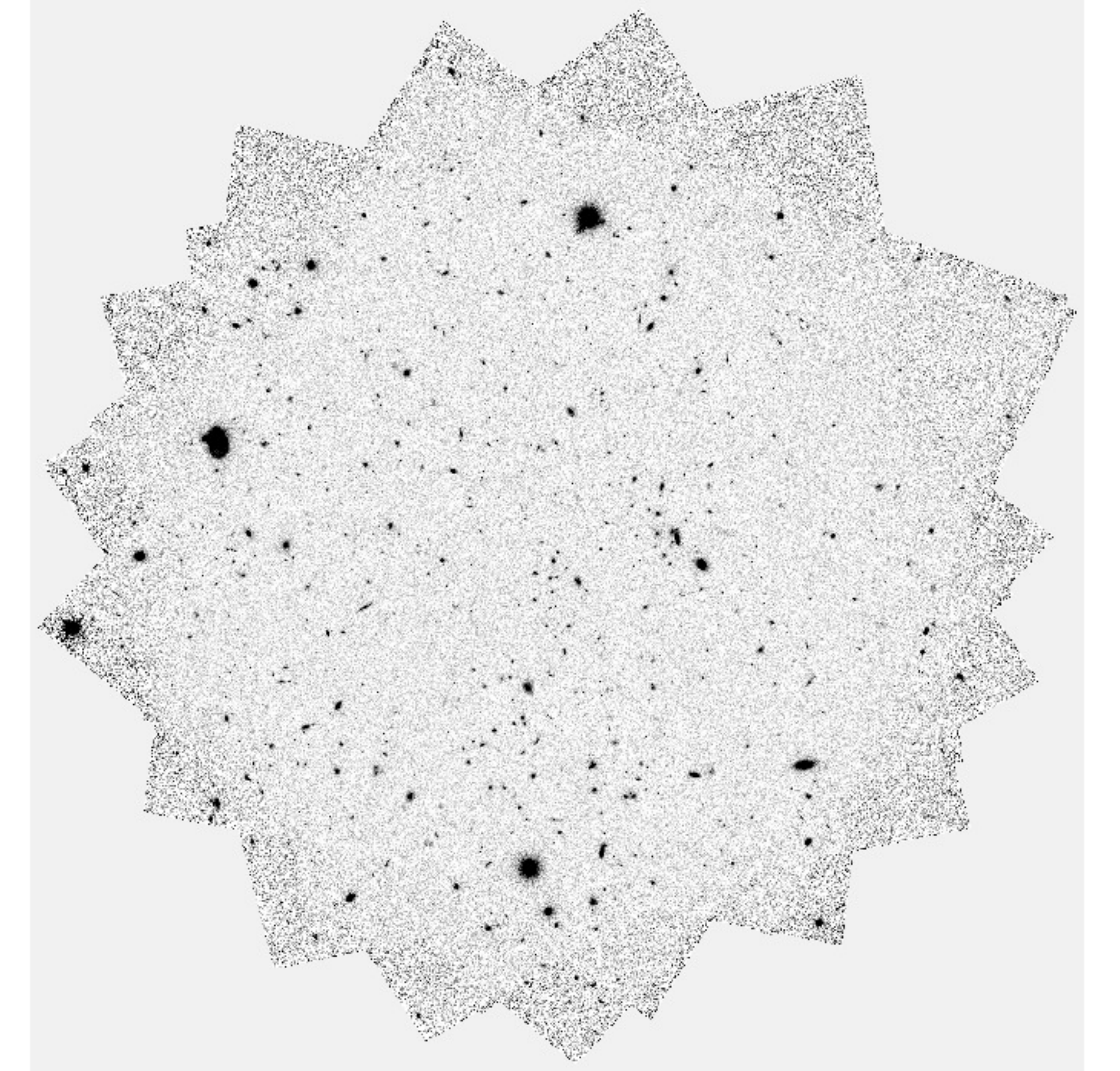}
\end{center}
\caption[F-cluster]{\label{fig:Fmosaic} Final $z_{850}$ mosaic for
  cluster ISCS~J1432.4+3332.  The differences in background noise as a
  function of position are apparent towards the edges of the mosaic.}
\end{figure}

Twenty-five massive high-redshift $(0.9 < z < 1.46)$ galaxy clusters
selected from X-ray, optical, and IR surveys were chosen for the {\it
  HST} Cluster SN Survey \citep{Dawson2009}.  Clusters were each
visited by {\it HST} four to nine times between July 2005 and December
2006.  Each visit typically consisted of four $\sim500$ second
exposures in the F850LP filter (hereafter $z_{850}$) of the Advanced
Camera for Surveys (ACS) wide field camera (WFC) and one $\sim500$
second exposure in the F775W filter (hereafter $i_{775}$) of the ACS
WFC.  The $i_{775}$ filter roughly matches rest-frame $U$-band for
clusters with $0.9 < z < 1.25$, with the best match occuring at
$z=1.1$.  The $z_{850}$ filter roughly matches the rest-frame $B$-band
in this redshift range with its closest match occuring at $z=1.05$.
For more distant clusters with $1.25 < z < 1.46$, the $z_{850}$ filter
more closely matches rest-frame $U$-band, with the best overlap at
$z=1.45$.  In this paper we look principally at the deep coadditions
of exposures from all observation epochs.  Due to gyroscopic
constraints, {\it HST} visits to individual clusters necessarily
occurred at different position angles, resulting in coadditions in
which pixels near the edges have smaller effective exposure time than
pixels near the center (Figure \ref{fig:Fmosaic}).  Each exposure
specifically targeted the cluster core, so total integration time is
nearly constant in the central region of each coaddition.  Four
clusters, RDCS~J0910+54 \citep{Mei2006b}, RDCS~J0848+44
\citep{Postman2005}, RDCS~J1252-29\citep{Blakeslee2003} and
XMMU~2235.3-2557 \citep{Jee2009}, had been previously targeted by ACS
in $i_{775}$ and $z_{850}$ (PID9290 and PID9919), and we have included
these additional exposures in our coadded images.  Cluster
CL~1604+4304 \citep{Postman2005} had also been observe with ACS, but
not in $i_{775}$ and $z_{850}$.  The individually sky subtracted
exposures were stacked using {\sc MultiDrizzle}
\citep{Fruchter2002,Koekemoer2002} with a square kernel,
$\mathrm{pixfrac}=0.8$ and the native output pixel scale of
$0.05\arcsec$.

We similarly processed images obtained by {\it HST} ACS as part of the
{\it HST} Great Observatories Origins Deep Survey (GOODS) Treasury
program \citep{Giavalisco2004}.  The {\it HST} GOODS program targeted
two high galactic latitude fields to obtain deep multiband images for
studies of galaxy evolution and included a ``piggy-back'' SN program
to follow suspected $z>1$ SNe with ACS and the Near Infrared Camera
and Multi-object Spectrograph (NICMOS).  The original search consisted
of five epochs over 15 ACS pointings for each of the two GOODS fields.
A subsequent extension contributed an additional 14 epochs.  The
survey used nearly identical exposures in $i_{775}$ and $z_{850}$ for
SN discovery and follow-up as our program.  The initial GOODS SN
survey yielded 11 $z>0.9$ SNe~Ia and its extension yielded an
additional 13 $z>0.9$ SNe~Ia \citep{Riess2004,Riess2007}.  We also add
the host of one $z>0.9$ SN~Ia discovered in a previous search of these
fields -- SN2002dd \citet{Blakeslee2003B}.  One other $z>0.9$ SN has
been discovered in these fields -- SN1997ff, but its most likely
redshift, $z = 1.76$ is tentative and far beyond that of the clusters
studied here.  Lacking a comparison cluster red-sequence we do not
analyze this SN host.

Galaxies that hosted SNe required special attention to prevent the SN
light from biasing photometric and morphological statistics of the
galaxies themselves.  For each SN we created postage stamp images of
each epoch in both bands.  From these images, we subtracted a PSF
model (described below) scaled to the flux of the light curve fit
(details in \citet{Suzuki2011}) of the SN in each epoch.  The postage
stamp images are then stacked and processed in the same manner as the
full mosaics.  When the final catalogs are created, the values derived
from the SN-subtracted postage stamp images of SN hosts are
substituted for the values derived from the full mosaics.  To test the
accuracy of the PSF subtractions, we also generated postage-stamp
images of SN hosts by stacking only the subset of epochs in which the
SN was far from its peak.  Except for a few cases in which only one or
two such epochs were available, we obtained consistent photometric and
morphological results to those obtained from the PSF-subtracted
images.

\subsubsection{Object detection}\label{subsubsec:detect}

To form initial object catalogs we used {\sc SExtractor} version 2.8.6
\citep{Bertin1996} in dual image mode, relying on the $z_{850}$ images
for object detection.  Determining appropriate parameters for the
extraction proved challenging due to galaxy crowding near cluster
cores.  One needs to simultaneously be able to deblend neighboring
galaxies while avoiding the dissection of single complexly structured
galaxies into multiple catalog entries.  We found that the two-pass
Cold/Hot method \citep{Rix2004} produced the best catalogs,
particularly near cluster cores.  This method works by aggressively
deblending relatively bright objects (such as cluster core galaxies)
in an initial pass of {\sc SExtractor}, and subsequently filling in
the object catalog with fainter objects in a second pass with the
deblending parameters set less aggressively.  The specific parameters
for the two steps were optimized by trial and error judged by the
successful identification and segmentation of galaxies near cluster
cores.

\subsubsection{PSF construction}\label{subsubsec:PSF}

In order to accurately measure the colors of galaxies we require PSF
estimates for both the $i_{775}$ and $z_{850}$ bands.  To identify
stars in each field with which to construct a PSF we used the {\sc
  SExtractor} FLUX\_RADIUS statistic.  Objects with FLUX\_RADIUS in
the range $[1.4,1.72]$ pixels are likely to be stars.  For fields with
many stars, we selected all stars within $1.3\arcmin$ of the cluster
center and $z_{850}$ MAG\_AUTO between $18$ and $24$.  For more
sparsely populated fields, we selected the $10$ stars closest to the
cluster center falling in the same magnitude range.  For each band of
each field, an initial PSF was constructed by subtracting a local
background (described below) from each star, oversampling each pixel
$9\times9$ times with the IDL procedure CONGRID, renormalizing each
oversampled image by the flux near the stellar core, aligning the
oversampled images to the nearest subpixel and taking the subpixel by
subpixel median image.  This PSF model was then refined by iteratively
fitting it to the original stars and adding the oversampled median
residual back into the model.  The final PSF was trimmed to
$31\times31$ pixels, which contains more than $92$\% of the encircled
energy in both $i_{775}$ and $z_{850}$ bands \citep{Sirianni2005}.

For more shallowly exposed fields, the wings of the PSF derived this
way are relatively noisy and tend to be biased high when compared with
PSFs from more deeply exposed fields.  To reduce this bias, we match
the radial profile of each PSF to the radial profile of the PSF
derived from a reference field, that of cluster RDCS~J1252.9-2927,
which was both exposed more deeply than other fields and contained
many stars.  To measure the radial profiles of each PSF, we fit a 1D
B-spline to the subpixel values as a function of radius.  We then
scale the PSF subpixel values by the ratio of the target field
B-spline and the reference field B-spline.  This procedure produces
PSF grids that are consistent with the encircled energy functions of
\citet{Sirianni2005} to $\sim1\%$.  They also individually retain the
correct azimuthal structure for coadditions of exposures which occured
at many different position angles.  Finally, we note that while a more
precise PSF model would vary across the field of each exposure, ours
is sufficient for correcting the effects of differential blurring
between the $i_{775}$ and $z_{850}$ filters when computing a galaxy
color.

\subsubsection{Galaxy magnitudes and colors}\label{subsubsec:magcolor}

Nonstellar objects with $19 < z_{850} < 26$ (measured with MAG\_AUTO)
and $-1 < i_{775} - z_{850} < 2$ (measured with MAG\_APER with a 10
pixel radius), were then selected as galaxies of interest for further
processing.  For each of these galaxies, postage stamp images were cut
out of the main $z_{850}$ mosaic for processing with GALFIT version
3.0 \citep{Peng2010}.  We used GALFIT to obtain two key statistics:
the $z_{850}$ magnitude, and the half-light radius $R_e$.  We also
used GALFIT to produce an interloper-subtracted image used later when
measuring quantitative morphology parameters.

Initial parameters for GALFIT were chosen using a variant of the
GALAPAGOS algorithm \citep{Haussler2007}.  Galaxies were modeled as
\citet{Sersic1968} profiles with their Sérsic indices constrained
between one and four.  Interloper galaxies near each target galaxy
were either masked out or simultaneously fit depending on the degree
of overlap.  For each galaxy, an adaptively sized elliptical annulus
(typically with $6$\arcsec\ semimajor axis and $\sim 3$\arcsec\ width)
with all galaxies and stars aggressively masked out was used to find
the local sky background level, which was held fixed during the fit.
We provided GALFIT with our estimated $z_{850}$ PSF to internally
convolve with its model before fitting to the actual image.  The
Sérsic profile magnitudes (and magnitude uncertainties) from the
GALFIT fits, corrected for Milky Way extinction with the dust maps of
\citep{Schlegel1998}, are the $z_{850}$ magnitudes and uncertainties
reported throughout the rest of this paper.

Measurements of galaxy color are complicated by the fact that the
$z_{850}$ PSF is $\sim 10\%$ broader than the $i_{775}$ PSF.  To
account for this difference we implemented a PSF matching scheme
where, when measuring color, the $z_{850}$ image is convolved with the
$i_{775}$ PSF and vice-versa.  To fairly treat any color gradients in
galaxies of different angular sizes, we measure color within a
circular aperture with radius equal to the {\it apparent} (as opposed
to intrinsic) half-light radius $R_e$ of the cross-convolved $z_{850}$
image.  Measuring this radius requires us to run GALFIT a second time
on each galaxy, this time using the cross-convolved $z_{850}$ image
and {\it not} providing a PSF for GALFIT to internally convolve with
its model.  A minimum color aperture radius of 3 pixels, corresponding
to $1.2$ kpc at $z \sim 1$, was enforced.

\begin{figure}
\begin{center}
\epsscale{1.175}
\plotone{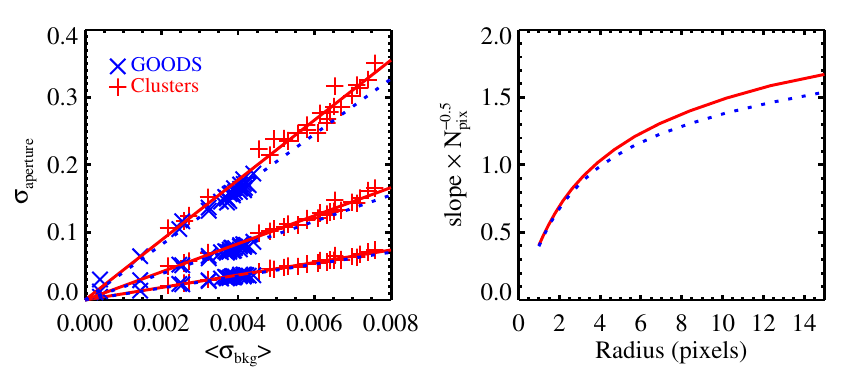}
\end{center}
\caption[ibkgerror]{ \label{fig:ibkgerror} {\bf Left:} Empirically
  measured photometric error contribution from background fluctuations
  in the $i_{775}$ filter, including sky level, readout noise, and
  {\sc MultiDrizzle} artifacts.  The x-axis is the average of
  $\sigma_{\mathrm{bkg}}$ for galaxies in the central region of each
  GOODS tile or cluster field, where $\sigma_{\mathrm{bkg}}$, as
  described in the text, is the standard deviation of background
  pixels (in the pre -- cross-convolved images) surrounding an
  individual galaxy.  The y-axis is the standard deviation of circular
  sky apertures (in the cross-convolved images) of fixed radius in the
  central region of each GOODS tile or cluster field.  The different
  curves from bottom to top are computed for aperture radii of $4.7$,
  $8.4$, and $15.0$ pixels; solid red for cluster fields and dashed
  blue for GOODS tiles.  The empirical relation is slightly different
  for the GOODS tiles compared to the cluster fields, presumably due
  to a different mosaic pattern and subsequent {\sc MultiDrizzle}
  artifacts.  {\bf Right:} The slope of the relation in the left-hand
  panel scaled by $1/\sqrt{N_{pix}}$ as a function of aperture radius.
  The solid red curve indicates cluster fields and dashed blue curve
  indicates GOODS tiles.  For uncorrelated pixels, this relation
  should be horizontal.  The deviation from horizontal confirms that
  the pixels are correlated, justifying our empirical calibration.}
\end{figure}

\begin{figure}
\begin{center}
\epsscale{1.175}
\plotone{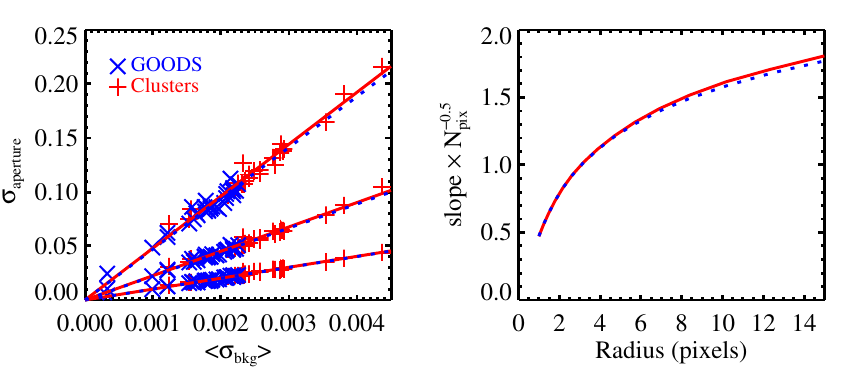}
\end{center}
\caption[zbkgerror]{ \label{fig:zbkgerror} Same as Figure
  \ref{fig:ibkgerror} but for the $z_{850}$ filter. }
\end{figure}

We modeled the errors for each band's flux as the sum in quadrature of
the Poisson error from the object flux and the error in the background
(e.g. sky, CCD readout noise).  Because some of the clusters are not
evenly exposed over the analysis region, we estimated the background
contribution to the error for each galaxy using
$\sigma_{\mathrm{bkg}}$, the local pixel-by-pixel standard deviation
within the same elliptical annulus previously defined in order to
measure the local background level.  If the pixels in the coadded
images were uncorrelated, then the aperture uncertainty
$\sigma_{\mathrm{aperture}}$ would be related to the pixel-by-pixel
uncertainty $\sigma_{\mathrm{bkg}}$ by a factor of the square-root of
the number of pixels in the aperture.  However, since both {\sc
  MultiDrizzle} and cross-convolution introduce correlations between
nearby pixels, we need to empirically calibrate the relation between
$\sigma_{\mathrm{aperture}}$ and $\sigma_{\mathrm{bkg}}$.  We measured
$\sigma_{\mathrm{bkg}}$ (in the unconvolved coadded images) and the
local background subtracted flux (in the cross-convolved coadded
images) of $1000$ sky apertures (selected to avoid galaxies and stars)
in each cluster field and GOODS tile.  For each aperture center we
measured the flux at $15$ different radii logarithmically spaced from
$1$ pixel to $15$ pixels.  For a given radius, the average value of
$\sigma_{\mathrm{bkg}}$ for each field is tightly correlated with the
standard deviation of the $1000$ aperture fluxes for that field
(Figures \ref{fig:ibkgerror}, \ref{fig:zbkgerror}).  The correlation
is slightly different between the cluster fields and the GOODS tiles,
presumably due to different average exposure times, different mosaic
patterns and subsequent {\sc MultiDrizzle} artifacts.  The right-hand
panels of Figures \ref{fig:ibkgerror} and \ref{fig:zbkgerror} show the
deviations from the scalings expected for uncorrelated noise, which
would show up as horizontal lines.

To apply these data to galaxies we interpolate the slope of the
$\sigma_{\mathrm{aperture}}$ -- $\sigma_{\mathrm{bkg}}$ relation from
the data plotted in the right-hand panels of Figures
\ref{fig:ibkgerror} and \ref{fig:zbkgerror} at the measured galaxy
half-light radius $R_e$.  In cases where $R_e$ is larger than $15$
pixels, we extrapolate using a linear fit to the data with radius
between $8$ and $15$ pixels.  This leads to an over-estimate of the
background uncertainties for very bright galaxies.  Fortunately, less
than $6\%$ of the red-sequence galaxies in our sample have $R_e > 15$
pixels, and only $1\%$ have $R_e > 25$ pixels.  In this way the
background error is dependent on each individual object's exposure
depth (quantified by $\sigma_{\mathrm{bkg}}$), even when the exposure
depth varies across the image.  Errors in flux are then converted to
errors in magnitude, and the cataloged color error is the sum in
quadrature of the $i_{775}$ magnitude error and the $z_{850}$
magnitude error.  Exposure depth varies from field to field, but
typical uncertainties in $i_{775} - z_{850}$ (for red galaxies) are
$0.01$ mag at $z_{850}=21$ mag and $0.05$ mag at $z_{850}=24$ mag.

\subsection{Quantitative morphology}\label{subsec:morphology}

We rely on quantitative morphology measurements to identify likely
red-sequence members.  The two parameters we use are the Gini
coefficient \citep{Abraham2003} and asymmetry \citep{Abraham1996}.
The Gini coefficient is a measure of the inequality in the
distribution of pixel fluxes belonging to a galaxy, independent of
their positions.  The minimum Gini coefficient of zero indicates
perfect equality; i.e. the flux in each pixel is identical.  The
maximum Gini coefficient of one indicates maximum inequality, in which
the total flux of all pixels within the chosen aperture is actually
contained in just one pixel (and all other pixels contain zero flux).
The concentrated cores of elliptical galaxies typically generate
higher Gini coefficients for these galaxies than in later-type
galaxies.

The asymmetry measures how poorly a galaxy image matches itself when
rotated 180 degrees about its center.  It is defined as half the sum
of the absolute values of the 180 degree subtraction residual pixels
divided by the sum of the unsubtracted pixels in a given aperture.
Elliptical galaxies typically have smaller asymmetry than later-type
galaxies.

Just as the crowding in cluster cores is an issue when forming {\sc
  SExtractor} catalogs, it is also an issue when selecting apertures
for morphology measurements.  To overcome this difficulty, we
construct interloper-subtracted images by using GALFIT to subtract
models of nearby interfering galaxies for each target galaxy.  For
each interloper-subtracted image, a preliminary aperture is created by
collecting all pixels exceeding $1.5\sigma_{\mathrm{bkg}}$ contiguous
to the pixel at the target galaxy's center.  Using this aperture, we
then compute the galaxy's quasi-Petrosian flux \citep{Abraham2007}.
If the calculation converges, then a new isophotal aperture is created
by selecting pixels from the preliminary aperture which exceed the
quasi-Petrosian flux.  The quasi-Petrosian flux fails to converge for
a small number of galaxies which are then excluded from the subsequent
analysis.  These galaxies are usually quite faint and would have had
large uncertainties had the fits converged.

The Gini coefficient is measured within the isophotal aperture.  Since
the Gini coefficient calculation does not depend on the relative
geometry of the aperture pixels but only on the pixel values, the
error can be estimated from bootstrap resampling \citep{Abraham2003}
(i.e. by repeatedly resampling from the original pixel values, with
replacement, and recomputing the Gini coefficient on these
resamplings.  We recompute the Gini coefficient this way 1000 times to
determine the Gini coefficient probability distribution function for
each galaxy and record the standard deviation of this distribution as
the Gini coefficient error).

We measure the asymmetry within a symmetrized aperture consisting of
the intersection of the isophotal aperture with its 180 degree
rotation.  The center of the rotation is fit iteratively to minimize
the asymmetry (so the symmetrized aperture changes with each
iteration, depending on the current candidate center).  Since the
random fluctuations present in the background contribute some
asymmetry (positive by definition), a correction must be applied.  To
estimate this correction, we generate 1000 Gaussian background images
with standard deviation $\sigma_{\mathrm{bkg}}$ and measure their
asymmetry (but without normalization) in the same symmetrized aperture
as the galaxy.  We can then subtract this background contribution to
the galaxy's asymmetry.  The standard deviation of these background
measurements is our estimated error for each galaxy's asymmetry (note
that galaxy photon noise is much smaller than the background noise).

\subsection{Spectroscopy}\label{subsec:spectroscopy}

As the {\it HST} Cluster SN Survey produced SN candidates, they were
spectroscopically targeted using prescheduled observing time on DEIMOS
on Keck II \citep{Faber2003} and FOCAS on Subaru \citep{Kashikawa2002}
and with ToO requests on FORS1 and FORS2 on Kueyen and Antu at the VLT
\citep{Appenzeller1998}.  The FORS1, FORS2 and DEIMOS observations are
described in \citet{Dawson2009}; the FOCAS observations are described
in \citet{Morokuma2010}.  By observing SN candidates through slitmasks
we were able to simultaneously target likely cluster members and form
spectroscopic catalogs of many galaxies in the cluster fields.  Galaxy
redshifts were found through cross-correlation with template
eigenspectra derived from SDSS spectra \citep{Agol2011}.

The flux and equivalent width of the \OII\ $3727$\AA~emission line
doublet was measured by fitting the simple stellar population (SSP)
templates from \citet[herefacter BC03]{Bruzual2003} assuming a
\citet{Chabrier2003} initial mass function to data on both sides of
the feature to define the continuum.  The flux and equivalent width
were computed by integrating from $500$ km/s blueward of the line
centered at $3726.032$\AA~to $500$ km/s redward of the line centered
at $3728.815$\AA.  To mitigate slit losses in the spectroscopy, we
normalized the spectra to the observed $i_{775}$ photometric
magnitude.

\begin{figure*}[p]
  \centering \subfigure{ \label{fig:SpecPanel:a}
    \includegraphics[scale=0.47]{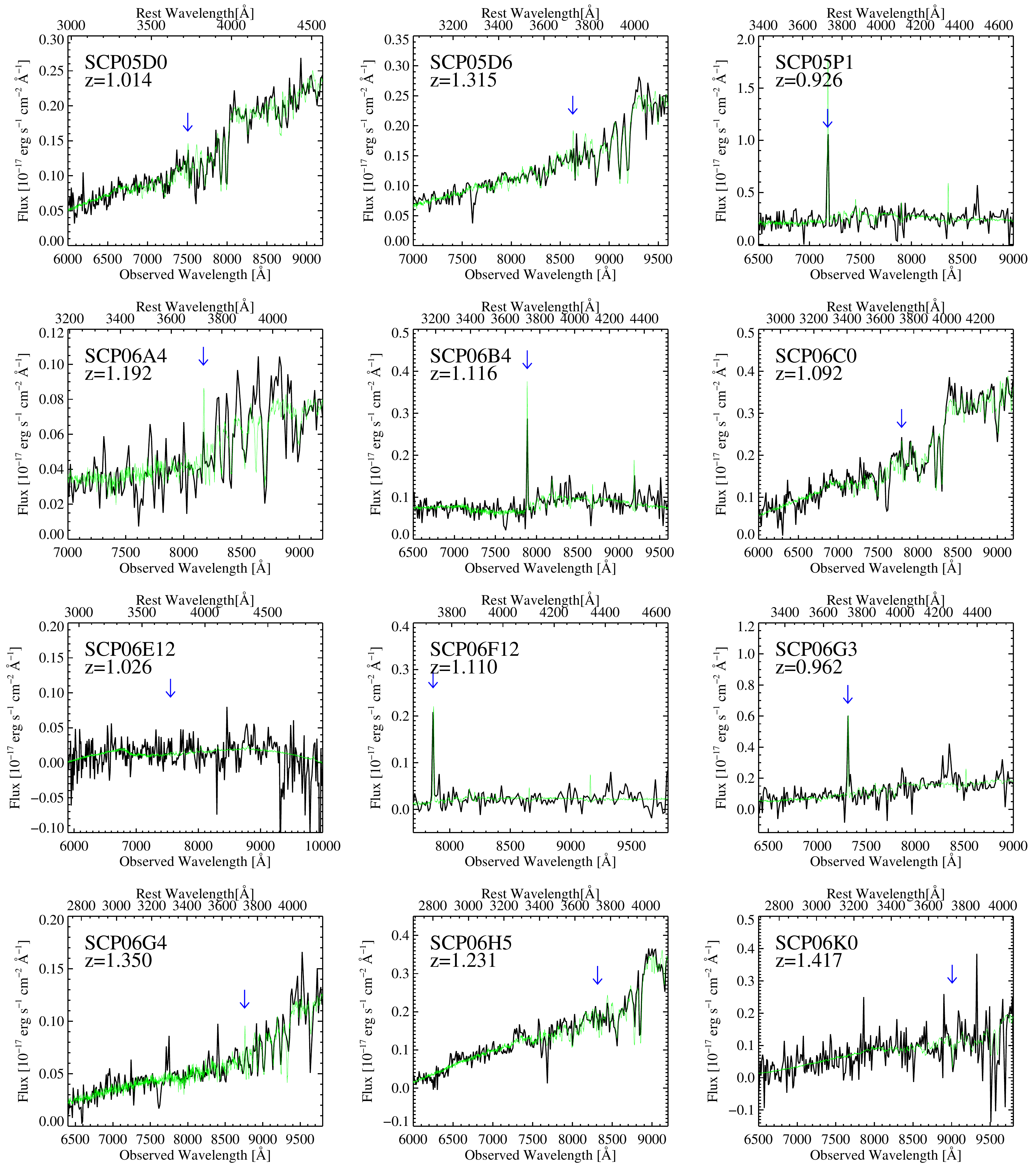}} \caption{SCP
    supernovae host spectroscopy.  For each host, data are in black,
    and the best fitting linear combination of eigenspectra at the
    fitted redshift is in green.  The blue arrow indicates the
    wavelength of the \OII\ 3727 emission line doublet.  Note that the
    green line, used here for determining the redshift, is not the
    same as the BC03 template (which does not contain \OII\ emission)
    used in fitting the \OII\ equivalent
    width.}  \label{fig:SpecPanel}
\end{figure*}

Additional literature redshifts and \OII\ equivalent width
measurements were used as available
\citep{Andreon2008,Bremer2006,Brodwin2006,Demarco2007,Eisenhardt2008,
  Hilton2007,Hilton2009,Postman1998,Rosati1999,Stanford2002,
  Stanford2005}.  Spectra of $z>0.9$ SN hosts from the {\it HST}
Cluster SN Survey are presented in Figure \ref{fig:SpecPanel}.  A few
of these spectra merit individual discussion.  The redshift of
SN~SCP06T1 is determined from a single well-detected emission line,
which we assume to be \OII.  Likewise, the redshift of SN~SCP06X26 is
determined from a single \OII\ emission line, however the detection of
this line is much more tentative than for SN~SCP06T1 and is only
barely visible in the 2D spectrum.  Finally, we have not identified
any features in the spectrum of SN~SCP06E12 that would allow us to
determine a redshift.  The color of this galaxy is consistent with the
color of the red sequence of the targeted cluster in the same field of
view, so it may be a cluster member.  However, we note that the colors
of the four early-type field SN hosts from the {\it HST} Cluster SN
Survey identified later are also roughly consistent with the colors of
the red sequences of the targeted clusters in their fields of view.
Though there is considerable uncertainty in the redshift of
SN~SCP06E12 and its host, we follow through with our analysis of this
galaxy assuming it is a cluster member.  We note that it is not used
for cosmological analyses or analyses of SN correlations with their
hosts.  All of the remaining SN hosts have secure redshifts.

\addtocounter{figure}{-1}
\begin{figure*}
  \addtocounter{subfigure}{1}
  \centering \subfigure{ \label{fig:SpecPanel:b}
    \includegraphics[scale=0.47]{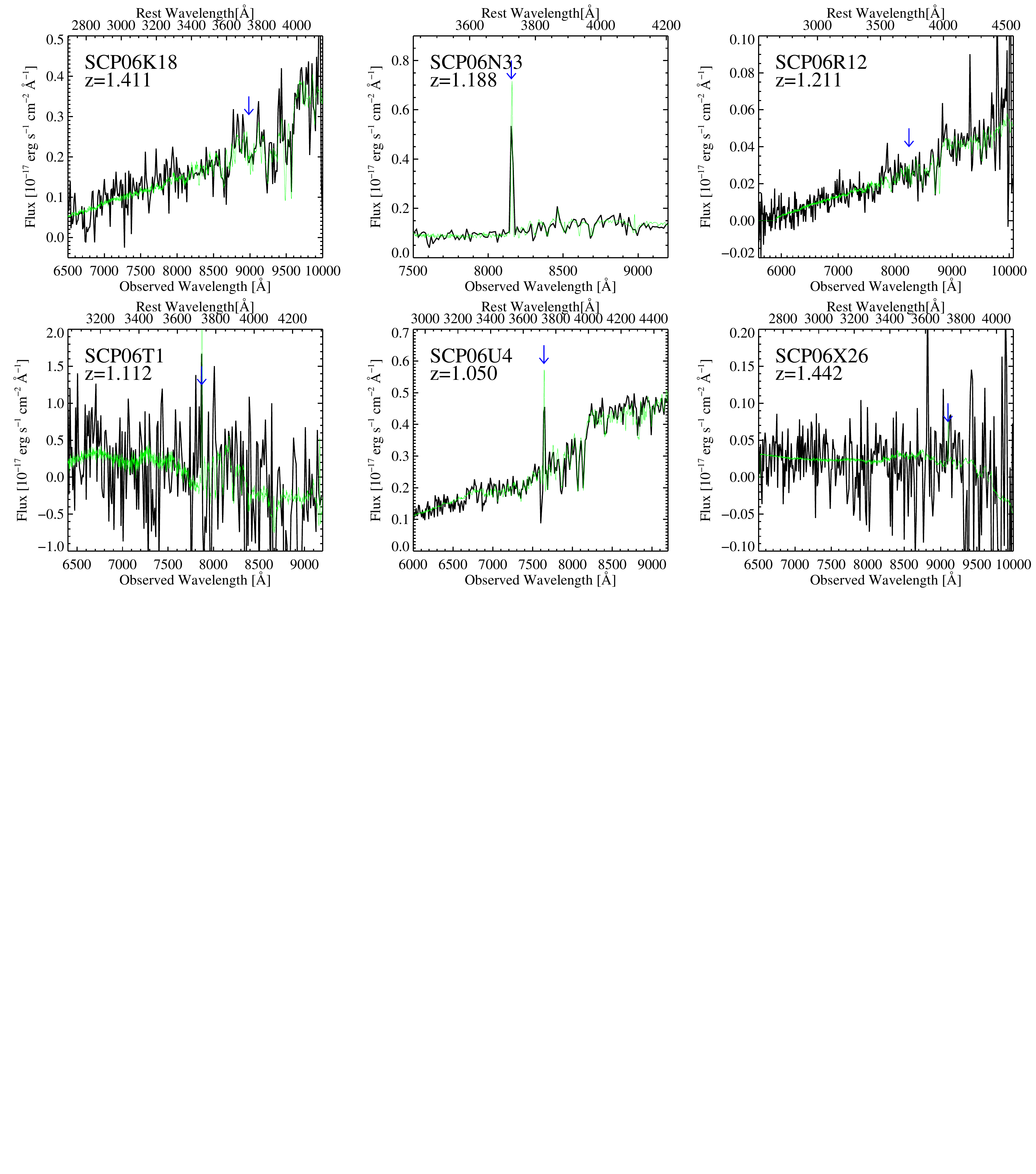}} \caption{
    SCP supernovae host spectroscopy (continued)}
\end{figure*}

\section{Analysis}\label{sec:analysis}
In this section we describe our determination of the red sequence in
clusters and in the field, which we later use to characterize SN host
galaxies.  To consistently identify cluster red-sequence galaxies,
especially in poorer clusters where the contrast between bluer
foreground or late-type galaxies and the cluster red sequence is less
distinct, we generate a composite red sequence using spectroscopically
confirmed members of multiple clusters with colors loosely consistent
with the red sequence.  When there is no cluster red sequence with
which to compare SN host galaxies, as is the case for the field hosted
SNe in the {\it HST} Cluster SN Survey and for all the GOODS fields
SNe, this composite red sequence will serve as the reference red
sequence.  We use an evolving SED model to {\it K}-correct and
evolution correct red-sequence members from their original redshifts
to a target redshift.  In this way we predict the location of the red
sequence across the redshift range of the clusters in the survey.

\subsection{Quantitative morphology calibration}\label{subsec:qmorph}

\begin{figure}
\begin{center}
\epsscale{1.175}
\plotone{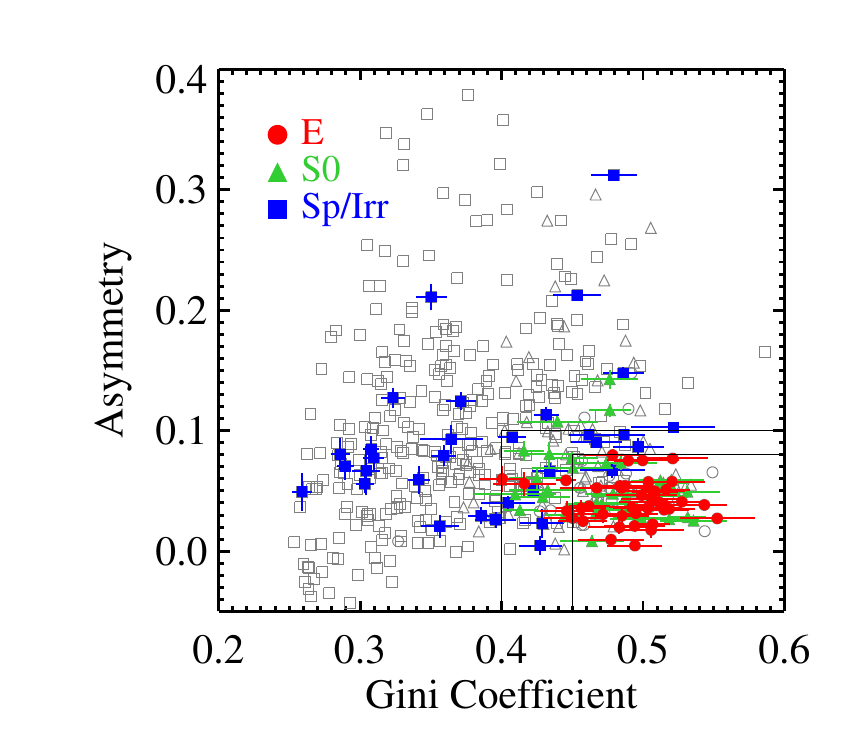}
\end{center}
\caption[vismorphs]{\label{fig:vismorphs} Comparison of visually
  derived morphologies to quantitative morphology measurements for
  three clusters.  Blue boxes represent visually classified late-type
  galaxies with T-type from 0 to 6.  Green triangles represent S0
  galaxies with visual T-type from $-2$ to $-1$.  Red circles
  represent elliptical galaxies with visual T-type from $-6$ to $-3$.
  Colored filled symbols represent galaxies with $i_{775}-z_{850} >
  0.75$, consistent with the red sequence, light grey open symbols
  represent galaxies bluer than the red sequence.  The two boxes in
  the lower-right show the position of our {\it broader} and {\it
  narrower} morphology cuts.  The horizontal and vertical lines
  extending from plotting symbols indicate $1\sigma$ uncertainties.}
\end{figure}

Cluster red sequences are dominated by early-type galaxies.  To
enhance the contrast of the red sequence against the background color
magnitude diagram, we use the Gini coefficient and asymmetry
from \S \ref{subsec:morphology} to select these galaxies.  We
calibrate our selection against visually derived morphology catalogs
(by co-author MP) for three clusters: ISCS~J1432.4+3332,
RDCS1252.9-2927, and ISCS~J1438.1+3414 at redshifts 1.10, 1.24, and
1.41 respectively.  Figure \ref{fig:vismorphs} shows the locus of {\it
red} $(i_{775}-z_{850} > 0.75)$ visually classified elliptical
galaxies compared to other visually classified morphological types in
the Gini coefficient -- asymmetry plane.  The color cut eliminates
many of the bluer galaxies (which we are not interested in here) but
is not so red (for these clusters) that red-sequence members are cut.
To select visually classified ellipticals we can evidently choose
galaxies with Gini coefficient greater than 0.45 and asymmetry less
than 0.08, cuts which we refer to as our {\it narrower} morphological
cuts.  With these cuts, $\sim 90\%$ of all red visually classified
elliptical galaxies are selected.  S0 galaxies make up about 40\% of
the objects selected. The contamination from late-type galaxies is
$\lesssim 5\%$.  Later, when we fit the CMRs of clusters we choose to
use somewhat more lenient cuts of Gini $>0.4$ and asymmetry $<0.1$,
which we refer to as our {\it broader} morphological cuts.  This
increases the number of red-sequence early-type galaxies included by
$\sim 50\%$ but only increases the late-type contamination to $\sim
10\%$.  The {\it broader} cuts further increase the contrast of the
red sequence but, critically, we find they only marginally increase
the scatter of the red sequence (from $0.045 \pm 0.003$ to $0.048 \pm
0.003$ when clusters are stacked; \S \ref{sec:typing}).

\begin{figure}
\begin{center}
\epsscale{1.175}
\plotone{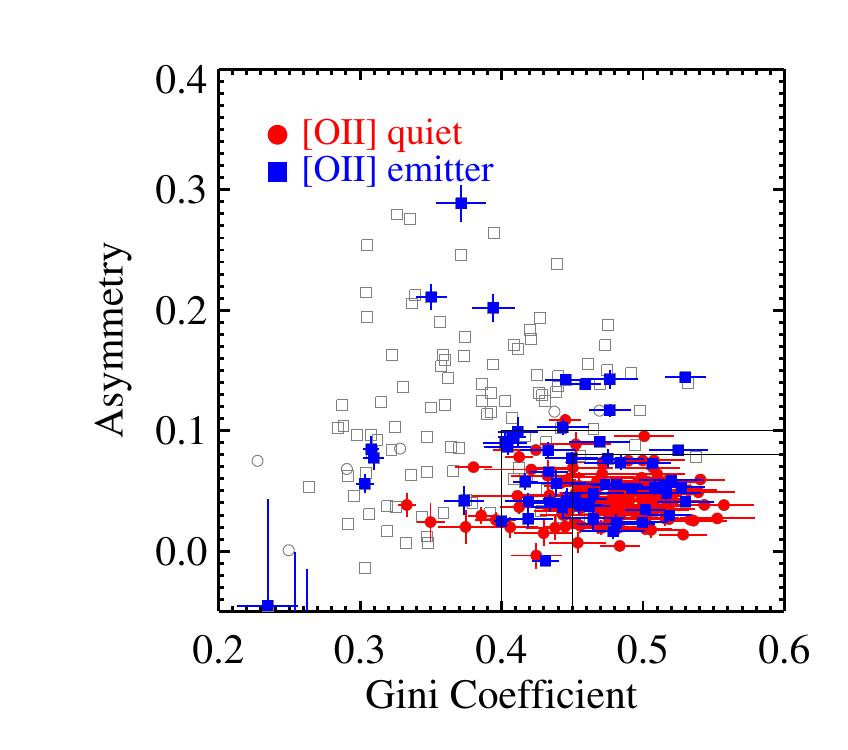}
\end{center}
\caption[o2morphs]{\label{fig:o2morphs} Spectroscopically confirmed
  galaxies' asymmetry and Gini coefficient distributions.  Red circles
  indicate galaxies with an \OII\ equivalent width greater than
  $-5$\AA\ (small or no emission). Blue squares indicate galaxies with
  strong \OII\ emission with an equivalent width less than $-5$\AA.
  Colored filled symbols represent galaxies with $i_{775}-z_{850} >
  0.75$, consistent with the red sequence, light grey open symbols
  represent galaxies bluer than the red sequence.  The two boxes in
  the lower-right show the position of our {\it broader} and {\it
  narrower} morphology cuts. The horizontal and vertical lines
  extending from plotting symbols indicate $1\sigma$ uncertainties. }
\end{figure}

As an independent test of the reliability of our cuts, we also
investigate the strength of \OII\ line emission in the Gini
coefficient -- asymmetry plane.  Figure \ref{fig:o2morphs} shows the
locus of red galaxies with weak or no \OII\ emission (equivalent width
greater than $-5$\AA).  Our broader cuts select $\sim 90\%$ of \OII\
quiet galaxies, though the contamination of \OII\ emitting galaxies is
higher ($\sim 35\%$) than the contamination of late-type galaxies in
the visual morphology case.  This is partly a selection effect:
redshifts are easier to obtain from emission lines than from
absorption lines so \OII\ emitting galaxies are more likely to be
spectroscopically confirmed as cluster members and hence appear in the
plot.  The observed fraction of \OII\ emitting galaxies is compatible
with results reported in \citet{Postman2001}, in which 45\% of
galaxies within the central $1.0 h_{65}^{-1}$ Mpc were observed to
have \OII\ equivalent width less than $-15$\AA.  Also, \citet{Yan2006}
have analyzed $\sim 55000$ SDSS low-redshift spectra and determined
that while nearly 38\% of red galaxies show \OII\ emission, only 9\%
of these show the emission line ratios characteristic of star
formation.  The remaining \OII-emitting galaxies show line ratios
indicative of LINERS or AGN.  \citet{Lemaux2010} have recently
extended this analysis to high-redshift by obtaining near-infrared
spectroscopy of $z>0.8$ cluster galaxies.  Of the five \OII-emitting
red-sequence early-type galaxies in their sample, four were found to
be consistent with AGN/LINERS, with some evidence that the fifth might
contain AGN/LINER activity as well.  We are unable to directly check
for AGN/LINER-like line ratios in our spectra because at high redshift
the necessary comparison lines are shifted out of the wavelength range
of our spectroscopic coverage.  However, we suspect that most of
the \OII-emitting morphologically early-type red galaxies in our
sample are not actively forming stars.

\subsection{The composite red sequence}\label{subsec:compositeRS}

\begin{figure}
\begin{center}
\epsscale{1.175}
\plotone{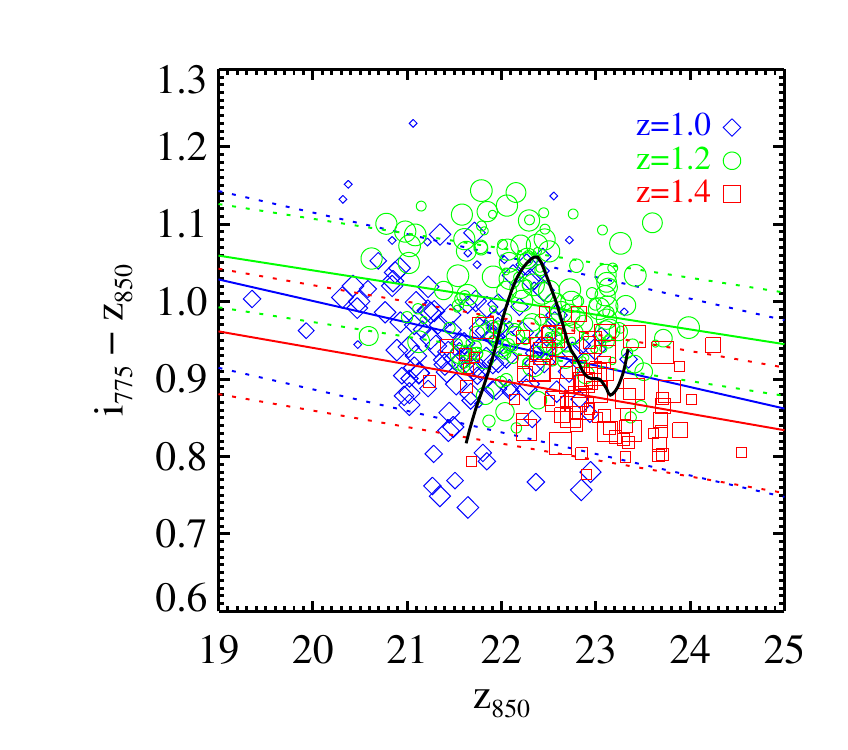}
\end{center}
\caption[Composite Red Sequence]{\label{fig:compositeRS} The
    composite red sequence at redshifts 1.0 (blue diamonds), 1.2
    (green circles), and 1.4 (red squares), with linear fits. The size
    of the plotting symbol is proportional to the weight of the galaxy
    in the linear fit.  The dashed lines indicate the measured
    residual scatter about the fits.  The black line shows the
    predicted evolution from $z=0.9$ to $z=1.5$ of an $L^*$ cluster
    member.  The redward then blueward evolution of the red sequence
    with redshift is the result of the rest-frame $4000$\AA~Balmer
    break crossing the gap between the $i_{775}$ and $z_{850}$
    filters.}
\end{figure}

The identification of red-sequence galaxies in clusters with
incomplete spectroscopy can be challenging, especially for poorer
clusters where it is difficult to distinguish red-sequence members
from interloping galaxies in the foreground or background of the
cluster.  We therefore use the spectroscopically confirmed
red-sequence members from all of our clusters to construct a composite
red sequence which serves as both a guide for fitting the CMR of
individual cluster red sequences and also as a reference red sequence
for field elliptical SN hosts.  We start by assuming that our sample
of spectroscopically confirmed cluster members with quantitative
morphology measurements consistent with elliptical galaxies (passing
our broader morphology cuts) and colors loosely consistent with an old
stellar population are each members of their own cluster's red
sequence.  By using a suitable {\it K}-correction and evolution
correction, we can then project each galaxy's magnitude and color from
their measured values at the galaxy's original redshift to a target
redshift.  To accomplish this we assume that the each sample galaxy's
SED takes the form:
\begin{equation}
  F_\lambda=
  \frac{M_{\mathrm{gal}}}
       {M_\odot}
  \frac{\mathrm{BC03}_\lambda\left(T\left(z\right),Z,z\right)}
       {\left(D_L\left(z\right)\right)^2}
  \label{eqn:kcorrSED}
\end{equation}
where $D_L(z)$ is the luminosity distance to redshift $z$,
$\mathrm{BC03}_\lambda\left(T\left(z\right), Z, z\right)$ indicates
the luminosity density of the BC03 SSP template SED of age
$T\left(z\right)$ and metallicity $Z$ with wavelength redshifted by
$\left(1+z\right)$, and $M_{\mathrm{gal}}$ is the initial stellar mass
of the BC03 template.  We only have two band photometry for galaxies,
so we can only constrain two parameters affecting the SED.  One
parameter needs to be the size of the galaxy ($M_{\mathrm{gal}}$),
leaving the choice of either constraining the galaxy age or
metallicity as the second parameter.  Since these two parameters are
largely degenerate, we simply fix the age $T(z)$ such that galaxies
all form at the same redshift $z_{\mathrm{form}}$ and fit for the
metallicity $Z$.  In a few cases, the range of colors produced by
Equation \ref{eqn:kcorrSED} for the available BC03 metallicities was
not red enough to cover the observed galaxy color.  This could happen
either because of a chance positive fluctuation of the galaxy color or
because the particular galaxy happens to have formed earlier than our
chosen $z_{\mathrm{form}}$.  Since we are not interested in the
metallicity values themselves, but only in using them to project
colors and magnitudes from one redshift to another, we allow this
parameter to be extrapolated outside of the BC03 range.  In practice,
we extrapolated metallicities for 27 out of 198 galaxies to a maximum
of $Z$ = 0.07 (the range of BC03 metallicities is $0.0001$ to $0.05$)
when choosing $z_{\mathrm{form}}=3.0$.  To then project each galaxy to
a target redshift, we evaluate the best-fit BC03 template at age
$T\left(z\right)$ given the target redshift $z$, which automatically
captures luminosity evolution.  We calculate $i_{775}$ and $z_{850}$
from this template.  This procedure is imperfect, and in particular
can introduce errors when {\it K}-correcting and evolution correcting
over large differences in redshift.  To mitigate this effect we
de-weight galaxies with redshifts far from the target redshift.
Specifically, we assign a Gaussian weight
$w \propto \mathrm{exp}(-(z-z_{targ})^2/(2\cdot 0.15^2)$ to each
galaxy.  The characteristic size of the redshift window in this
weighting, 0.15, is arbitrary but we have confirmed that our results
do not significantly change with window sizes ranging from 0.10 to
0.25.  The composite red sequences at redshifts 1.0, 1.2, and 1.4 are
shown in Figure \ref{fig:compositeRS} for $z_{\mathrm{form}}=3.0$.

\subsection{Individual cluster red sequences}\label{subsec:clusterRS}

To determine the red sequences of individual clusters we assume a
model in which galaxies are normally distributed with intrinsic
scatter $\sigma_{\mathrm{int}}$ in color about the mean
color-magnitude relation (characterized by slope $m$ and intercept at
$z_{850}=22$, $b_{22}$) and have their magnitudes distributed like a
\citet{Schechter1976} function (described by parameters $\phi^*$,
$L^*$ and $\alpha$).  Additionally, foreground and background galaxies
will contaminate the color-magnitude diagram.  We model the density of
these galaxies in color-magnitude space as a plane characterized by
its average value (${\rm B}$) and derivatives with respect to
magnitude and color ($d{\rm B}/dz_{850}$ and $d{\rm
B}/d(i_{775}-z_{850})$).  To break degeneracies between the
red-sequence parameters and the foreground/background parameters, we
simultaneously fit each cluster field with the GOODS fields which
serve as a control.

We consider only galaxies that pass our broader morphology cuts and
fall within a redshift dependent color-magnitude rectangle for this
analysis.  The allowed color range is $b_{22}^{\rm comp} \pm 0.3$ mag
where $b_{22}^{\rm comp}$ is the intercept of the composite red
sequence.  The allowed magnitude range is $[M^*_{\rm
model}-2.2,M^*_{\rm model}+1.0]$ where $M^*_{\rm model}$ is determined
by passive evolution (Eqn. \ref{eqn:kcorrSED}) of a red-sequence
characteristic magnitude of $z_{850}=22.7$ at redshift
$z=1.24$ \citep{Blakeslee2003}.  We construct a likelihood function
for these galaxies and the GOODS fields control galaxies falling in
the same color-magnitude rectangle as in \citet{Andreon2005} and use a
Markov chain Monte Carlo algorithm to explore the nine-dimensional
space ($\sigma_{\mathrm{int}}$, $m$, $b_{22}$, $\phi^*$, $L^*$,
$\alpha$, ${\rm B}$, $d{\rm B}/dz_{850}$ and $d{\rm
B}/d(i_{775}-z_{850})$).  The techniques presented in
\citet{Dunkley2005} are used to determine the convergence of the
Markov chains.  The marginal distributions of fitting parameters can
then be extracted directly from the Markov chains.

\begin{figure}
\begin{center}
\epsscale{1.175}
\plotone{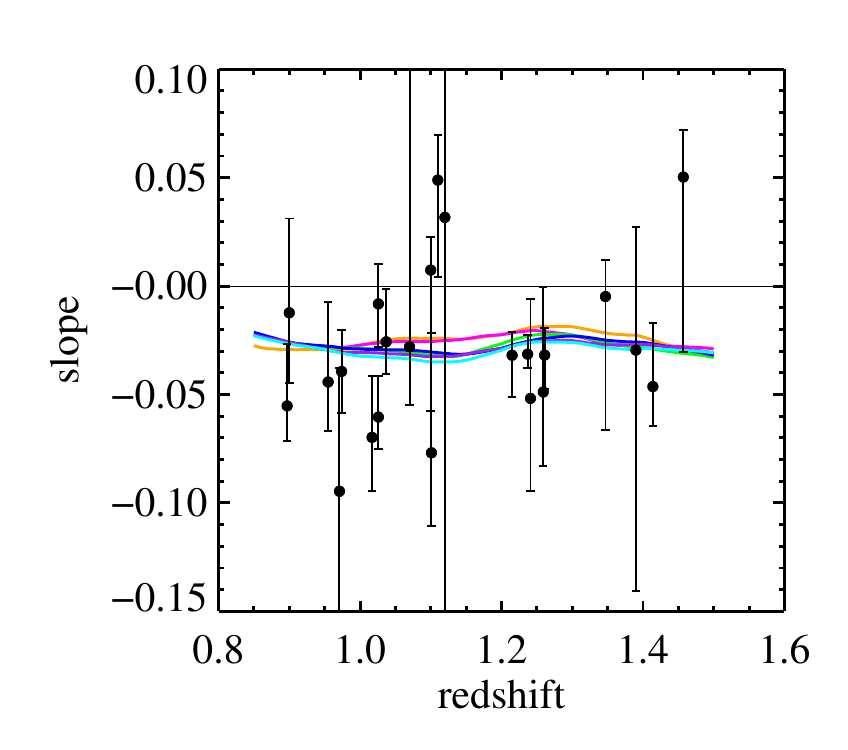}
\end{center}
\caption[Composite Red-Sequence Slope Comparison]
          {\label{fig:compositeComp} Comparison of the slopes of the
            composite red-sequence CMR fits to individual cluster CMR
            fits when allowing the slope to float.  The multicolor
            curves are fits to the composite red sequence assuming
            different values of $z_{\mathrm{form}}$ in the range
            [2.5,7.0].  The black points show the locations of maximum
            likelihood from the MCMC fits to 23 individual clusters.
            The errorbars are the smallest confidence intervals
            containing 68\% of the marginal posterior probability. }
\end{figure}

In Figure \ref{fig:compositeComp}, we compare the slopes of individual
cluster red-sequence CMRs to the slopes of the composite red-sequence
CMRs generated assuming a variety of values of $z_{\mathrm{form}}$
between $2.5$ and $7.0$.  The effects on the composite CMR slope of
changing $z_{\mathrm{form}}$ are much smaller than the statistical
uncertainties in individual CMR slope fits.  In fact, the slopes of
the individual cluster fits are statistically consistent with the
narrow range of slopes of the composite red sequence regardless of
$z_{\mathrm{form}}$ ($\chi^2 = 19$ for $23$ clusters assuming
$z_{\mathrm{form}}=3$ and changes negligibly for other values of
$z_{\mathrm{form}}$).  In order to consistently treat both field and
cluster hosted SNe and also to avoid using physically perplexing
positive slopes, we choose to fix the slope of each cluster to the
value of the composite red sequence.  We choose
$z_{\mathrm{form}}=3.0$ as a typical galaxy formation epoch and refit
each cluster red-sequence CMR.  The results of these fixed-slope fits
are presented in Table \ref{table:RS}.

\begin{deluxetable*}{llrrrccr}

\tablewidth{0pt}
\tabletypesize{\scriptsize}
\tablecaption{\label{table:RS} Color-Magnitude Relation fits }
\tablehead{ID & Cluster Name & Redshift & Intercept & Slope & $\sigma(i_{775}-z_{850})$ & $\sigma(U-V)_{z=0}$ & $N_{\mathrm{gal}}$ }
\startdata
A\phantom{\Large I} & XMMXCS    J2215.9-1738 &           1.457 &   0.923 & $-0.020$ & $0.051_{-0.020}^{+0.034}$ & $0.101_{-0.040}^{+0.068}$ &    23.3 \\
B\phantom{\Large I} &   XMMU    J2205.8-0159 & 1.12\phantom{1} &   1.028 & $-0.023$ & $0.040_{-0.002}^{+0.186}$ & $0.079_{-0.003}^{+0.372}$ &     9.7 \\
C\phantom{\Large I} &   XMMU    J1229.4+0151 &           0.974 &   0.906 & $-0.028$ & $0.066_{-0.014}^{+0.003}$ & $0.131_{-0.028}^{+0.007}$ &    54.8 \\
D\phantom{\Large I} &    RCS  J022144-0321.7 &           1.017 &   0.953 & $-0.026$ & $0.069_{-0.017}^{+0.017}$ & $0.137_{-0.035}^{+0.034}$ &    32.1 \\
E\phantom{\Large I} &  WARPS    J1415.1+3612 &           1.026 &   0.969 & $-0.025$ & $0.060_{-0.013}^{+0.015}$ & $0.119_{-0.026}^{+0.029}$ &    27.0 \\
F\phantom{\Large I} &   ISCS    J1432.4+3332 &           1.110 &   1.031 & $-0.023$ & $0.046_{-0.016}^{+0.016}$ & $0.093_{-0.031}^{+0.032}$ &    18.8 \\
G\phantom{\Large I} &   ISCS    J1429.3+3437 &           1.259 &   0.989 & $-0.019$ & $0.062_{-0.034}^{+0.019}$ & $0.125_{-0.068}^{+0.038}$ &    23.3 \\
H\phantom{\Large I} &   ISCS    J1434.4+3426 &           1.241 &   0.964 & $-0.018$ & $0.061_{-0.013}^{+0.037}$ & $0.122_{-0.026}^{+0.073}$ &    22.5 \\
I\phantom{\Large I} &   ISCS    J1432.6+3436 &           1.347 &   0.969 & $-0.023$ & $0.042_{-0.025}^{+0.014}$ & $0.083_{-0.050}^{+0.029}$ &    14.8 \\
J\phantom{\Large I} &   ISCS    J1434.7+3519 & 1.37\phantom{1} & \nodata &  \nodata &                   \nodata &                   \nodata & \nodata \\
K\phantom{\Large I} &   ISCS    J1438.1+3414 &           1.414 &   0.944 & $-0.021$ & $0.008_{-0.008}^{+0.011}$ & $0.016_{-0.016}^{+0.023}$ &    30.4 \\
L\phantom{\Large I} &   ISCS    J1433.8+3325 & 1.37\phantom{1} & \nodata &  \nodata &                   \nodata &                   \nodata & \nodata \\
M\phantom{\Large I} &     CL      J1604+4304 &           0.897 &   0.749 & $-0.026$ & $0.050_{-0.006}^{+0.019}$ & $0.100_{-0.012}^{+0.038}$ &    27.1 \\
N\phantom{\Large I} &    RCS  J022056-0333.4 &           1.026 &   0.982 & $-0.025$ & $0.043_{-0.004}^{+0.014}$ & $0.086_{-0.008}^{+0.027}$ &    30.4 \\
P\phantom{\Large I} &    RCS  J033750-2844.8 & 1.1\tablenotemark{a}\phantom{1} &   0.988 & $-0.023$ & $0.023_{-0.010}^{+0.026}$ & $0.047_{-0.020}^{+0.051}$ &    15.6 \\
Q\phantom{\Large I} &    RCS  J043934-2904.7 &           0.955 &   0.858 & $-0.027$ & $0.045_{-0.010}^{+0.021}$ & $0.091_{-0.019}^{+0.041}$ &    24.8 \\
R\phantom{\Large I} &   XLSS    J0223.0-0436 &           1.215 &   1.013 & $-0.019$ & $0.015_{-0.015}^{+0.017}$ & $0.029_{-0.029}^{+0.034}$ &    20.6 \\
S\phantom{\Large I} &    RCS  J215641-0448.1 & 1.07\phantom{1} &   1.022 & $-0.024$ & $0.005_{-0.005}^{+0.084}$ & $0.010_{-0.010}^{+0.168}$ &     7.1 \\
T\phantom{\Large I} &   RCS2  J151104+0903.3 &           0.971 &   0.899 & $-0.028$ & $0.087_{-0.023}^{+0.040}$ & $0.174_{-0.045}^{+0.080}$ &    15.7 \\
U\phantom{\Large I} &    RCS  J234526-3632.6 &           1.037 &   0.957 & $-0.025$ & $0.047_{-0.007}^{+0.015}$ & $0.093_{-0.014}^{+0.030}$ &    29.1 \\
V\phantom{\Large I} &    RCS  J231953+0038.0 &           0.900 &   0.766 & $-0.026$ & $0.051_{-0.007}^{+0.027}$ & $0.101_{-0.015}^{+0.054}$ &    30.3 \\
W\phantom{\Large I} &     RX    J0848.9+4452 &           1.261 &   0.985 & $-0.019$ & $0.028_{-0.005}^{+0.013}$ & $0.057_{-0.009}^{+0.026}$ &    20.1 \\
X\phantom{\Large I} &   RDCS      J0910+5422 &           1.101 &   1.001 & $-0.023$ & $0.060_{-0.012}^{+0.024}$ & $0.119_{-0.024}^{+0.047}$ &    18.4 \\
Y\phantom{\Large I} &   RDCS    J1252.9-2927 &           1.237 &   0.965 & $-0.018$ & $0.029_{-0.007}^{+0.003}$ & $0.058_{-0.014}^{+0.007}$ &    34.5 \\
Z\phantom{\Large I} &   XMMU    J2235.3-2557 &           1.390 &   0.926 & $-0.022$ & $0.031_{-0.016}^{+0.017}$ & $0.061_{-0.032}^{+0.033}$ &    17.4
\enddata
\tablenotetext{a}{Photometric cluster redshift.}
\tablecomments{Intercept indicates the CMR color at $z_{850}=22$.  The slope for each fit is fixed to the slope of the composite red sequence at the cluster redshift.  $\sigma(i_{775}-z_{850})$ $(\sigma(U-V)_{z=0})$ is the observer-frame (rest-frame) intrinsic scatter of morphologically selected red-sequence members.  The uncertainties reported indicate the smallest intervals containing 68\% of the posterior probability for $\sigma(i_{775}-z_{850})$ and $\sigma(U-V)_{z=0}$.  $N_{\rm gal}$ is the effective number of red-sequence galaxies in the fitted magnitude range computed by integrating a Schechter function with the best fit values of $\phi^*$, $L^*$, and $\alpha$ over the magnitude range of the fit.}

\end{deluxetable*}

\section{SN host classification}\label{sec:classify}
In this section, we describe how, for each SN host, we use the
color--magnitude, quantitative morphology, and spectroscopic
information to classify the host as either a passively evolving
early-type galaxy, or a potentially star-forming late-type galaxy.  We
restrict our attention to redshifts greater than $0.9$, which is the
lowest cluster redshift in the {\it HST} Cluster SN
Survey.  \citet{Dawson2009} lists 17 SNe with $z>0.9$.  From this list
we subtract SN~SCP06C1, a spectroscopically confirmed SN~Ia at
$z=0.98$ with an uncertain host, but for completeness add SN~SCP06E12
and SN~SCP06X26 for consideration, which are likely $z>0.9$ SNe but
lack conclusive spectroscopic redshifts.  The host of SN~SCP06E12 is
faint but has photometry consistent with the red sequence of the
cluster in the same field of view, and we assume here that its
redshift is that of this cluster: $1.026$.  Spectroscopy of the host
of SN~SCP06X26 shows a possible emission line at $9100$\AA, which
if \OII\ indicates $z=1.44$.  There are 26 $z>0.9$ SNe~Ia discovered
in surveys of the GOODS fields, 19 of which are spectroscopically
confirmed as Type~Ia \citep{Gilliland1999, Blakeslee2003B, Riess2004,
Riess2007}.  We were unable to identify the host galaxy of one of
these: HST04Gre.  This SN is approximately $1.7\arcsec$ from each of
two potential hosts.  Also, the host of SN1997ff has a tentative
redshift much greater than that of any of the clusters considered
here, preventing us from comparing its color and magnitude to the red
sequence.  We do not consider either of these SNe in the following
analysis.

The principle criteria that our hosts must pass to be considered
passively evolving early-types are: (A) they pass our broader
morphology cuts and (B) they lie within $2\sigma$ of the red sequence
where $\sigma =
\sqrt{\sigma_{\mathrm{int}}^2+\sigma_{\mathrm{col}}^2}$, and
$\sigma_{\mathrm{col}}$ is the color uncertainty.  For (B) we use the
cluster red sequence for cluster member hosts, and the composite red
sequence with an assumed intrinsic scatter of $0.10$ mag (greater than
the measured intrinsic scatter of any of our clusters) for field
hosts.  Where spectroscopy is available, we also take note of possible
star formation indicators.

In summary, from the {\it HST} Cluster SN Survey we classify six out
of nine cluster SN hosts and four out of nine field SN hosts as
early-type.  One host galaxy, that of SN~SCP06E12, we were unable to
classify due to its faintness.  The remaining {\it HST} Cluster SN
Survey hosts are classified as late-type.  From the GOODS SN Survey,
we classify seven out of 24 field SN hosts as early-type, and the
remaining 17 hosts as late-type.  Classifications are presented in
Tables \ref{table:EarlySNe} and \ref{table:LateSNe} and details are
below.

\begin{center}
{\it {\bf Early-type Hosts of SCP SNe}}
\end{center}

\begin{center}
{\it \small Cluster members}
\end{center}

{\it SN~SCP06H5, SN~SCP06K18, SN~SCP06R12}.  The hosts of these three
SNe~Ia are spectroscopically consistent with early-type galaxies, have
photometry placing them on their clusters' red sequences, and pass our
narrower morphology cuts.

{\it SN~SCP05D0}.  The host of this SN has photometry placing it on
its cluster's red sequence and passes our broader morphology cuts.
The somewhat large asymmetry of this galaxy appears to result from a
slightly skewed core light distribution, though not from any spiral
structure or clumps of star formation.  A small amount of
\OII\ emission is observed in its spectrum which might indicate star
formation but is also not uncommon in passively evolving galaxies with
LINERs in their cores.

{\it SN~SCP06K0}.  The host of this SN is spectroscopically consistent
with an early-type galaxy, has photometry placing it on its cluster's
red sequence, and passes our broader morphology cuts, though it has a
somewhat small Gini coefficient.  We note that the Gini coefficient is
biased small for low signal-to-noise ratio images \citep{Lotz2004}
(such as the image of SN~SCP06K0's host) and it is likely that with
increased exposure time the Gini coefficient would increase.

{\it SN~SCP06U4}.  The host of this SN has photometry placing it on
its cluster's red sequence and passes our narrower morphology cuts.
Its spectrum exhibits moderate \OII\ emission but otherwise is
consistent with an early-type spectrum.

\begin{center}
{\it \small Field galaxies}
\end{center}

{\it SN~SCP06C0, SN~SCP06D6}.  The hosts of these SNe have photometry
placing them on the composite red sequence and pass our narrower
morphology cuts.  Small amounts of \OII\ emission are observed in
their spectra.

{\it SN~SCP06G4}.  The host of this SN has photometry placing it on
the composite red sequence and passes our narrower morphology cuts.
Its spectrum shows relatively strong hydrogen Balmer absorption
features consistent with an ``E+A'' spectrum
\citep{Dressler1983,Dressler1992}, which may indicate it is younger
than the other early-type galaxies in our sample (though still likely
older than the cutoff timescale for SNe~CC).

{\it SN~SCP06A4}. The host of this SN has photometry placing it on the
composite red sequence and passes our narrower morphology cuts.  A
small amount of \OII\ emission may be present in its spectrum and
hydrogen Balmer absorption lines indicate an ``E+A'' classification.

\begin{center}
{\it {\bf Late-type Hosts of SCP SNe}}
\end{center}

\begin{center}
{\it \small Cluster members}
\end{center}

{\it SN~SCP06B4}.  The host of this SN fails our broader morphology
cuts, is significantly bluer than its cluster's red sequence, and
shows strong \OII\ emission in its spectrum.  It is likely that this
galaxy is star-forming.

{\it SN~SCP06F12}.  Although the host of this SN only barely fails our
broader Gini coefficient cut, its color is significantly bluer than
its cluster's red sequence, and it shows strong \OII\ emission in its
spectrum.

\begin{center}
{\it \small Field galaxies}
\end{center}

{\it SN~SCP05P1, SN~SCP06T1}. The hosts of these SNe fail both of our
broader morphology cuts, are significantly bluer than the composite
red sequence, and show strong \OII\ emission in their spectra.

{\it SN~SCP06N33}.  Although the host of this SN passes our broader
Gini coefficient cut, its color is significantly bluer than the
composite red sequence and it shows strong \OII\ emission in its
spectrum.

{\it SN~SCP06X26}. Although we do not detect strong \OII\ emission
from the host of this SN, it fails our broader Gini coefficient cut
and its color is significantly blue.

{\it SN~SCP06G3}.  Although the host of this SN passes our narrow Gini
coefficient cut, passes (barely) our broader asymmetry cut and is only
moderately bluer than the composite red sequence, its spectrum shows
very strong \OII\ emission.  We also note that the axis-ratio of the
$z_{850}$ GALFIT model for this galaxy ($b/a = 0.25$) is $3.5\sigma$
smaller than the model axis-ratios of the galaxies classified as
early-types above; it is likely an edge-on disk galaxy.

\begin{center}
{\it {\bf Unclassified Hosts of SCP SNe}}
\end{center}

{\it SN~SCP06E12}. We do not have a redshift for the host of this SN.
Its color is consistent with the red sequence of the cluster in the
same field of view.  Its Gini coefficient is small but this galaxy,
like SN~SCP06K0, has a low signal-to-noise ratio image and its Gini
coefficient is likely biased low.  We do not have enough information
to classify this galaxy, and do not include it in any subsequent
analysis in this paper.

\begin{center}
{\it {\bf Host Galaxies of GOODS SNe}}
\end{center}

From the GOODS SN Survey, we analyze the morphology and photometry
(but not spectroscopy) of 24 of the 26 $z>0.9$ SN hosts mentioned in
\citet{Riess2007}.  The hosts of seven SNe pass our morphological and
photometric cuts: SN2003az, SN2003XX, SN2003es, HST04Sas, HST04Tha,
HST05Lan and SN2002hp. \citet{Riess2007} classifies the hosts of all
of these but SN2003az as elliptical.  The remaining 17 GOODS SN~Ia
hosts analyzed here are classified as late-type.

\begin{deluxetable*}{lccccccccc}

\tablewidth{0pt}
\tabletypesize{\scriptsize}
\tablecaption{\label{table:EarlySNe} Early-type $z > 0.9$ SN Hosts}
\tablehead{
\colhead{Name} &
\colhead{$z_{\mathrm{host}}$} &
\colhead{$z_{\mathrm{cluster}}$} &
\colhead{[O~{\sc ii}] EW (\AA)} &
\colhead{Confidence Interval} &
\colhead{Gini Coefficient} &
\colhead{Asymmetry} &
\colhead{$i_{775}-z_{850}$} &
\colhead{CMR residual} &
\colhead{Notes} }
\startdata
\multicolumn{3}{l}{\phantom{0000}{\it SCP Cluster SN Hosts}} \\
 SN~SCP05D0 &  1.014 &   1.017 &   $-1.7$ & [ $ -2.6$, $-0.6$] & $0.525 \pm 0.021$ & $ 0.084 \pm 0.003$ & $0.939 \pm 0.021$ & $-0.016 \pm 0.072$ & s,h,x,c\\
 SN~SCP06H5 &  1.231 &   1.241 &   $-0.4$ & [ $ -1.5$, $-0.0$] & $0.541 \pm 0.017$ & $ 0.059 \pm 0.003$ & $1.075 \pm 0.022$ & $+0.106 \pm 0.065$ & L,h\\
 SN~SCP06K0 &  1.416 &   1.414 &   $-0.4$ & [ $ -2.4$, $-0.0$] & $0.419 \pm 0.018$ & $ 0.027 \pm 0.008$ & $0.951 \pm 0.066$ & $+0.020 \pm 0.066$ & L,h,x,c\\
SN~SCP06K18 &  1.412 &   1.414 &   $-0.1$ & [ $ -0.7$, $-0.0$] & $0.456 \pm 0.015$ & $ 0.021 \pm 0.006$ & $0.986 \pm 0.037$ & $+0.038 \pm 0.038$ & \\
SN~SCP06R12 &  1.212 &   1.215 &   $-0.2$ & [ $ -1.1$, $-0.0$] & $0.455 \pm 0.025$ & $ 0.054 \pm 0.011$ & $0.956 \pm 0.050$ & $-0.019 \pm 0.052$ & L,h\\
 SN~SCP06U4 &  1.050 &   1.037 &   $-9.6$ & [ $-10.9$, $-8.1$] & $0.481 \pm 0.018$ & $ 0.055 \pm 0.004$ & $0.926 \pm 0.017$ & $-0.053 \pm 0.050$ & s,c\\
\multicolumn{3}{l}{\phantom{0000}{\it SCP Field SN Hosts}} \\
 SN~SCP06A4 &  1.193 &   1.457 &   $-1.6$ & [ $ -3.6$, $-0.0$] & $0.468 \pm 0.021$ & $ 0.020 \pm 0.006$ & $0.976 \pm 0.044$ & $-0.013 \pm 0.109$ & L,h,x,c\\
 SN~SCP06C0 &  1.092 &   0.974 &   $-3.6$ & [ $ -5.6$, $-1.7$] & $0.476 \pm 0.021$ & $ 0.027 \pm 0.003$ & $1.072 \pm 0.017$ & $+0.014 \pm 0.101$ & L,h,x,c\\
 SN~SCP05D6 &  1.314 &   1.017 &   $-3.2$ & [ $ -4.2$, $-2.2$] & $0.516 \pm 0.017$ & $ 0.043 \pm 0.005$ & $0.960 \pm 0.027$ & $+0.044 \pm 0.103$ & L,h,x,c\\
 SN~SCP06G4 &  1.350 &   1.259 &   $-0.3$ & [ $ -1.5$, $-0.0$] & $0.492 \pm 0.022$ & $ 0.000 \pm 0.006$ & $1.004 \pm 0.040$ & $+0.103 \pm 0.108$ & s,h,x,c\\
\multicolumn{3}{l}{\phantom{0000}{\it GOODS SN Hosts}} \\
SN2002hp &  1.305 & \nodata &  \nodata &            \nodata & $0.501 \pm 0.025$ & $ 0.032 \pm 0.005$ & $0.919 \pm 0.019$ & $+0.015 \pm 0.102$ & h,x,c\\
SN2003az &  1.270 & \nodata &  \nodata &            \nodata & $0.435 \pm 0.023$ & $ 0.054 \pm 0.008$ & $0.927 \pm 0.048$ & $+0.018 \pm 0.111$ & s,h,x,c\\
SN2003es &  0.954 & \nodata &  \nodata &            \nodata & $0.497 \pm 0.018$ & $ 0.047 \pm 0.001$ & $0.942 \pm 0.006$ & $+0.039 \pm 0.100$ & s,x,c\\
SN2003XX &  0.935 & \nodata &  \nodata &            \nodata & $0.504 \pm 0.028$ & $ 0.025 \pm 0.001$ & $0.789 \pm 0.005$ & $-0.081 \pm 0.100$ & s,h,x,c\\
HST04Sas &  1.390 & \nodata &  \nodata &            \nodata & $0.431 \pm 0.023$ & $ 0.071 \pm 0.007$ & $0.792 \pm 0.032$ & $-0.081 \pm 0.105$ & s,h,x,c\\
HST04Tha &  0.954 & \nodata &  \nodata &            \nodata & $0.432 \pm 0.031$ & $ 0.058 \pm 0.001$ & $0.886 \pm 0.008$ & $+0.027 \pm 0.100$ & s,x,c\\
HST05Lan &  1.235 & \nodata &  \nodata &            \nodata & $0.492 \pm 0.022$ & $ 0.078 \pm 0.004$ & $0.997 \pm 0.016$ & $+0.050 \pm 0.101$ & s,h,x,c
\enddata
\tablecomments{The [O {\sc ii}] EW reported is the median of the posterior probability of the true EW given the observed EW, the spectroscopic uncertainties, and a prior that the EW be positive.  The reported confidence interval is the smallest interval containing 90\% of the posterior probability.  The CMR residual is computed using the fixed-slope fits to individual clusters for cluster member hosts and the composite red sequence fits for field hosts.  The CMR residual uncertainties reported are $\sqrt{\sigma_{\rm int}^2+\sigma_{\rm col}^2}$ where $\sigma_{\rm int}$ is the measured intrinsic scatter of the parent cluster for cluster member hosts and a conservative $0.10$ mag for field hosts.}
\tablenotetext{s}{SN is classified as Type~Ia from spectrum \citep{Blakeslee2003,Riess2004,Riess2007,Barbary2010}}
\tablenotetext{L}{SN is classified as Type~Ia from light curve \citep{Riess2004,Riess2007,Barbary2010}}
\tablenotetext{h}{SN passes Union2.1 cuts and has reliable Hubble residual measurement \citep{Suzuki2011}}
\tablenotetext{x}{SN has SALT2 X1 uncertainty less than 1.0 \citep{Suzuki2011}}
\tablenotetext{c}{SN has SALT2 color uncertainty less than 0.1 \citep{Suzuki2011}}

\end{deluxetable*}

\begin{deluxetable*}{lccccccccc}

\tablewidth{0pt}
\tabletypesize{\scriptsize}
\tablecaption{\label{table:LateSNe} Late-type and Unclassified $z > 0.9$ SN Hosts}
\tablehead{
\colhead{Name} &
\colhead{$z_{\mathrm{host}}$} &
\colhead{$z_{\mathrm{cluster}}$} &
\colhead{[O~{\sc ii}] EW (\AA)} &
\colhead{Confidence Interval} &
\colhead{Gini Coefficient} &
\colhead{Asymmetry} &
\colhead{$i_{775}-z_{850}$} &
\colhead{CMR residual} &
\colhead{Notes} }
\startdata
\multicolumn{3}{l}{\phantom{0000}{\it SCP Cluster SN Hosts}} \\
 SN~SCP06B4 &  1.116 &    1.12 &  $-48.2$ & [$ -60.0$, $-36.0$] & $0.362 \pm 0.010$ & $ 0.144 \pm 0.011$ & $0.281 \pm 0.045$ & $-0.722 \pm 0.060$ & \\
SN~SCP06E12\tablenotemark{a} & 1.026\tablenotemark{b} &   1.026 &  $-26.4$ & [$ -47.4$, $ -0.0$] & $0.358 \pm 0.017$ & $ 0.056 \pm 0.017$ & $0.985 \pm 0.143$ & $+0.077 \pm 0.154$ & L\\
SN~SCP06F12 &  1.110 &   1.110 &  $-80.0$ & [$-103.8$, $-59.2$] & $0.392 \pm 0.022$ & $ 0.031 \pm 0.014$ & $0.424 \pm 0.052$ & $-0.549 \pm 0.070$ & L,h\\
\multicolumn{3}{l}{\phantom{0000}{\it SCP Field SN Hosts}} \\
 SN~SCP06G3 &  0.962 &   1.259 &  $-69.1$ & [$ -75.6$, $-62.4$] & $0.473 \pm 0.018$ & $ 0.087 \pm 0.004$ & $0.689 \pm 0.024$ & $-0.200 \pm 0.103$ & L\\
SN~SCP06N33 &  1.188 &   1.026 &  $-45.5$ & [$ -47.5$, $-43.6$] & $0.426 \pm 0.013$ & $ 0.131 \pm 0.007$ & $0.606 \pm 0.033$ & $-0.402 \pm 0.105$ & L,h,x\\
 SN~SCP05P1 &  0.926 &     1.1\tablenotemark{d} &  $-49.8$ & [$ -53.0$, $-46.3$] & $0.327 \pm 0.006$ & $ 0.200 \pm 0.010$ & $0.239 \pm 0.035$ & $-0.612 \pm 0.106$ & L,x\\
 SN~SCP06T1 &  1.112 &   0.971 &  $-38.1$ & [$ -84.8$,  $-0.0$] & $0.333 \pm 0.021$ & $ 0.119 \pm 0.042$ & $0.167 \pm 0.262$ & $-0.824 \pm 0.280$ & \\
SN~SCP06X26 &   1.44 &   1.101 &  $ -4.0$ & [$ -13.5$,  $-0.0$] & $0.367 \pm 0.020$ & $ 0.024 \pm 0.020$ & $0.448 \pm 0.079$ & $-0.417 \pm 0.128$ & L\\
\multicolumn{3}{l}{\phantom{0000}{\it GOODS SN Hosts}} \\
SN2002dd &  0.950 & \nodata & \nodata &           \nodata & $0.345 \pm 0.016$ & $ 0.114 \pm 0.018$ & $0.187 \pm 0.037$ & $-0.589 \pm 0.106$ & s,h,x,c\\
SN2002fw &  1.300 & \nodata & \nodata &           \nodata & $0.365 \pm 0.012$ & $ 0.113 \pm 0.007$ & $0.647 \pm 0.043$ & $-0.215 \pm 0.109$ & s,h,x,c\\
SN2002fx &  1.400 & \nodata & \nodata &           \nodata & $0.304 \pm 0.018$ & $ 0.024 \pm 0.025$ & $0.278 \pm 0.107$ & $-0.550 \pm 0.147$ & L,h\\
SN2002ki &  1.141 & \nodata & \nodata &           \nodata & $0.376 \pm 0.009$ & $ 0.091 \pm 0.006$ & $0.522 \pm 0.039$ & $-0.513 \pm 0.107$ & s,h,x\\
SN2003eb &  0.900 & \nodata & \nodata &           \nodata & $0.396 \pm 0.011$ & $ 0.150 \pm 0.004$ & $0.267 \pm 0.019$ & $-0.523 \pm 0.102$ & s,x,c\\
SN2003aj &  1.307 & \nodata & \nodata &           \nodata & $0.419 \pm 0.021$ & $ 0.143 \pm 0.007$ & $0.589 \pm 0.031$ & $-0.294 \pm 0.105$ & L,h,x,c\\
SN2003ak &  1.551 & \nodata & \nodata &           \nodata & $0.469 \pm 0.026$ & $ 0.104 \pm 0.004$ & $0.333 \pm 0.026$ & $-0.684 \pm 0.103$ & L,c\\
SN2003dy &  1.340 & \nodata & \nodata &           \nodata & $0.405 \pm 0.012$ & $ 0.279 \pm 0.004$ & $0.492 \pm 0.014$ & $-0.417 \pm 0.101$ & s,h,x,c\\
HST04Eag &  1.019 & \nodata & \nodata &           \nodata & $0.389 \pm 0.008$ & $ 0.200 \pm 0.003$ & $0.370 \pm 0.013$ & $-0.585 \pm 0.101$ & s,h,x,c\\
HST04Mcg &  1.357 & \nodata & \nodata &           \nodata & $0.324 \pm 0.007$ & $ 0.243 \pm 0.009$ & $0.773 \pm 0.042$ & $-0.124 \pm 0.108$ & s,h,x,c\\
HST04Pat &  0.970 & \nodata & \nodata &           \nodata & $0.358 \pm 0.007$ & $ 0.140 \pm 0.007$ & $0.584 \pm 0.025$ & $-0.330 \pm 0.103$ & s,h,c\\
HST04Omb &  0.975 & \nodata & \nodata &           \nodata & $0.423 \pm 0.016$ & $ 0.287 \pm 0.002$ & $0.191 \pm 0.008$ & $-0.704 \pm 0.100$ & s,h,x,c\\
HST05Fer &  1.020 & \nodata & \nodata &           \nodata & $0.477 \pm 0.030$ & $ 0.049 \pm 0.008$ & $0.496 \pm 0.028$ & $-0.410 \pm 0.104$ & s,h,x,c\\
HST05Gab &  1.120 & \nodata & \nodata &           \nodata & $0.252 \pm 0.019$ & $-0.088 \pm 0.048$ & $0.618 \pm 0.187$ & $-0.356 \pm 0.212$ & s,h,x,c\\
HST05Koe &  1.230 & \nodata & \nodata &           \nodata & $0.284 \pm 0.007$ & $ 0.119 \pm 0.011$ & $0.625 \pm 0.117$ & $-0.318 \pm 0.154$ & s,x,c\\
HST05Red &  1.189 & \nodata & \nodata &           \nodata & $0.404 \pm 0.029$ & $ 0.079 \pm 0.004$ & $0.247 \pm 0.014$ & $-0.730 \pm 0.101$ & L,h,x,c\\
HST05Str &  1.027 & \nodata & \nodata &           \nodata & $0.336 \pm 0.006$ & $ 0.121 \pm 0.007$ & $0.496 \pm 0.038$ & $-0.457 \pm 0.107$ & s,h,c
\enddata
\tablecomments{The [O {\sc ii}] EW reported is the median of the posterior probability of the true EW given the observed EW, the spectroscopic uncertainties, and a prior that the EW be positive.  The reported confidence interval is the smallest interval containing 90\% of the posterior probability.  The CMR residual is computed using the fixed-slope fits to individual clusters for cluster member hosts and the composite red sequence fits for field hosts.  The CMR residual uncertainties reported are $\sqrt{\sigma_{\rm int}^2+\sigma_{\rm col}^2}$ where $\sigma_{\rm int}$ is the measured intrinsic scatter of the parent cluster for cluster member hosts and a conservative $0.10$ mag for field hosts.}
\tablenotetext{a}{Unclassified galaxy.}
\tablenotetext{b}{Redshift undetermined; assumed to be a cluster member at $z=1.026$}
\tablenotetext{d}{Photometric cluster redshift.}
\tablenotetext{s}{SN is classified as Type~Ia from spectrum \citep{Blakeslee2003,Riess2004,Riess2007,Barbary2010}}
\tablenotetext{L}{SN is classified as Type~Ia from light curve \citep{Riess2004,Riess2007,Barbary2010}}
\tablenotetext{h}{SN light curve passes Union2.1 cuts \citep{Suzuki2011}}
\tablenotetext{x}{SN has SALT2 X1 uncertainty less than 1.0 \citep{Suzuki2011}}
\tablenotetext{c}{SN has SALT2 color uncertainty less than 0.1 \citep{Suzuki2011}}

\end{deluxetable*}

\section{SN typing by host galaxy}\label{sec:typing}
The classification of the supernovae discovered in the {\it HST}
Cluster SN Survey uses several approaches simultaneously
(see \citet{Barbary2010} for a discussion of the photometric and light
curve constraints).  Here we discuss the typing constraints that come
from just the host galaxy information.  These constraints are already
quite strong.

As described in \S \ref{subsec:typing}, classifying a SN as Type~Ia
when hosted by an early-type galaxy is a robust alternative to
spectroscopic typing.  However, it is natural to wonder if
misclassification of the host galaxy and subsequently the SN is more
frequent at high redshift, where morphological classification is more
difficult and star formation more prevalent.  To investigate this
potential shortcoming, we estimate the {\it a priori} relative rates
of detecting SNe~Ia and SNe~CC in the host galaxies from the {\it HST}
Cluster SN Survey which are classified as early-type
in \S \ref{sec:classify} (we also include the host of SN~SCP06E12 for
analysis in this section, as it is plausibly early-type, although we
leave it as unclassified in \S \ref{sec:classify}).

A rough estimate of the {\it intrinsic} rate of SNe~Ia in an
individual galaxy can be made using the popular A+B SN~Ia rate
parameterization in which the rate is the sum of a term proportional
to the ongoing star-formation rate and a term proportional to the
stellar mass:
\begin{equation}
  \label{eqn:R_Ia}
  R_{SN_{Ia}} = A M + B \dot{M}
\end{equation}
Several choices of $A$ and $B$ are available in the literature.  We
choose the values derived from photometric estimates of SN host masses
and star formation rates in the SNLS: $A=5.3\times10^{-14}{\rm
yr}^{-1}{\rm M_{\odot}}^{-1}$, $B=3.9\times10^{-4}{\rm
M_\odot}^{-1}$ \citep{Sullivan2006}.  The values for $A$ and $B$
depend on which initial mass function (IMF) is assumed.  The values
above are derived assuming a \citet{Kroupa2001} IMF, which is very
similar to the Chabrier IMF with which the BC03 SSPs we have been
using are generated.

The cosmic SN~CC rate has been shown to be proportional to the cosmic
star formation rate.  By dividing the cosmic SN~CC rate found in
\citet{Bazin2009} at $z=0.3$ ($R_{SN_{CC}} = 1.63 \times 10^{-4}
(h_{70}^{-1} \mathrm{Mpc})^{-3}$) by the cosmic star formation rate at
$z=0.3$ found in \citet{Hopkins2006} of $0.03M_\odot \mathrm{yr}^{-1}
\mathrm{Mpc}^{-3}$ (and converting from the modified Salpeter IMF used
there to a Kroupa IMF) we can obtain a similar formula for SNe~CC:
\begin{equation}
  \label{eqn:R_CC}
  R_{SN_{CC}} = C \dot{M}
\end{equation}
with $C=6.52 \times 10^{-3}{\rm M_\odot}^{-1}$.

To estimate the star formation rate in these galaxies we rely on the
the spectroscopic \OII\ luminosity.  As mentioned earlier, while \OII\
may indicate star formation, it may also indicate LINER activity in an
otherwise passive galaxy.  The \OII-inferred star formation rate
should thus be viewed as an upper limit to the true star formation
rate, which may be much less or even zero.  We investigate
the \OII-inferred star formation rate parameterization
of \citet{Kennicutt1998} scaled to a Kroupa IMF:
\begin{equation}
  \label{eqn:sfr}
  \mathrm{SFR}_{\mathrm{[O~\textsc{ii}]}}(M_\odot \mathrm{yr}^{-1})
  = 2.65 \times 10^{-41}
  L_{\mathrm{[O~\textsc{ii}]}}(\mathrm{erg~s}^{-1})
\end{equation}
The \OII-inferred star formation rate is sensitive to galactic
metallicity and dust.  These are in turn correlated with the galaxy's
stellar mass. \citet{Gilbank2010} have investigated the correlation of
the nominal \OII-inferred star formation rate with other more robust
star formation rate indicators in different mass bins and derived an
empirical correction as a function of galaxy stellar mass that
accounts for trends in galaxy metallicity and dust:
\begin{equation}
        \mathrm{SFR}_\mathrm{corr}
        = \frac{\mathrm{SFR}_\mathrm{nom}}{a\; \mathrm{tanh}[(x-b)/c]+d}
\end{equation}
where $\mathrm{SFR}_\mathrm{corr}$ is the corrected star formation
rate, $\mathrm{SFR}_\mathrm{nom}$ is the nominal star formation rate
from Equation \ref{eqn:sfr}, $x = \mathrm{log}(M_*/M_{\odot})$,
$a=-1.424$, $b=9.827$, $c=0.572$, and $d=1.700$.  Since the cluster
early-type SN hosts are quite massive, this correction is significant.
We carry out our analysis both with and without this mass correction.

No \OII\ is detected in the spectroscopy of many of the host galaxies
in the {\it HST} Cluster SN Survey.  Rather than infer that the star
formation is simply zero for these galaxies, we use Bayes' theorem to
derive the probability distribution of the true \OII\ luminosity given
the observed value, taking into account the observational
uncertainties, and enforcing a prior that the \OII\ luminosity must be
positive (i.e. we perform a likelihood analysis).  The median star
formation rates (both nominal and mass-corrected) from the posterior
distributions are reported in Table \ref{table:SNerate}.

To estimate the host galaxy stellar mass, we fit BC03 SSP templates as
in Equation \ref{eqn:kcorrSED} to the observed $i_{775}$ and $z_{850}$
magnitudes, assuming $z_{\rm form} = 3.0$.  The best fitting value of
$M_{\mathrm{gal}}$ is then the {\it initial} stellar mass of the
template, but the {\it current} stellar mass, factoring in the rapid
deaths of massive stars, is also supplied by BC03.  With only two
photometric bands available, we only attempt this mass measurement for
galaxies which are classified as early-types in \S \ref{sec:classify}.
The more complicated SEDs of late-type galaxies will generally require
information from more photometric bands and a more sophisticated
approach to mass measurements.  The statistical uncertainty of this
mass measurement is limited by the uncertainty in our photometric
measurements, both in $z_{850}$ magnitude and in $i_{775}-z_{850}$
color.  To this uncertainty, we also add a systematic uncertainty (in
quadrature) to capture our ignorance of the precise star-formation
histories and metallicities of these galaxies.  By fitting BC03
composite stellar spectra of a variety of exponential and delayed
exponential star-formation histories and various metallicities to the
$i_{775}$ and $z_{850}$ photometry, we estimate that this systematic
uncertainty in mass is around 25\% or $0.1$ dex.  The host galaxy
masses are reported in Table \ref{table:SNerate}.

With the masses and star formation rates derived above, we apply
Eqns. \ref{eqn:R_Ia} and \ref{eqn:R_CC} to estimate the {\it
intrinsic} rates of SNe~Ia and SNe~CC in each host galaxy.  One
interesting consequence of Eqn. \ref{eqn:R_Ia} is that the expected
increase in the SN~Ia rate due to recent star-formation (the B
component) is only $\sim$ 1\% - 5\% for all of our early-type hosts
except that of SN~SCP06U4 for which the increase is $\sim$ 30\%.  The
true enhancement may, in fact, be smaller if the observed \OII\
luminosity contains LINER emission.  This implies that our early-type
hosted SNe~Ia probably have old progenitors and hence are not
associated with dust from recent star formation.

The intrinsic rates are a start, however, since we know {\it a priori}
that the particular galaxies we are analyzing each hosted a {\it
detectable} SN, it is more informative to estimate the {\it apparent}
SN rates.  For example, the apparent rate of a SN of type $X$ given
its intrinsic rate can be written:
\begin{equation}
R_{SN_X}^\mathrm{apparent} = R_{SN_X}^\mathrm{intrinsic}
f_{SN_X}^\mathrm{detectable}
\end{equation}
where $f_{SN_X}^{\rm detectable}$ is the fraction of detectable SNe of
type $X$ for a particular galaxy.  The $z_{850}$ detection threshold
for a point source in a single epoch of the {\it HST} Cluster SN
Survey is about 25.3 \citep{Barbary2010}.  However, since the SNe in
question have passed quality cuts requiring detection in multiple
epochs and multiple bands we adopt an effective $z_{850}$ detection
threshold for SN peak brightness of $\sim 24.7$.  For each host
galaxy, we estimate the fraction of SNe~Ia and SNe~CC with peak
$z_{850}$ magnitude brighter than this detection threshold using the
rest frame {\it R}-band luminosity functions derived from the Lick
Observatory Supernova Search (LOSS) \cite{Li2011}.  Using the maximum
light spectral templates of \citet{Hsiao2007} for normal SNe~Ia
and \citet{Nugent2002} for SN1991bg-like and SN1991T-like SNe~Ia,
SNe~Ibc, SNe~II-N, SNe~II-P, and SNe~II-L we {\it K}-correct our
$z_{850}$ detection threshold to the rest frame {\it R}-band for each
SN subtype.  We estimate the fractions of detectable SNe~Ia and SNe~CC
as the fractions of LOSS SNe~Ia and SNe~CC that exceed these {\it
K}-corrected thresholds:
\begin{equation}
f_{SN_X}^\mathrm{detectable} = N_{SN_X}^>/N_{SN_X}
\end{equation}
where $N_{SN_X}^>$ and $N_{SN_X}$ are the number of SNe of type $X$
brighter than their respective $R$-band threshold and the total number
of SNe of type $X$, respectively.  We find that even at $z=0.9$, none
of the 25 SNe~Ibc and only four (three SNe~II-N and one SN~II-L) out
of 80 SNe~II from LOSS would have been detected in our survey.  On the
other hand, our survey would have discovered 55 out of 74 LOSS SNe~Ia,
or 53 out of 59 if sub-luminous SN1991bg-like and SN2002cx-like SNe~Ia
are excluded.  We again employ Bayes' theorem to derive the
probability distribution of the expected value of $N_{SN_X}^>$ (and
hence the probability distribution of $f_{SN_X}^{\rm detectable}$)
assuming the observed number is Poisson distributed and a positive
flat prior.

With probability distribution functions for the masses (assumed to be
log-normally distributed), \OII\ luminosities, and detectable
fractions of SNe~Ia and SNe~CC for each galaxy, we can use a Monte
Carlo simulation to sample from these distributions and derive a
probability distribution function of the apparent rates of SNe~Ia and
SNe~CC.  Given the apparent rates, the probability that a SN is a
SN~CC is:
\begin{equation}
P(CC) = \frac{R_{SN_{CC}}^{\rm apparent}}{R_{SN_{CC}}^{\rm
apparent}+R_{SN_{Ia}}^{\rm apparent}}
\end{equation}
The results of this analysis are shown in Table \ref{table:SNerate}.
Even under the assumption that \OII\ traces star formation and not
LINER activity, five of eleven SNe have $P(CC) \leq 0.02$, and four
others have $P(CC) \leq 0.15$.  One of the two remaining SNe,
SN~SCP06U4, is a spectroscopically confirmed Type~Ia.  The final
remaining SN, SN~SCP06E12, was already considered uncertain due to its
lack of spectroscopic redshift, though its light curve is consistent
with a Type~Ia at the redshift of the cluster in the same field of
view.  In fact, for almost all of the early-type hosted SCP SNe, the
already strong type constraints derived here from just the host galaxy
information can be supplemented with additional constraints by
considering the full SN light curve shapes, colors and magnitudes.
Ten out of the eleven early-type hosted SCP SNe analyzed in this
section have evidence independent of their hosts or spectra indicating
that they are SNe~Ia \citep[for details, see paper II of this
series:][]{Barbary2010}.  (These SNe are indicated with a ``L'' in the
Notes column of Table \ref{table:SNerate}).  The one remaining SN,
SN~SCP06K18, does not have sufficient early time coverage to constrain
its type through its light curve.  However, as indicated in
Table \ref{table:SNerate}, its massive early-type host and lack
of \OII\ emission strongly suggest it is a Type~Ia.

\begin{deluxetable*}{lccrcccrc}

\tablewidth{0pt}
\tabletypesize{\scriptsize}
\tablecaption{\label{table:SNerate} A Priori Rate Estimates for Early-type SCP SN Host Galaxies}
\tablehead{
  \colhead{Name} &
  \colhead{log(Mass)} &
  \colhead{median SFR} &
  \multicolumn{2}{c}{median $R_{SN_{CC}}$\tablenotemark{a}} &
  \multicolumn{2}{c}{median $R_{SN_{Ia}}$\tablenotemark{a}} &
  \colhead{$P(CC)$} &
  \colhead{Notes} \\
  &
  &
  \colhead{$ \mathrm{M}_\odot \mathrm{yr}^{-1}$ } &
  \multicolumn{2}{c}{$ 10^{-3} \mathrm{yr}^{-1}$ } &
  \multicolumn{2}{c}{$ 10^{-3} \mathrm{yr}^{-1}$ }
}
\startdata
\multicolumn{3}{l}{\phantom{0000}\textit{Nominal Star Formation Rates}} \\
 SN~SCP05D0 &            10.9 &  \phantom{0}0.53 &       0.13 &  (\phantom{00}3.43) &    \phantom{0}2.73 &  (\phantom{0}3.95) &       0.04 &      s,L \\
 SN~SCP05D6 &            11.4 &  \phantom{0}2.56 &       0.23 &  (\phantom{0}16.44) &    \phantom{0}7.82 &            (14.71) &       0.03 &        L \\
 SN~SCP06A4 &            10.7 &  \phantom{0}0.29 &       0.04 &  (\phantom{00}1.85) &    \phantom{0}1.58 &  (\phantom{0}2.53) &       0.02 &        L \\
 SN~SCP06C0 &            11.2 &  \phantom{0}2.20 &       0.42 &  (\phantom{0}14.18) &    \phantom{0}6.60 &  (\phantom{0}9.85) &       0.06 &        L \\
SN~SCP06E12\tablenotemark{b} &  \phantom{0}9.9 &  \phantom{0}0.50 &       0.11 &  (\phantom{00}3.19) &    \phantom{0}0.42 &  (\phantom{0}0.63) &       0.21 &      L \\
 SN~SCP06G4 &            11.2 &  \phantom{0}0.12 &    $<$0.01 &  (\phantom{00}0.75) &    \phantom{0}4.66 &  (\phantom{0}9.02) &    $<$0.01 &      s,L \\
 SN~SCP06H5 &            11.6 &  \phantom{0}0.32 &       0.04 &  (\phantom{00}2.06) &              12.11 &            (20.77) &    $<$0.01 &        L \\
 SN~SCP06K0 &            11.4 &  \phantom{0}0.32 &       0.02 &  (\phantom{00}2.09) &    \phantom{0}4.84 &            (14.08) &    $<$0.01 &        L \\
SN~SCP06K18 &            11.8 &  \phantom{0}0.17 &       0.01 &  (\phantom{00}1.08) &              11.00 &            (32.12) &    $<$0.01 &          \\
SN~SCP06R12 &            10.4 &  \phantom{0}0.02 &    $<$0.01 &  (\phantom{00}0.12) &    \phantom{0}0.78 &  (\phantom{0}1.33) &    $<$0.01 &        L \\
 SN~SCP06U4 &            11.1 &  \phantom{0}7.45 &       1.90 &  (\phantom{0}47.95) &    \phantom{0}6.82 &            (10.22) &       0.22 &      s,L \\
\multicolumn{3}{l}{\phantom{0000}\textit{Mass-Corrected Star Formation Rates}} \\
 SN~SCP05D0 &            10.9 &  \phantom{0}1.93 &       0.47 &  (\phantom{0}12.44) &    \phantom{0}3.11 &  (\phantom{0}4.51) &       0.13 &      s,L \\
 SN~SCP05D6 &            11.4 &  \phantom{0}9.26 &       0.84 &  (\phantom{0}59.57) &    \phantom{0}9.20 &            (17.35) &       0.08 &        L \\
 SN~SCP06A4 &            10.7 &  \phantom{0}1.04 &       0.13 &  (\phantom{00}6.70) &    \phantom{0}1.79 &  (\phantom{0}2.87) &       0.06 &        L \\
 SN~SCP06C0 &            11.2 &  \phantom{0}7.99 &       1.51 &  (\phantom{0}51.37) &    \phantom{0}8.13 &            (12.13) &       0.15 &        L \\
SN~SCP06E12\tablenotemark{b} &  \phantom{0}9.9 &  \phantom{0}1.80 &       0.41 &  (\phantom{0}11.55) &    \phantom{0}0.77 &  (\phantom{0}1.18) &       0.35 &      L \\
 SN~SCP06G4 &            11.2 &  \phantom{0}0.42 &       0.03 &  (\phantom{00}2.71) &    \phantom{0}4.78 &  (\phantom{0}9.27) &    $<$0.01 &      s,L \\
 SN~SCP06H5 &            11.6 &  \phantom{0}1.16 &       0.15 &  (\phantom{00}7.46) &              12.42 &            (21.32) &       0.01 &        L \\
 SN~SCP06K0 &            11.4 &  \phantom{0}1.18 &       0.09 &  (\phantom{00}7.57) &    \phantom{0}5.09 &            (14.86) &       0.02 &        L \\
SN~SCP06K18 &            11.8 &  \phantom{0}0.61 &       0.05 &  (\phantom{00}3.92) &              11.14 &            (32.58) &    $<$0.01 &          \\
SN~SCP06R12 &            10.4 &  \phantom{0}0.07 &    $<$0.01 &  (\phantom{00}0.44) &    \phantom{0}0.81 &  (\phantom{0}1.38) &       0.01 &        L \\
 SN~SCP06U4 &            11.1 &            27.01 &       6.87 &            (173.73) &              11.94 &            (17.90) &       0.36 &      s,L
\enddata
\tablenotetext{a}{Rates in parentheses are \textit{intrinsic} quantities.  Rates not in parentheses are \textit{apparent} and factor in SN detectability.}
\tablenotetext{b}{Redshift undetermined; assumed to be a cluster member at $z=1.026$.  This galaxy is left unclassified in \S6.2.}
\tablenotetext{s}{Spectroscopically confirmed SN~Ia.}
\tablenotetext{L}{Light curve shape, color and magnitude consistent with Type~Ia \citep{Barbary2010}.}

\end{deluxetable*}

\section{Early-type host dust constraints}\label{sec:dust}
In this section we use measurements of the red sequence scatter to
place constraints on the scatter of reddening affecting these
galaxies.  Under the assumption that at least some red-sequence
galaxies have very low dust content, this scatter constraint also sets
an absolute scale for dust.

\subsection{Red-sequence scatter}\label{subsec:rsdust}

\begin{figure*}
\begin{center}
\epsscale{1.175}
\plotone{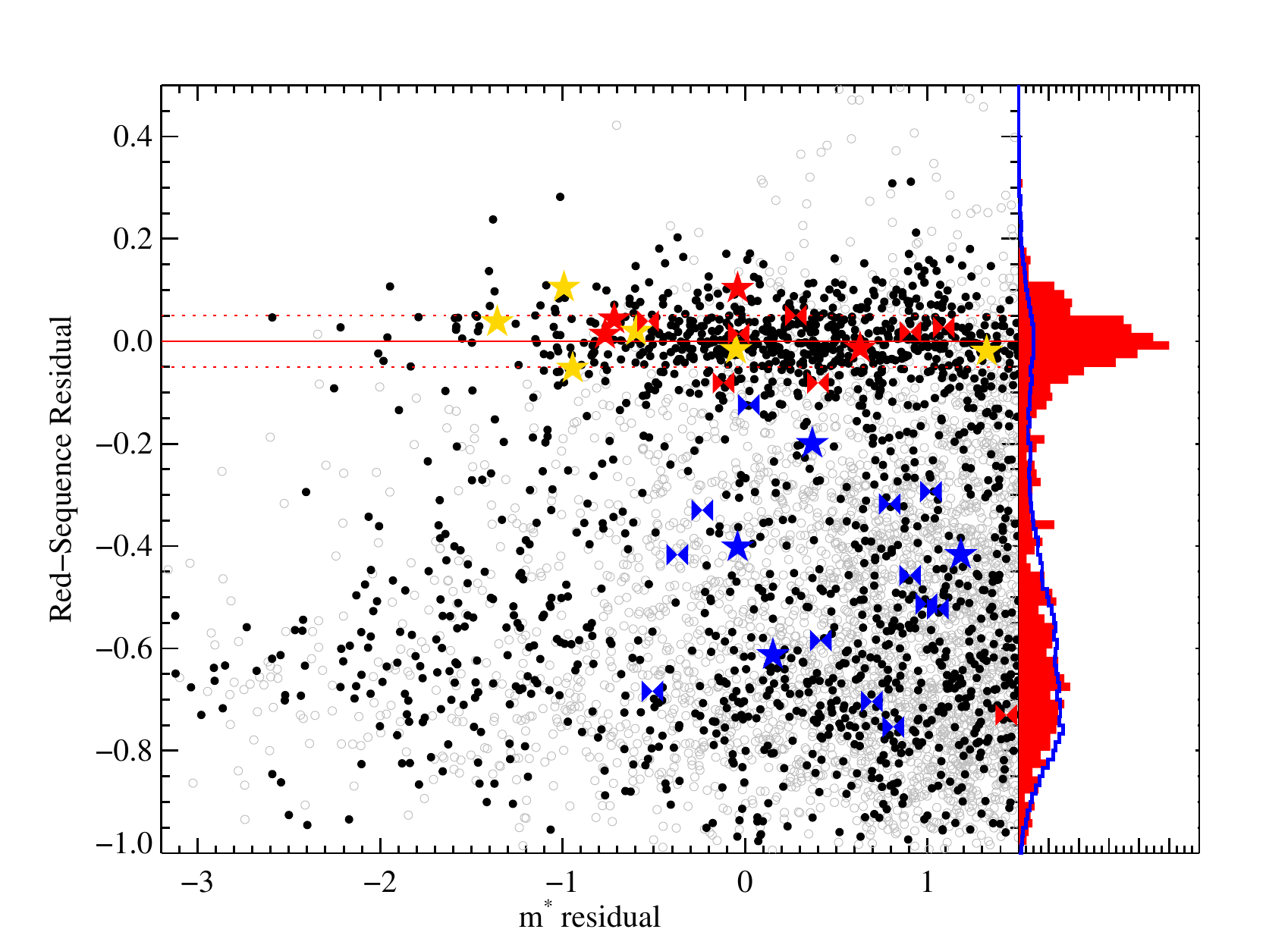}
\end{center}
\caption[StackCMD]{\label{fig:StackCMD} Stacked ``Color-Magnitude''
  Diagram.  The x-axis is galaxy $z_{850} - m^*$ (where the
  characteristic magnitude, $m^*$, is modeled to passively evolve with
  redshift as described in the text) while the y-axis is galaxy
  color-magnitude relation residual: $(i_{775}-z_{850})
  - \mathrm{CMR}$.  Open grey circles indicate galaxies which do not
  pass our broader morphological cuts; solid black circles do.  The
  solid red line indicates zero CMR residual.  The dashed red lines
  indicate the $1\sigma$ intrinsic scatter of the stacked red sequence
  CMR.  The red histogram on the right is the projection of CMR
  residuals for the stacked cluster galaxies.  The blue histogram is
  the projection of CMR residuals for the GOODS control fields.  The
  $z>0.9$ SN~Ia hosts brighter than $m^*+1.5$ are overplotted as stars
  (HST Cluster SN Survey) or bowties (GOODS field SN surveys).  Red
  (gold) stars and bowties indicate field (cluster) SN~Ia hosts that
  meet our morphological and spectroscopic early-type criteria as
  described in \S \ref{sec:classify}.  Blue stars and bowties indicate
  field or cluster SN~Ia hosts that do not meet these criteria.  Both
  the cluster and field early-type SN~Ia hosts are consistent with the
  red sequence and may have similar properties such as low dust
  content and old stellar populations.}
\end{figure*}

Color-magnitude diagrams of each of our clusters are presented in
Figure \ref{fig:CMDpanel} in Appendix A.  For most clusters, the red
sequence is clearly visible as an overdensity of morphologically
early-type galaxies with $i_{775}-z_{850} \approx 1$.  For 23 of the
25 clusters, the marginal likelihood distributions for the cluster CMR
intrinsic scatters show clear maxima between 0.0 and 0.1 magnitudes.
The intrinsic scatter marginal likelihood distributions for clusters
ISCS~J1434.7+3519 and ISCS~J1433.8+3325 are essentially flat from 0.0
up to 0.3 magnitudes (the largest scatter for which we sampled the
posterior likelihood) because the clusters contain too few red
early-type galaxies to constrain the fit.  Using the CMR fits as a
baseline with which to combine clusters, we find that the intrinsic
scatter of the stacked ``color-magnitude'' diagram is $0.046$ mag
(Figure \ref{fig:StackCMD}).  Turning to individual clusters, we find
that the best-fit intrinsic scatters of three clusters are consistent
with zero. The best-fit intrinsic scatters of the remaining clusters
range from $0.023$ mag to $0.087$ mag (see Table \ref{table:RS} for
the best-fit values and uncertainties of each cluster's intrinsic
scatter).

To compare these observer-frame CMR scatter measurements to published
scatter measurements at other redshifts, we first convert to the
rest-frame $\sigma(U-V)_{z=0}$ scatter.  Using BC03 SSPs, we construct
a library of mock galaxy spectra with a variety of formation
redshifts, exponentially declining star-formation timescales, and
metallicities.  For each cluster redshift, we determine the relation
between observer-frame $i_{775}-z_{850}$ color and rest-frame $U-V$
color through synthetic photometry of these library spectra.  The
slope of this relation evaluated at the observed CMR color ranges from
$\sim 1.9$ to $\sim 3.2$ depending on the redshift; this is the
desired multiplicative conversion factor from
$\sigma(i_{775}-z_{850})$ to $\sigma(U-V)_{z=0}$.  Our rest-frame
$\sigma(U-V)_{z=0}$ estimates lie in the range $\sim 0.02 - 0.19$ mag
with a typical value of $\sim 0.10$ mag and are listed in
Table \ref{table:RS}.  This range and typical value are consistent
with the compilation of rest-frame cluster scatters presented
in \citet{Jaffe2011}, which span the redshift range $0.0 < z < 1.46$.
In the rest of this section we use the properties of cluster red
sequences to constrain the amounts of dust present in cluster
galaxies.

A precision upper limit on dust along the line of sight to each SN~Ia
seen in the early-type hosts analyzed here is not possible with this
particular data set (this rest-frame color, in particular, is not as
sensitive to dust as a broader color baseline would be); nonetheless,
it is useful to outline a rough analysis to see that dust limits on
these $z>1$ cluster-hosted SNe are consistent with the low dust values
expected for red and dead hosts as outlined
in \S \ref{subsec:nearbyETGdust}.

We can estimate such limits using the expectation that dust will both
redden the intercept of the CMR and broaden the scatter of CMR
residuals.  While it is conceivable that one could construct dust
distributions such that blue star-forming galaxies are reddened to
\emph{precisely} fall on the CMR (and subsequently decrease the CMR
scatter), we take this to be an unlikely scenario requiring fine
tuning.  In fact, since the effects of dust and age show different
redshift evolution (see Figure \ref{fig:scatterz}), the amount of dust
required in such a scenario would need to be fine-tuned to the
specific cluster redshift.  Furthermore, the typical timescale for
dust destruction, $\sim 10^7-10^8$\ yr \citep{Jones1996, Temi2007,
Clemens2010}, is significantly shorter than the typical ages of even
the youngest $z \sim 1.2$ red-sequence ellipticals ($\sim 10^9$\
yr)\citep{Rettura2010}, so we do not expect a difference in dust
between blue and red extremes of the red sequence to play a role in
the CMR residual scatter.  Finally, a dust distribution in which each
galaxy is reddened the same amount is a similarly unlikely fine-tuning
scenario, and is inconsistent with findings in nearby ellipticals, in
which infrared-inferred dust masses vary by several orders of
magnitude \citep{Tran2001,Temi2004,Temi2007}.

\subsection{Dust scatter}\label{subsec:reldust}

We begin by considering the dust constraints obtainable from the CMR
scatter and then return below to the constraints from the CMR
intercept.  Although both age and metallicity affect the
unextinguished colors of galaxies, following the
literature \citep{Bower1992, vanDokkum1998, Blakeslee2003, Mei2009,
Jaffe2011}, we assume that age is the dominant variable affecting the
unextinguished CMR residual scatter (and that metallicity is primarily
responsible for the CMR slope).  We therefore assume a constant
metallicity model in which differences in age and differences in dust
reddening create the observed CMR scatter.

\begin{figure}
\begin{center}
\epsscale{1.175}
\plotone{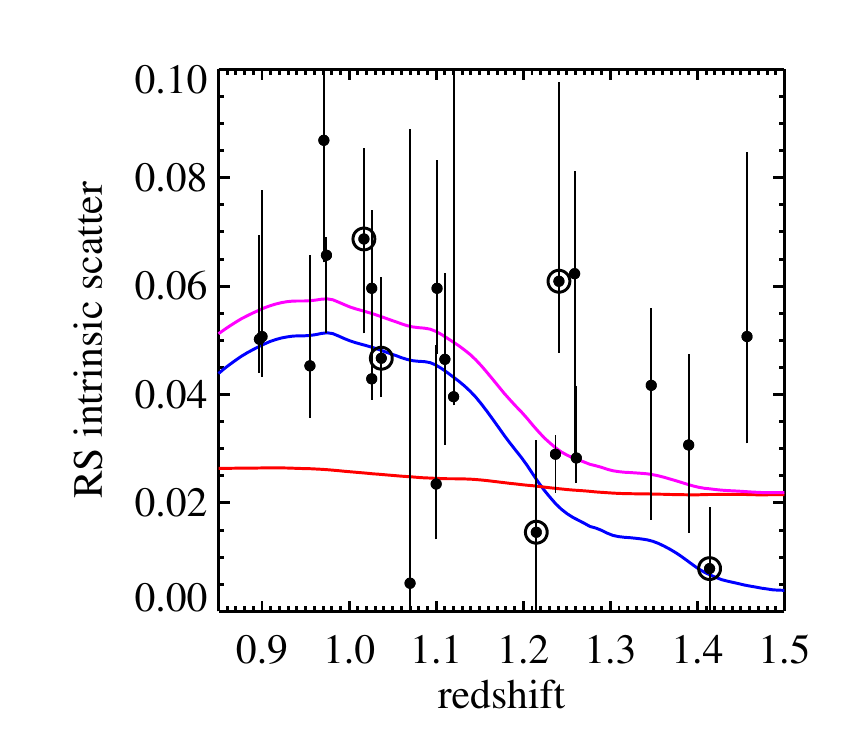}
\end{center}
\caption[scatterz]{\label{fig:scatterz} Measured early-type
  red-sequence scatter and fit assuming that CMR scatter is produced
  by scatter in galaxy ages and dust. The data points and error bars
  are our measurements and confidence intervals for the 23 clusters
  with scatter constraints.  The blue line indicates the scatter
  attributable to age differences in our model, the red line indicates
  the scatter attributable to dust, and the magenta line indicates the
  combined scatter that is fit to the data.  Clusters which hosted
  SNe~Ia in early-type galaxies are circled.  The steep evolution of
  the age-related scatter at $z\sim1.15$ is due to the shifting of the
  4000\AA\ Balmer break through the $z_{850}$ filter.  The SEDs of
  passive galaxies of different ages have smaller $i_{775}-z_{850}$
  color differences above this redshift because the filters no longer
  bracket the break.  In contrast, the smooth dust extinction law
  produces scatter roughly independent of redshift.}
\end{figure}

As mentioned above, the observer-frame CMR scatter
$\sigma(i_{775}-z_{850})$ attributable to differences in galaxy ages
evolves with redshift (Figure \ref{fig:scatterz}) due to the shifting
overlap of filters with spectral features.  In particular, above
$z \sim 1.2$, the $i_{775}$ and $z_{850}$ filters no longer bracket
the redshifted age-sensitive 4000\AA\ Balmer break and the predicted
scatter for a given age distribution consequently sharply decreases.
On the other hand, because the \citet{Cardelli1989} reddening law is
smooth over the relevant wavelengths, the scatter attributable to a
given distribution of dust is nearly constant with redshift.  This
complementary behavior permits us to simultaneously fit both the
average cluster star-formation history and average cluster dust
distribution using the scatter -- redshift data gathered
in \S \ref{subsec:clusterRS}.

We investigate the age-scatter model first described
by \citet{vanDokkum1998} in which early-type galaxies are formed in
delta function starbursts at times uniformly distributed in the
interval ($t(z_0), t_z-t_{\rm delay}$). Here $t(z_0)$ indicates the
epoch when cluster galaxies first form at redshift $z_0$, $t_z$
indicates the epoch at which the cluster is observed, and $t_{\rm
delay}$ corrects for {\it progenitor bias} \citep{vanDokkum2001} by
allowing time for galaxies which recently ceased star formation to
become red and evolve morphologically so that they will be identified
as red-sequence early-types. At any redshift, the scatter due to age
variation can be computed from a population of synthetic BC03 galaxies
generated with ages drawn from the distribution defined by $z_0$ and
$t_{\rm delay}$.  The amplitude of the effects of age variation may in
general depend on the assumed metallicity.  To test this dependency,
we have carried out the present analysis for both $Z=0.02$ and
$Z=0.05$, which are appropriate for the masses of early-type galaxies
studied here \citep{Trager2000b}.  The final results are essentially
the same with either choice of metallicity.  To additionally account
for the effects of dust, we use the parameter $\sigma(E(B-V)_{z=0})$,
the rest-frame scatter of $B-V$ reddening.  This is related to
$\sigma(E(i_{775}-z_{850}))$ by a redshift dependent multiplier, which
we compute through synthetic photometry of typical early-type galaxy
SEDs with and without dust, similar to the analysis above in which we
compare our scatter measurements to the literature.  The measured
cluster scatters are fit with three parameters -- $z_0$, $t_{\rm
delay}$, and $\sigma(E(B-V)_{z=0})$ -- using a Markov Chain Monte
Carlo.  We find that the value of $z_0$ has little effect on the model
likelihood within the range $4.0 < z_0 < 10.0$.  The most likely value
for $t_{\rm delay}$ is $2.0$ Gyr, which corresponds nicely to youngest
spectroscopically measured red-sequence galaxy age in cluster
RDCS~J1252-2927 of $1.9 \pm 0.5$ Gyr \citep{Rettura2010}.  The maximum
likelihood value for $\sigma(E(B-V)_{z=0})$ is 0.031 mag, with 68\%
and 95\% upper limits of 0.035 mag and 0.042 mag, respectively.

If clusters with higher intrinsic CMR residual scatter at a given
redshift (such as ISCS~J1434.4+3426 which hosts SN~SCP06H5) in fact
contain dustier galaxies, then their CMR intercepts should also be
redder.  We have further investigated this possibility by searching
for a correlation between $\Delta \sigma(i_{775}-z_{850})$ and $\Delta
b_{22}$ using pairs of clusters with $|\Delta z| < 0.02$.  We find
that the pair members with higher scatter are as likely to have bluer
intercepts as redder intercepts.  Thus the difference between the
measured intrinsic scatter and the age-attributed intrinsic scatter
should be interpreted only as an upper limit to the dust content of a
cluster, and not as a dust detection.

\subsection{Absolute scale for dust}\label{subsec:absdust}

So far we have estimated approximately how much CMR residual {\it
scatter} is attributable to dust.  However, for SN measurements we are
principally interested in the corresponding amount of {\it reddening}
or {\it extinction}.  To convert from residual scatter to reddening,
we consider a toy model for dust.  In a mid-infrared study of the
cluster RXJ1716.4+6708, \citet{Koyama2008} found that only $\sim 10$\%
of red cluster galaxies showed emission at $15\mu$m from dust
(compared to $\sim 50$\% of blue cluster galaxies).  As discussed
in \S \ref{subsec:nearbyETGdust}, studies of nearby early-type
galaxies also suggest that at least some have very small levels of
extinction.  We therefore assume that most cluster early-type galaxies
are unaffected by dust and that the distribution of dust in the
remaining galaxies follows a half-Gaussian (we have also considered
exponential, tophat and Dirac delta function distributions for the
remaining galaxies; the differences are small).  This highly
simplified model captures the main features we wish to study here: a
fraction of dusty galaxies (and hence a remaining fraction of nearly
dust-free galaxies) and a characteristic scale for the amount of dust.
The full probability distribution function for our toy dust model is:
\begin{equation}
\small
f(X;f_d,\mu_d)=\left(1-f_d\right)\delta\left(X\right) +
f_d\frac{2\Theta\left(X\right)}{\pi\mu_d}
\mathrm{exp}\left(\frac{-X^2}{\pi\mu_d^2}\right)
\label{eqn:dustfmu}
\end{equation}
where the parameters $f_d$ and $\mu_d$ specify the fraction of dusty
galaxies and the mean dust content of the subset of dusty galaxies,
respectively; $\delta(\cdot)$ is the Dirac delta function, which
represents the dust-free galaxies (this term ensures that the integral
over the full distribution is one), $\Theta(\cdot)$ is the Heaviside
step function, and $X = E(B-V)_{z=0}$ quantifies the rest-frame galaxy
reddening due to dust.

\begin{figure}
\begin{center}
\epsscale{1.175}
\plotone{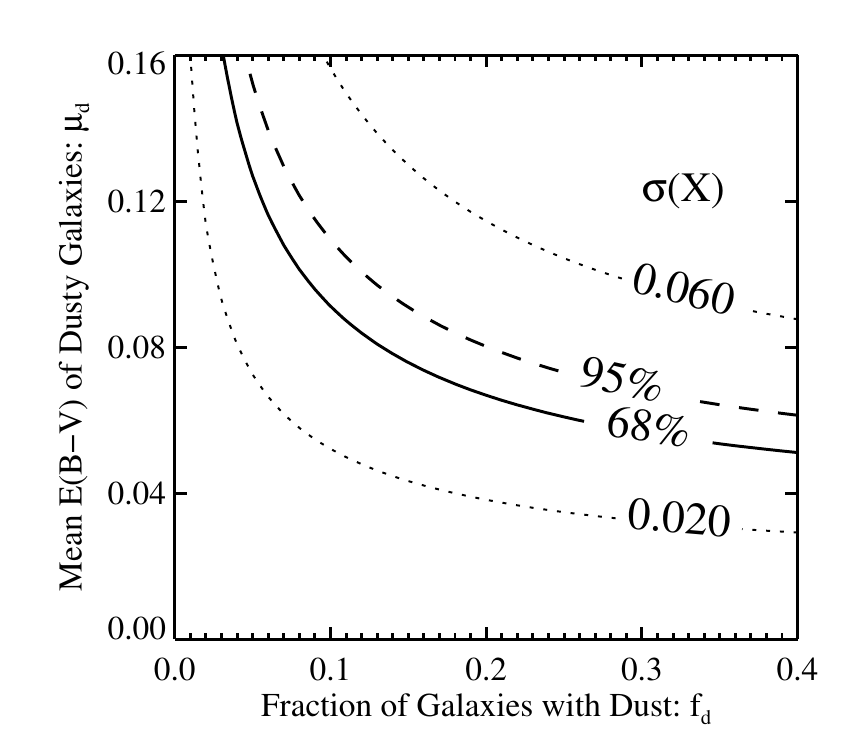}
\end{center}
\caption[dustvar]{\label{fig:dustvar} Contours of constant CMR
  residual scatter attributable to dust, $\sigma(X)$, in our toy
  model, in magnitudes.  The model parameters $f_d$ and $\mu_d$
  specify the fraction of dusty galaxies and the mean dust extinction
  of the subset of dusty galaxies, respectively.  The solid and dashed
  contours indicate the $1\sigma$ and $2\sigma$ upper limits on
  $\sigma(X)$ respectively.  The $\mu_d$ -- $f_d$ parameter space
  above these curves are excluded at the indicated level, implying
  that either the amount of dust in an individual cluster member is
  low ($\mu_d$ is small), or that the probability that an individual
  cluster member is dusty is small ($f_d$ is small).}
\end{figure}

The scatter of our toy model, which can be calculated analytically, is
$\sigma(X) = \mu_d\sqrt{(\frac{\pi}{2}-f_d)f_d}$
(Figure \ref{fig:dustvar}).  Thus, for $\sigma(X) = 0.035$ (the 68\%
upper limit) we obtain constraints of $f_d \lesssim 15\%$ if $\mu_d$
is large ($>0.08$) or $\mu_d \lesssim 0.06$ if $f_d$ is large
($>0.3$).  The implication for SNe hosted by red-sequence galaxies is
that either the probability of being affected by dust at all is small
or alternatively that the amount of dust affecting the SN is small.
Turning our attention to individual SN host clusters, we see that the
scatters of clusters RCS~J234526-3632.6 (hosts SN~SCP06U4),
XLSS~J0223.0-0436 (hosts SN~SCP06R12), and ISCS~J1438.1+3414 (hosts
SCP06K0 and SCP06K18) at redshifts 1.037, 1.215, and 1.414
respectively, are below or just above the age-related scatter curve in
Figure \ref{fig:scatterz}.  Cluster RCS~J022144-0321.7 (host of
SCP06D0) at redshift 1.017 has more scatter than the three other
clusters within $|\Delta z < 0.02|$, but has the bluest intercept.
Similarly, cluster ISCS~J1434.4+3426 (host of SCP06H5) at redshift
1.241 has large intrinsic scatter but no more red an intercept than
cluster RDCS~J1252.9-2927 at redshift 1.237, which has only half as
much scatter.

The dust constraints derived to this point apply to cluster early-type
galaxies.  However, Figure \ref{fig:StackCMD} demonstrates that the
field early-type SN hosts have similar red-sequence residuals as
cluster early-type SN hosts.  The weighted mean residual of cluster
early-type SN hosts is $0.03 \pm 0.03$ whereas the weighted mean
residual of field early-type SN hosts is $-0.01 \pm 0.02$.  Thus, the
field early-types are actually slightly bluer than cluster early-types
and unlikely to have significant reddening.

\section{SN~Ia correlations with host galaxies}\label{sec:correlations}
In this section we look at correlations between the properties of
SN~Ia light curves and the properties of SN~Ia host galaxies at high
redshift.

\subsection{Stretch and color}\label{subsec:x1c}

One easily observed distance independent property of SNe~Ia is the
rate at which their light curves rise and fall.  This property has
been parameterized several ways in the literature: as the stretch,
$s$, which linearly stretches the time axis of a template light curve
to match the observed (rest-frame) light curve \citep{Perlmutter1997},
$\Delta m_{15}(B)$, which is the decline in $B$-band magnitudes from
peak to 15 rest-frame days later \citep{Phillips1993}, or $x_1$, which
is roughly the coefficient of the first component in a principle
component analysis of SN~Ia spectral time series \citep{Guy2007}.  The
light curve shape has been shown to correlate with SN~Ia host type at
lower redshifts with the light curves of SNe~Ia hosted by early-type
galaxies rising and falling more quickly than the light curves of
SNe~Ia hosted by late-type galaxies (i.e. early-type galaxies host
SNe~Ia with smaller stretch) \citep{Hamuy1996, Gallagher2005,
  Sullivan2006}.  Here we compare SN stretch as a function of host
type for our $z>0.9$ SN~Ia dataset.

\begin{figure}[h]
\begin{center}
\epsscale{1.175}
\plotone{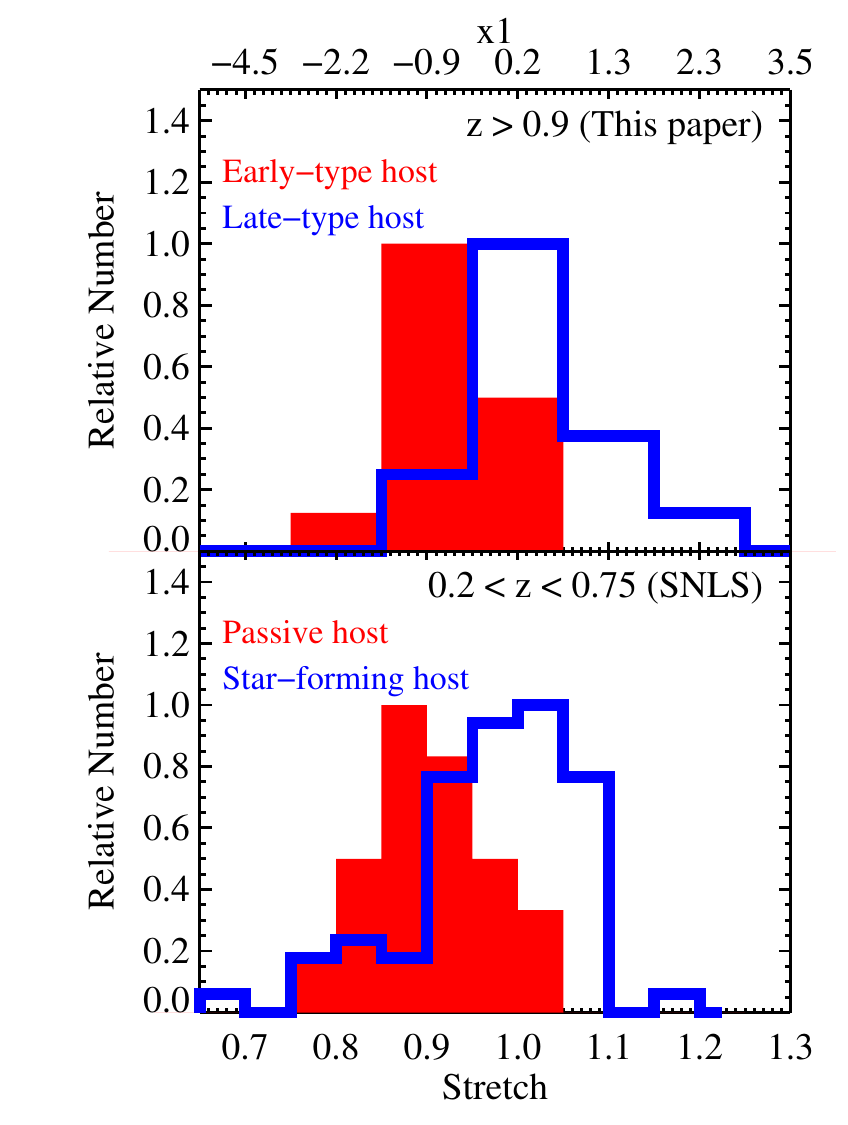}
\end{center}
\caption[doublestretchplot]{ \label{fig:doublestretchplot}
  Distribution of SN light curve shape parameters stretch (bottom
  axis) and $x_1$ (top axis).  The $x_1$ axis labels are computed
  using the cubic polynomial conversion from stretch given in
  \protect{\citet{Guy2007}}.  Both axes apply to both panels.  {\bf
    Top:} Light curve shape distribution for the $HST$-observed
  $z>0.9$ SNe~Ia discussed in the current paper.  The red (blue)
  histogram corresponds to SNe~Ia with hosts classified as early
  (late) type.  {\bf Bottom:} Light curve shape distribution for SNe
  drawn from the first year of SNLS as given in
  \protect{\citet{Sullivan2006}}.  The red (blue) histogram
  corresponds to SNe~Ia with hosts classified as passively-evolving
  (star-forming). }
\end{figure}

In \citet{Suzuki2011} we use SALT2 to fit light curves of $z>0.9$
SNe~Ia from the {\it HST} Cluster SN Survey and the GOODS SN Surveys
in the ACS $i_{775}$, $z_{850}$ filters and also the F110W and F160W
filters of NICMOS.  To convert the SALT2 light curve shape parameter
$x_1$ to stretch, we use the cubic relation from \citet{Guy2007}.  The
distributions of stretch (and $x_1$) for SN~Ia subsets split by host
type are shown in Figure \ref{fig:doublestretchplot}.  To prevent
poorly measured SNe~Ia from influencing our results, we have only
included SNe~Ia with $x_1$ uncertainty less than $1.0$, which is
roughly the size of the histogram bins.  The SNe~Ia whose hosts we
classify as passively evolving early-type galaxies show smaller
stretch than the SNe~Ia whose hosts we classify as late-type galaxies,
consistent with lower redshift results.  A Kolmogorov-Smirnov (K-S)
test reveals that the probability that the stretch values from the two
host subsets are drawn from the same distribution is $<0.01$.

We have also compared the light curve shape distributions of these
$z>0.9$ SNe~Ia to that of the first year of SNLS SNe~Ia at $0.2
\lesssim z \lesssim 0.8$ presented in \citet[Figure
  \ref{fig:doublestretchplot}]{Sullivan2006}.  We assume that the
individual values of stretch in the histogram plotted there are
uniformly distributed within their bins.  We find that the K-S
probability that the $s$ values of the early-type hosted SNe~Ia
analyzed in this paper and the $s$ values of the passively hosted
first year SNLS SNe~Ia are drawn from the same distribution is quite
high at $0.91$.  The K-S probability comparing the $s$ distribution in
$z>0.9$ late-type galaxies to that from the star-forming galaxies of
first year SNLS SNe is also suggestive of similar distributions at
$0.63$.  Finding similar demographics across such a wide redshift
baseline instills confidence that SNe~Ia are truly standardizable
candles over this range.

\begin{figure}[h]
\begin{center}
\epsscale{1.175}
\plotone{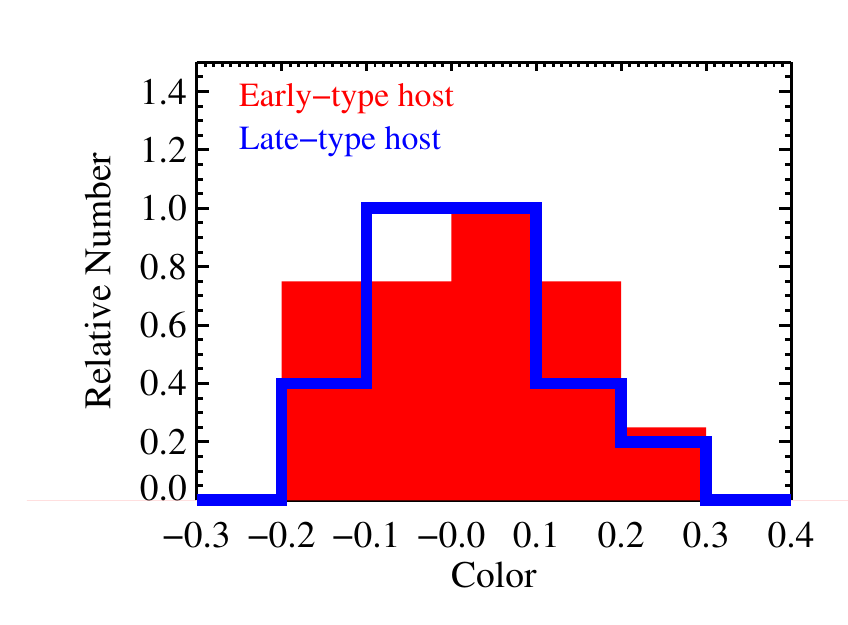}
\end{center}
\caption[cdist]{\label{fig:cdist} Distribution of $z > 0.9$ SN~Ia
  color for SNe with hosts classified as either early-type (red) or
  late-type (blue).}
\end{figure}

Likewise, we can compare the distribution of the color parameter in
both host subsets.  In Figure \ref{fig:cdist} we compare the
distribution of color for SN subsets of different host types.  The K-S
test reveals that the probability that the $c$ values from the two
host subsets are drawn from the same distribution is $0.75$, also
consistent with findings at lower redshift \citep{Lampeitl2010}.  One
possible explanation for this result is that the dominant source of
color for these SNe~Ia is intrinsic or circumstellar and that host
galaxy extinction plays a smaller role.

\subsection{ SN~Ia Hubble residual vs. host mass }\label{subsec:HR}

\begin{figure}[h]
\begin{center}
\epsscale{1.175}
\plotone{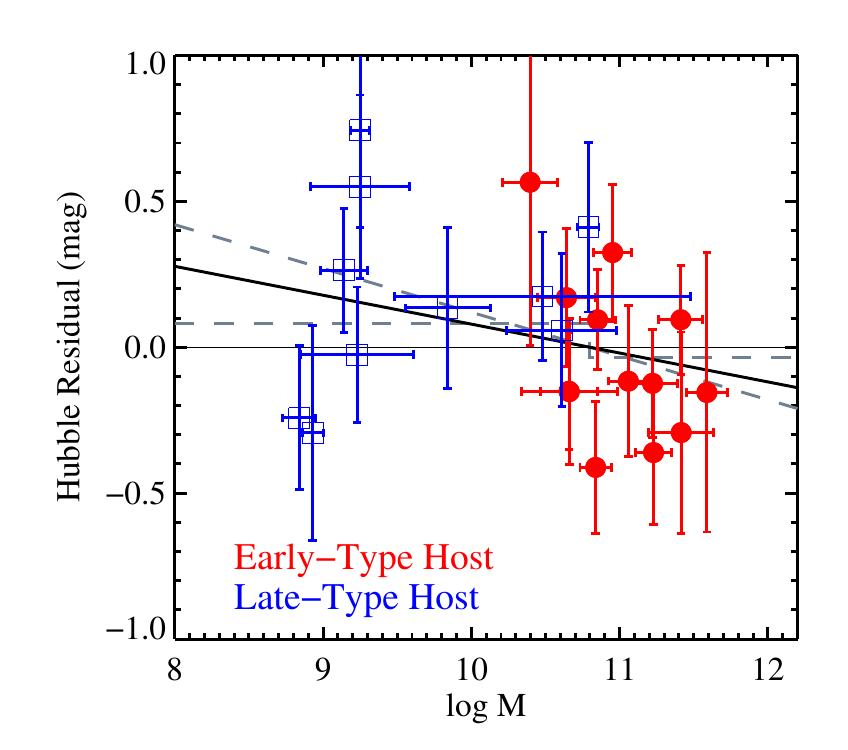}
\end{center}
\caption[HR_vs_mass]{\label{fig:HR_vs_mass} Hubble residual as a
  function of host galaxy stellar mass.  Open blue squares are
  late-type galaxies with masses from \citet{Thomson2011}.  Filled red
  circles are early-type galaxies with masses derived from SED fits to
  $i_{775}$ and $z_{850}$ as described in \S \ref{sec:correlations}.
  The solid line is the best-fit linear relationship with slope
  $-0.098 \pm 0.085$\ mag\ dex$^{-1}$. The dashed line and step
  function are the relations plotted in \protect{\citet{Kelly2010}}
  when using the SALT2 light curve fitter.}
\end{figure}

Recent analyses of SN~Ia hosts have indicated that after corrections
for stretch and color have been applied, SNe~Ia are brighter in
galaxies with more stellar mass \citep{Kelly2010, Sullivan2010,
  Lampeitl2010}.  Kelly et al. have parameterized this effect both as
a step function at $\log\frac{M}{M_\odot} = 10.8$ and as a linear
function.  In \S \ref{sec:typing} we argued that the relatively simple
SEDs of early-type galaxies (as opposed to the generally more
complicated SEDs of late-type galaxies) allow us to measure stellar
masses of early-type hosts using just the $i_{775}$ and $z_{850}$
photometry collected here.  Thirteen of our $z>0.9$ SNe have both host
mass measurements and Hubble residuals (uncorrected for host mass)
from \citet{Suzuki2011}.  To extend this sample, we have also used the
masses of $z>0.9$ GOODS SN hosts from \citet{Thomson2011}, which are
derived from SED fits to ACS optical, IRAC near infrared and MIPS 24
$\mu m$ photometry.  The Hubble residuals for these SNe are also
available from \citet{Suzuki2011}, increasing the sample from 13 to
23.  Five galaxies with mass estimates are in common between the
\citet{Thomson2011} dataset and the early-type hosts analyzed here.
For four of these galaxies, the difference in mass is less than $1
\sigma$, and it is less than $2 \sigma$ for the fifth.  In Figure
\ref{fig:HR_vs_mass}, we plot Hubble residuals against host stellar
masses, using our own mass estimates for the galaxies with two mass
estimates.  A suggestion of the trend of negative residuals for higher
mass hosts is apparent.  A linear fit to these data results in a slope
measurement of $-0.098 \pm 0.085$\ mag\ dex$^{-1}$, (i.e. a
$1.1\sigma$ detection significance).  The next generation of SN~Ia
studies (such as the SNe~Ia discovered and followed with Wide Field
Camera 3 as part of the {\it HST} Multi-Cycle Treasury Programs) may
have sufficiently better data to confirm this trend at high redshift.

\section{Discussion and conclusions}\label{sec:conclusion}
In this paper, we have studied the relation between SNe~Ia and their
host galaxies at high redshift $(z>0.9)$.  In total, we study 16
SNe~Ia (seven in clusters, eight in the field, and one with uncertain
cluster membership) from the ACS cluster fields and 24 SNe~Ia from the
GOODS fields \citep{Dawson2009, Gilliland1999, Blakeslee2003B,
Riess2004, Riess2007}.  Using color-magnitude diagrams constructed
from $i_{775}$ and $z_{850}$ photometry and measurements of
quantitative morphology, we have developed a technique to segregate
high-redshift, early-type galaxies from high-redshift, late-type
galaxies in these fields.  Additional measurements of the \OII\
emission line strength in the spectra of SN host galaxies were used to
estimate galaxy star formation rates and the rate of core collapse
SNe.  This classification scheme was applied to the hosts of the 40
SNe, first in conjunction with the \OII\ emission line strengths to
determine the likelihood of early-type hosted SNe being Type~Ia, then
to identify trends in the supernova properties with host galaxy type.
Of the SNe~Ia studied here, 17 have host galaxies classified as
early-type, 22 have host galaxies classified as late-type, and one
galaxy has a host which we have left unclassified.

From our analysis of these data, our principle results are:

{\bf (i) SN Classification} A review of the
literature \citep{Hakobyan2008} shows that the confidence level is
quite high for the classification of a SN~Ia based on the
identification of its host galaxy as early-type.  We here confirm, on
the basis of host morphology and spectroscopic limits on star
formation, that SNe hosted by early-type galaxies in our current
sample are likely Type~Ia.  Using the A+B SN~Ia rate parameterization
from \citet{Sullivan2006} (Eqn. \ref{eqn:R_Ia}), and by deriving a
similar formula for SNe~CC (Eqn. \ref{eqn:R_CC}), we estimate the
intrinsic rate of SNe~Ia and SNe~CC in individual early-type galaxies.
By multiplying the intrinsic rates by the detectable fractions
(derived from the detection threshold of the survey and the luminosity
functions of different types of SNe) we calculate the rate of {\it
detectable} SNe~Ia and SNe~CC for each early-type SN host.  For most
of the early-type hosted SNe in our survey, the probability of
misclassification is very small, even if relying on only the host
galaxy information.  The three examples where the purely host galaxy
classification is weakest ($ P(Ia) \lesssim 90\%$) can be traced back
to observations of \OII\ emission.  If some or all of early-type
galaxy \OII\ emission is due to LINER activity instead of star
formation, then these misclassification probabilities may be
significantly overestimated.  Fortunately, in this particular case two
of these three SNe have been spectroscopically confirmed as a Type~Ia.
Of course, our combined classification for cosmological analysis
includes the full light curve data and thus the classsifications
become much stronger \citep{Barbary2010,Suzuki2011}.  In the {\it HST}
Cluster SN Survey, ten $z>0.9$ SNe in all can be typed by their
early-type hosts, six more than a comparable blank field search would
have found since this many were found in the clusters themselves.

{\bf (ii) Contamination from Dust} The narrow color scatter about
cluster red sequences directly constrains the amount of dust
extinction affecting SNe in red-sequence hosts.  We measure a typical
intrinsic scatter in the observed $i_{775}-z_{850}$ color of
morphologically selected early-type galaxies of $\sim 0.045$
magnitudes.  While other authors have modeled this scatter as a
distribution of galaxy ages \citep[e.g.][]{Jaffe2011}, we have
investigated extinction due to dust as an additional parameter.  By
simultaneously fitting the average cluster galaxy formation history
and dust content, we conclude that the rest-frame $B-V$ scatter of red
sequence residuals attributable to dust is likely $\lesssim 0.04$
magnitudes.  By modeling this distribution as the sum of a dust-free
galaxy population (with fraction $1-f_d$) and a dusty galaxy
population (with fraction $f_d$ and mean rest-frame $E(B-V)$ reddening
$\mu_d$), we show that either the probability that any particular
red-sequence member is reddened by dust is small ($f_d \lesssim 15\%$)
or alternatively that the amount of reddening is small
($\mu_d \lesssim 0.06$).  These limits are comparable to early-type
galaxy reddening limits set at low redshift obtained by exploiting the
correlation of the Mg$_2$ absorption line and intrinsic $B-V$ colors
of early-type galaxies \citep{Ferguson1993,Schlegel1998}.  We also
find that the clusters with higher scatter have the same CMR
intercepts as clusters with low scatter, contrary to expectation if
dust were the primary factor in producing the color scatter.

{\bf (iii) SN Light Curve Properties} The light curves of $z>0.9$
SNe~Ia rise and fade more quickly (have smaller stretch) in early-type
hosts than in late-type hosts.  The distribution of SN colors is
similar in both host subsets, suggesting that SN~Ia color is largely
intrinsic or due to circumstellar dust and not dominated by host
galaxy extinction.  These trends are well known at lower redshifts
\citep{Hamuy1996,Gallagher2005,Sullivan2006,Lampeitl2010}, but have not
been demonstrated before at $z>0.9$.

{\bf (iv) Galaxy Mass and Hubble Residuals} We have obtained mass
estimates of the early-type classified SN host subset by fitting
passively-evolving SEDs to their $i_{775}$ and $z_{850}$ photometry.
Together with infrared-derived mass estimates of a subset of the
late-type classified SN hosts from \citet{Thomson2011}, we have found
a suggestion of a correlation between SN~Ia host mass and
stretch-and-color-corrected Hubble residual ($1.1 \sigma$).  This
effect has been seen with more significance in three datasets at lower
redshifts \citep{Kelly2010, Sullivan2010, Lampeitl2010} and may be
related to the metallicity or age of SN progenitors or to extinction
by dust, all of which correlate with galaxy mass.

While the analysis in this paper focuses on the astrophysical
properties of SNe~Ia and their host galaxies, the results clearly have
implications for the cosmological interpretation of SN observations.
The evolution of host demographics with survey strategy or with
redshift can lead to biases in cosmological parameter fits if not
accounted for.  In particular, many low redshift surveys explicitly
target massive galaxies to improve their discovery rates.  The cluster
galaxies targeted by the {\it HST} Cluster SN Survey are also more
massive on average than those in an untargeted survey and may produce
SNe~Ia which lead to biases in cosmological interpretation.  A
detailed analysis of the cosmological impact of host dependent
properties is discussed in \citet{Suzuki2011}.  In that work, we
present several other trends in SN~Ia parameters as a function of host
type, including the values of color-magnitude relation coefficient
$\beta = \Delta M_B/\Delta c$ in star-forming galaxies and in passive
galaxies, and the residual scatter about the Hubble diagram for SNe~Ia
hosted by star-forming galaxies and by early-type galaxies.  These
studies are compared to similar studies at lower redshift
\citep{Sullivan2003,Jha2007,Sullivan2010,Lampeitl2010}.

The ability to identify early-type galaxies, and their overdensity in
clusters, provides many advantages in SN cosmology studies at $z>0.9$.
SNe hosted by early-type galaxies are essentially all Type~Ia,
reducing the need for potentially expensive confirmation spectroscopy
that must be obtained during the narrow time window that the SN is
active.  Cluster early-type galaxies are minimally contaminated with
dust (either less than 15\% are dusty or the reddening is less than
0.06 mag), and can be directly checked for dust via the cluster
red-sequence scatter.  The cluster galaxies provide additional
potential SN~Ia progenitors and hence yield a higher rate of SNe~Ia
than a comparable blank field search, especially of early-type hosted
SNe~Ia.  Finally, the relatively simple SEDs expected for early-type
galaxies permit SN host mass measurements with fewer photometric bands
than for late-type hosts.  Targeting clusters is therefore not only
the most efficient method to discover and type SNe~Ia at $z>0.9$, but
is likely the most efficient way to collect the host galaxy
information necessary for the next generation of $z>0.9$ SN~Ia
studies.

\acknowledgements
We would like to thank a very helpful referee for suggestions that
improved the quality of this paper.  We would also like to thank
Pasquale Temi for comments on the infrared dust properties of nearby
early-type galaxies.  Financial support for this work was provided by
NASA through program GO-10496 from the Space Telescope Science
Institute, which is operated by AURA, Inc., under NASA contract NAS
5-26555.  This work was also supported in part by the Director, Office
of Science, Office of High Energy and Nuclear Physics, of the
U.S. Department of Energy under Contract No. AC02-05CH11231, as well
as a JSPS core-to-core program ``International Research Network for
Dark Energy'' and by a JSPS research grant (20040003).  The authors
wish to recognize and acknowledge the very significant cultural role
and reverence that the summit of Mauna Kea has always had within the
indigenous Hawaiian community.  We are most fortunate to have the
opportunity to conduct observations from this mountain.  Finally, this
work would not have been possible without the dedicated efforts of the
daytime and nighttime support staff at the Cerro Paranal Observatory.

{\it Facilities:} \facility{HST (ACS)}, \facility{Subaru
     (FOCAS)}, \facility{Keck:II (DEIMOS)}, \facility{VLT:Kueyen
     (FORS1)}, \facility{VLT:Antu (FORS2)}

\bibliographystyle{apj}
\bibliography{clhosts}

\appendix

Here we present the color-magnitude diagrams for the 25 clusters in
the {\it HST} Cluster SN Survey.
\begin{figure*}[H]
  \centering \subfigure{ \label{fig:CMDpanel:a} \plotone{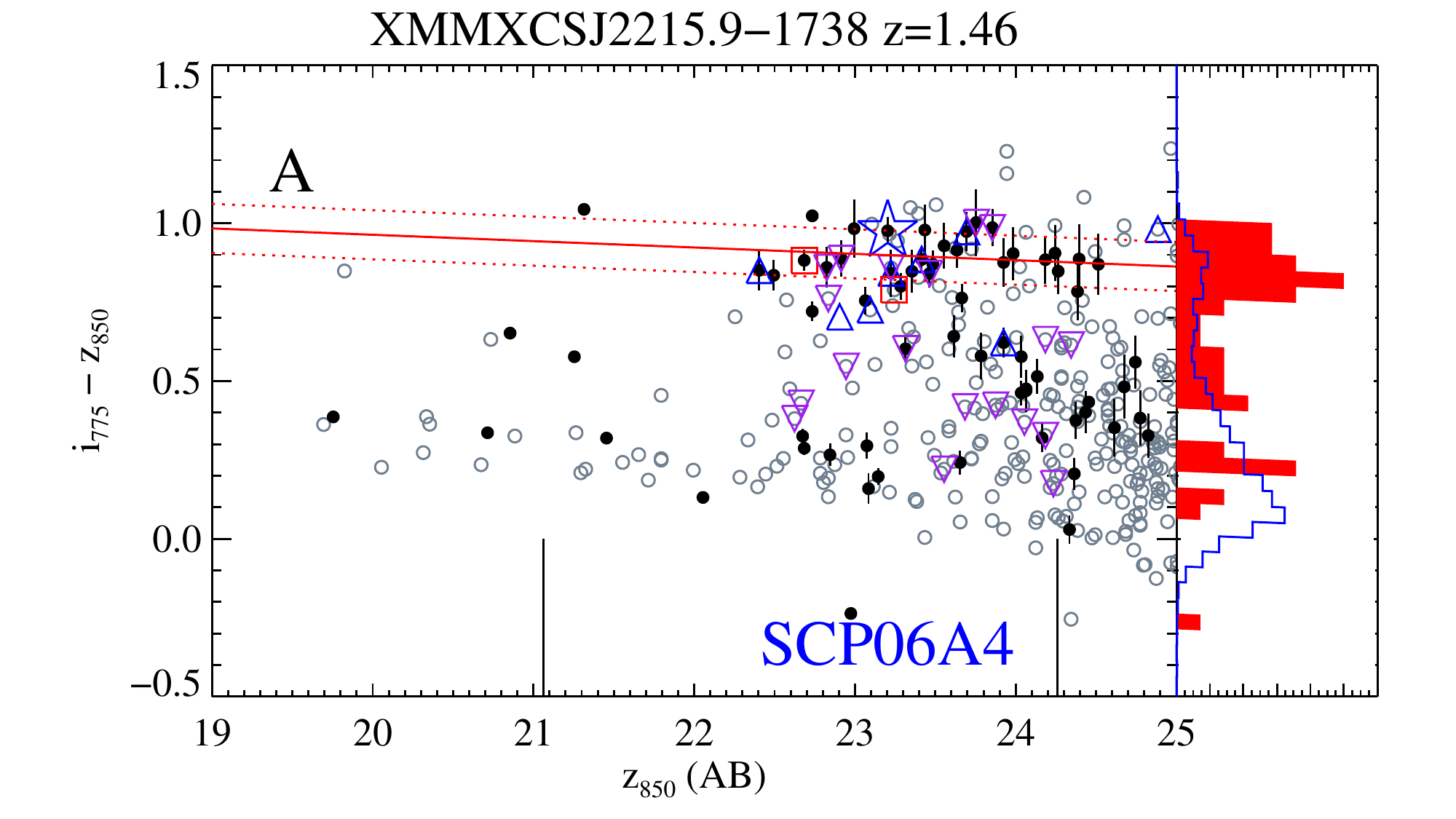}
  } \caption{Cluster color-magnitude diagrams.  Small circles indicate
  galaxies located within $0.65 \mathrm{Mpc}$ of the cluster center.
  Dark solid circles pass our broader quantitative morphology cuts,
  light open circles do not.  The slightly larger squares and
  triangles indicate galaxies spectroscopically confirmed to be
  cluster members, including those that lie outside of the
  $0.65 \mathrm{Mpc}$ aperture. Red squares indicate spectroscopically
  confirmed members lacking significant \OII\ emission ($EW > -5\AA$.)
  Blue upward pointing triangles indicate spectroscopically confirmed
  members with significant \OII\ emission.  Purple downward pointing
  triangles indicate spectroscopically confirmed members found in the
  literature but with unpublished \OII\ EWs.  Stars indicate SN hosts.
  Blue stars are foreground SN hosts, gold stars are cluster members
  SN hosts, and red stars are background SN hosts.  The solid red line
  is the best-fit linear color-magnitude relation, and the dashed
  lines indicate the measured scatter (including contributions from
  both measurement uncertainties and the intrinsic scatter).  The red
  histogram indicates CMR residuals of cluster galaxies passing our
  broader morphology cuts falling in the magnitude range $M^*-2.2 <
  z_{850} < M^*+1.0$, which is indicated by the long tick marks on the
  x-axis.  The blue histogram is the solid-angle-scaled analog for the
  GOODS control fields.}
\label{fig:CMDpanel}
\end{figure*}

\addtocounter{figure}{-1}
\begin{figure*}[H]
  \addtocounter{subfigure}{1}
  \centering \subfigure{ \label{fig:CMDpanel:b}
    \plotone{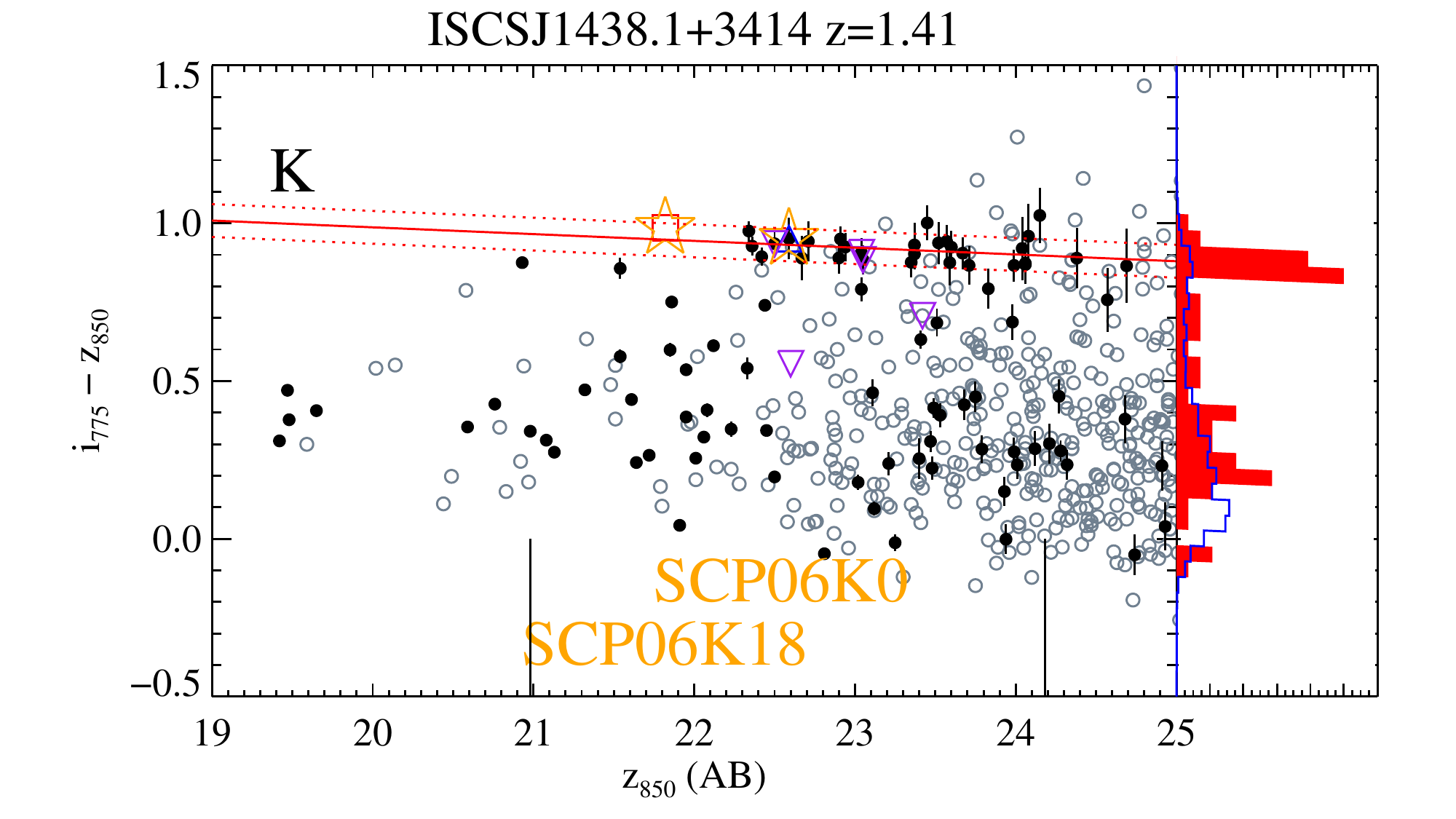}
  } \caption{Color magnitude diagrams (continued).}
\end{figure*}

\addtocounter{figure}{-1}
\begin{figure*}[H]
  \addtocounter{subfigure}{1}
  \centering \subfigure{ \label{fig:CMDpanel:c}
    \plotone{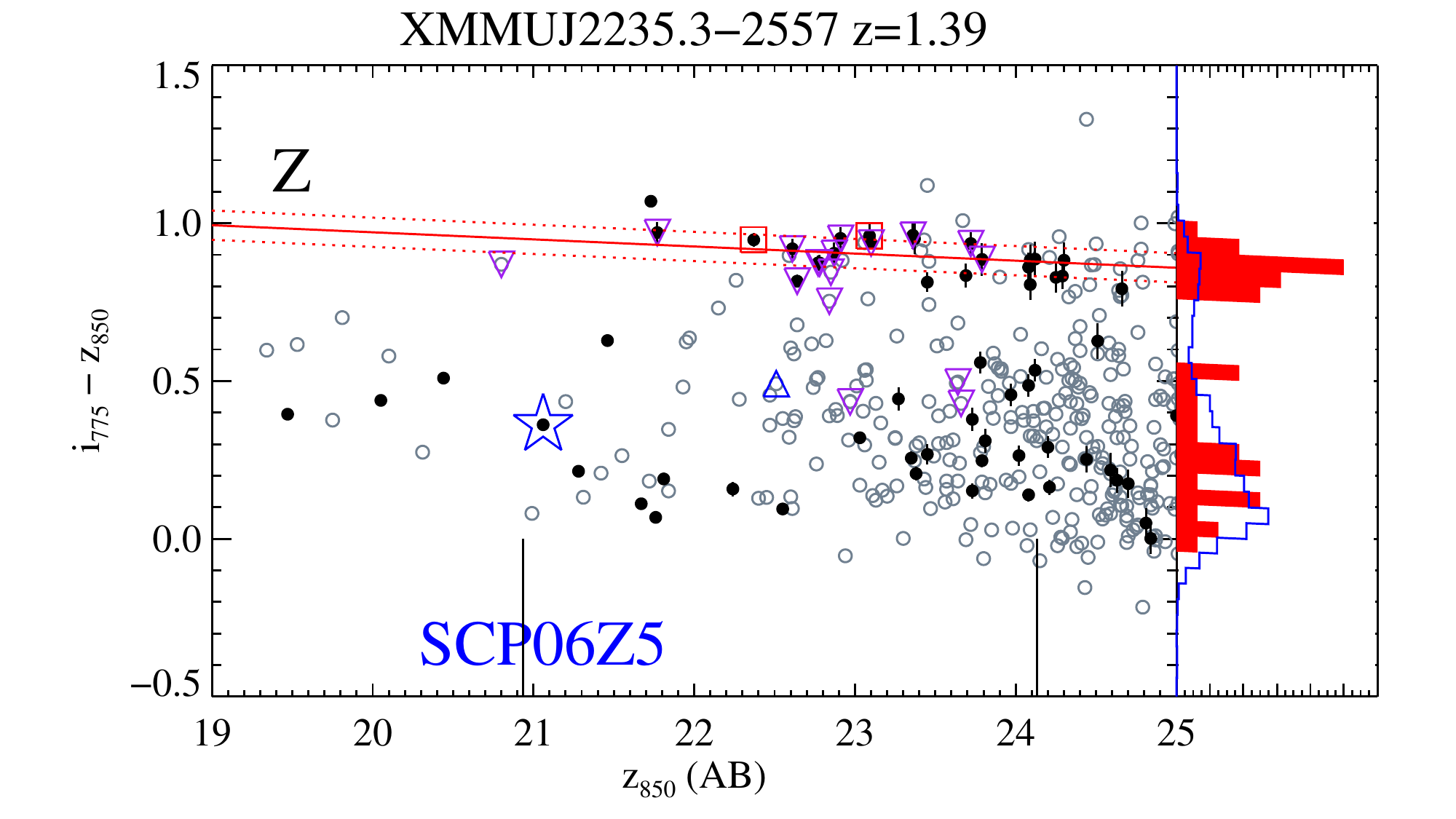}
  } \caption{Color magnitude diagrams (continued).}
\end{figure*}

\addtocounter{figure}{-1}
\begin{figure*}[H]
  \addtocounter{subfigure}{1}
  \centering \subfigure{ \label{fig:CMDpanel:d}
  \epsscale{0.905}
    \plotone{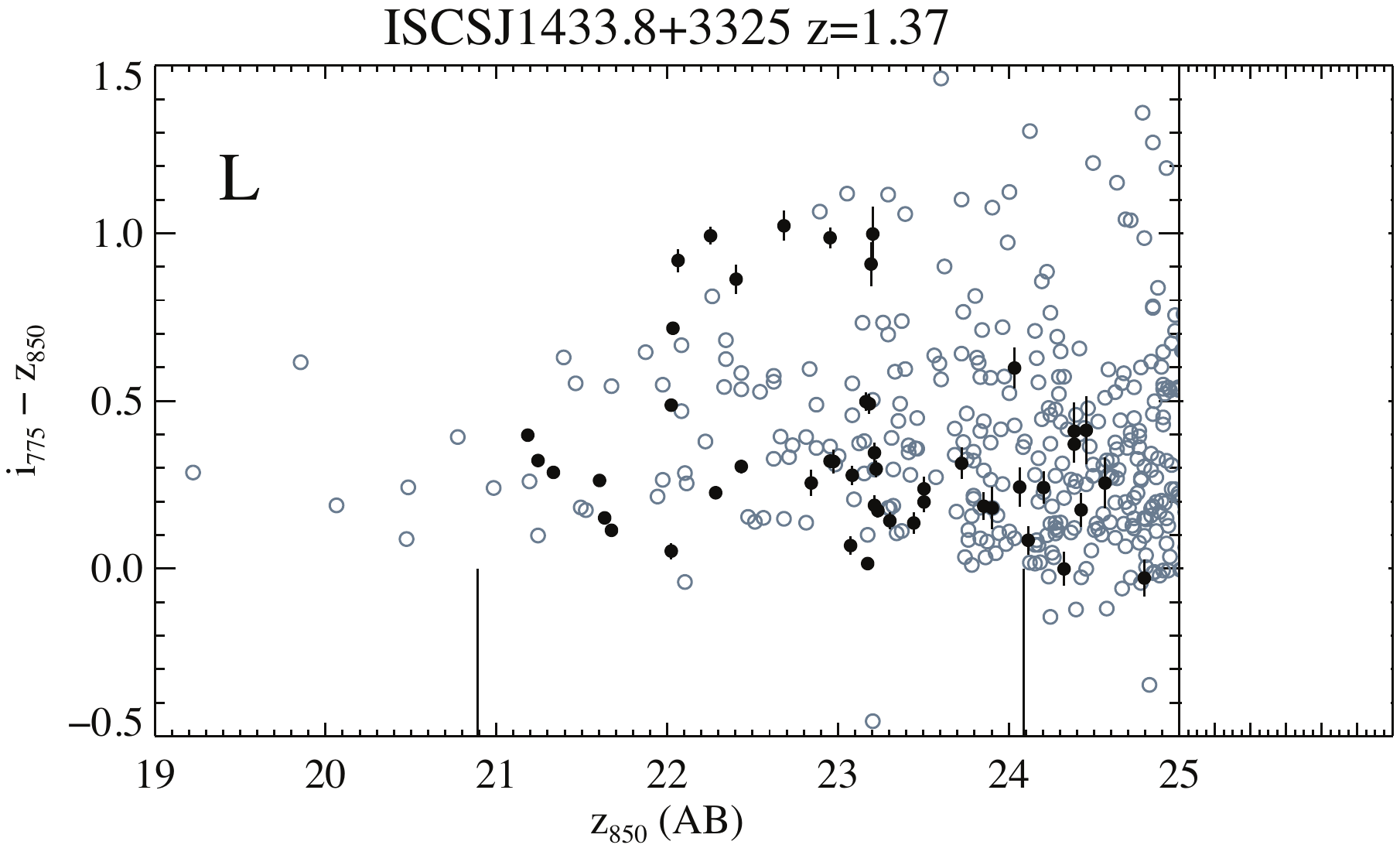}
  } \caption{Color magnitude diagrams (continued).}
\end{figure*}

\addtocounter{figure}{-1}
\begin{figure*}[H]
  \addtocounter{subfigure}{1}
  \centering \subfigure{ \label{fig:CMDpanel:e}
  \epsscale{0.905}
    \plotone{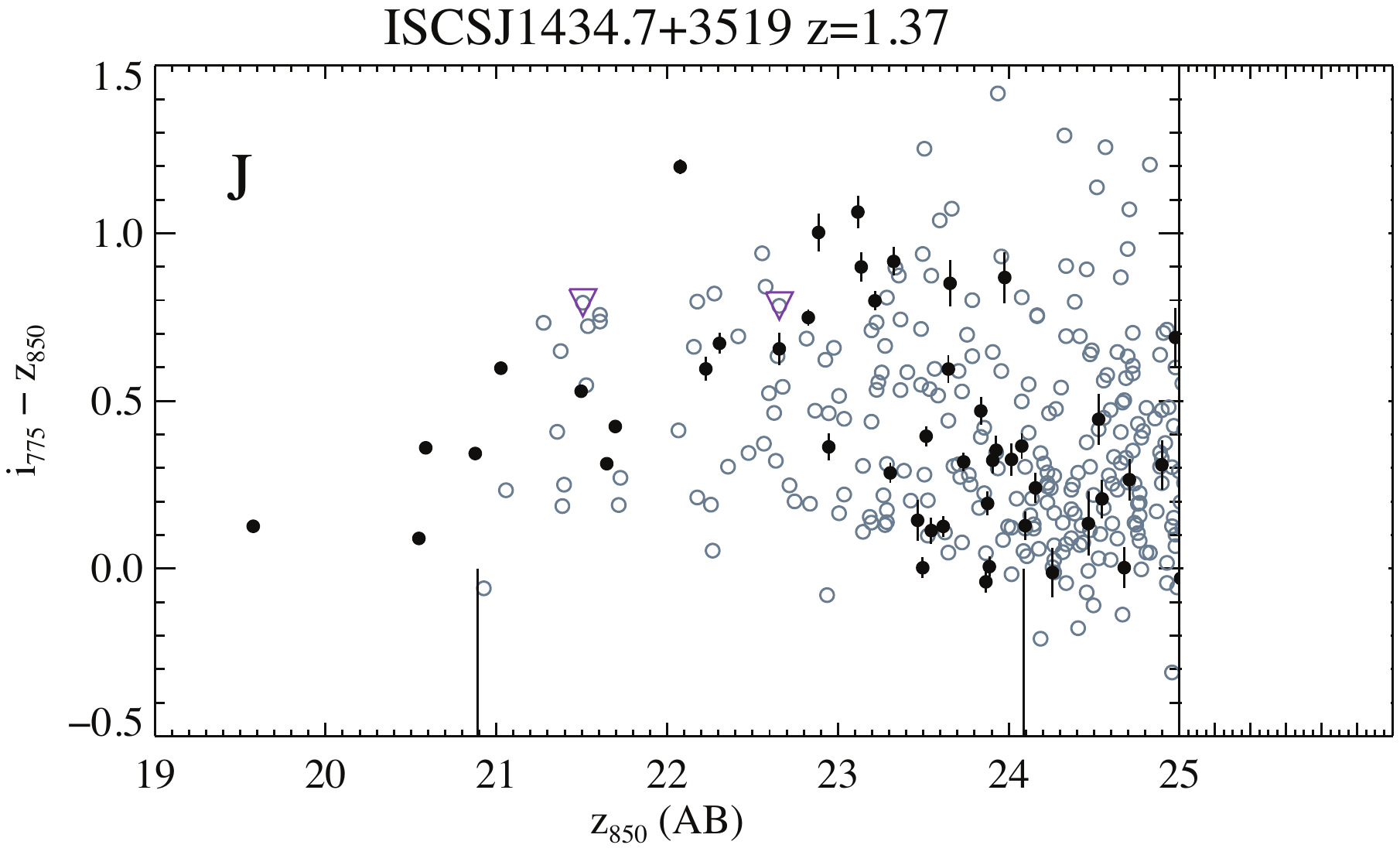}
  } \caption{Color magnitude diagrams (continued).}
\end{figure*}

\addtocounter{figure}{-1}
\begin{figure*}[H]
  \addtocounter{subfigure}{1}
  \centering \subfigure{ \label{fig:CMDpanel:f}
    \plotone{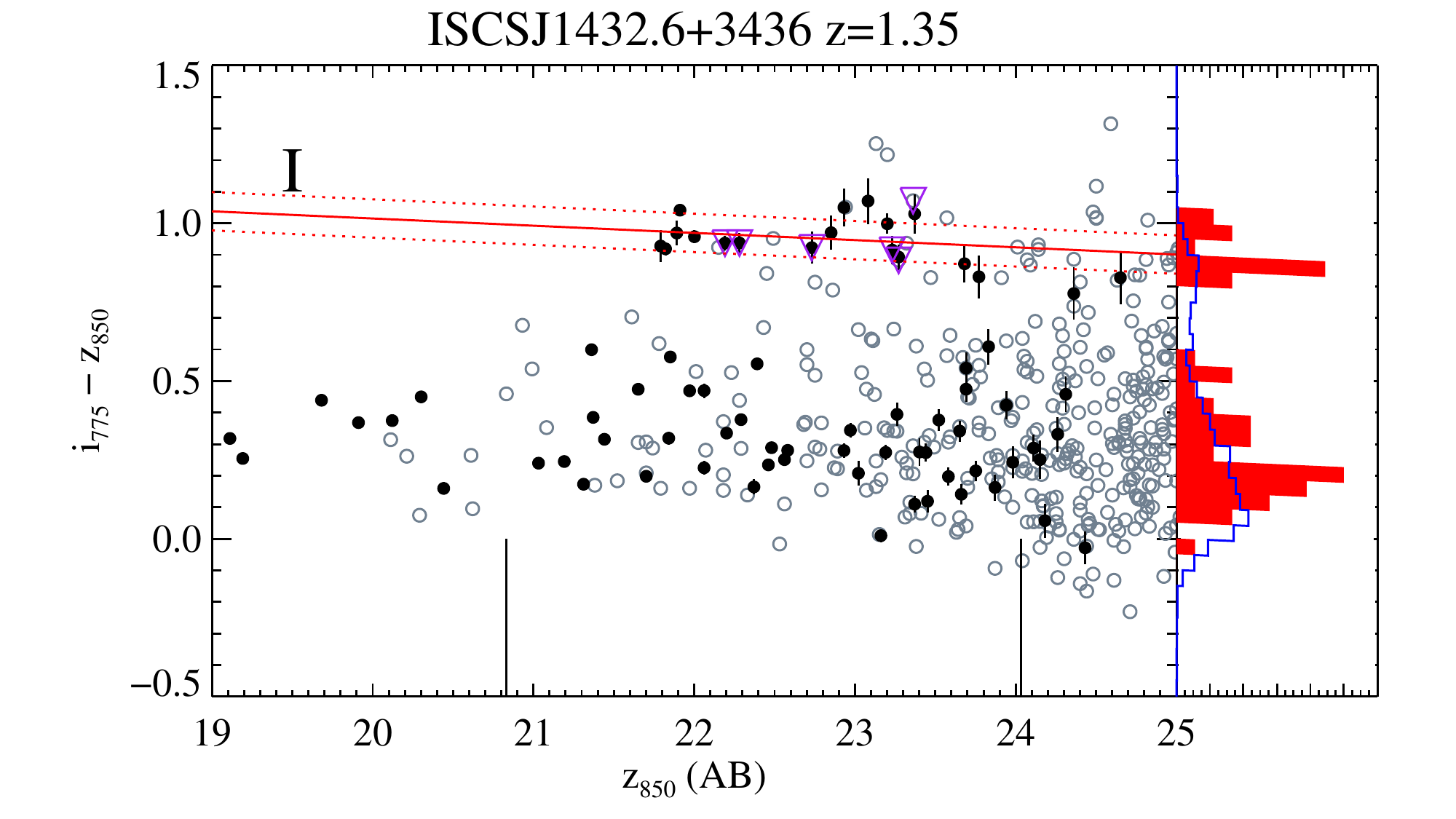}
  } \caption{Color magnitude diagrams (continued).}
\end{figure*}

\addtocounter{figure}{-1}
\begin{figure*}[H]
  \addtocounter{subfigure}{1}
  \centering \subfigure{ \label{fig:CMDpanel:g}
    \plotone{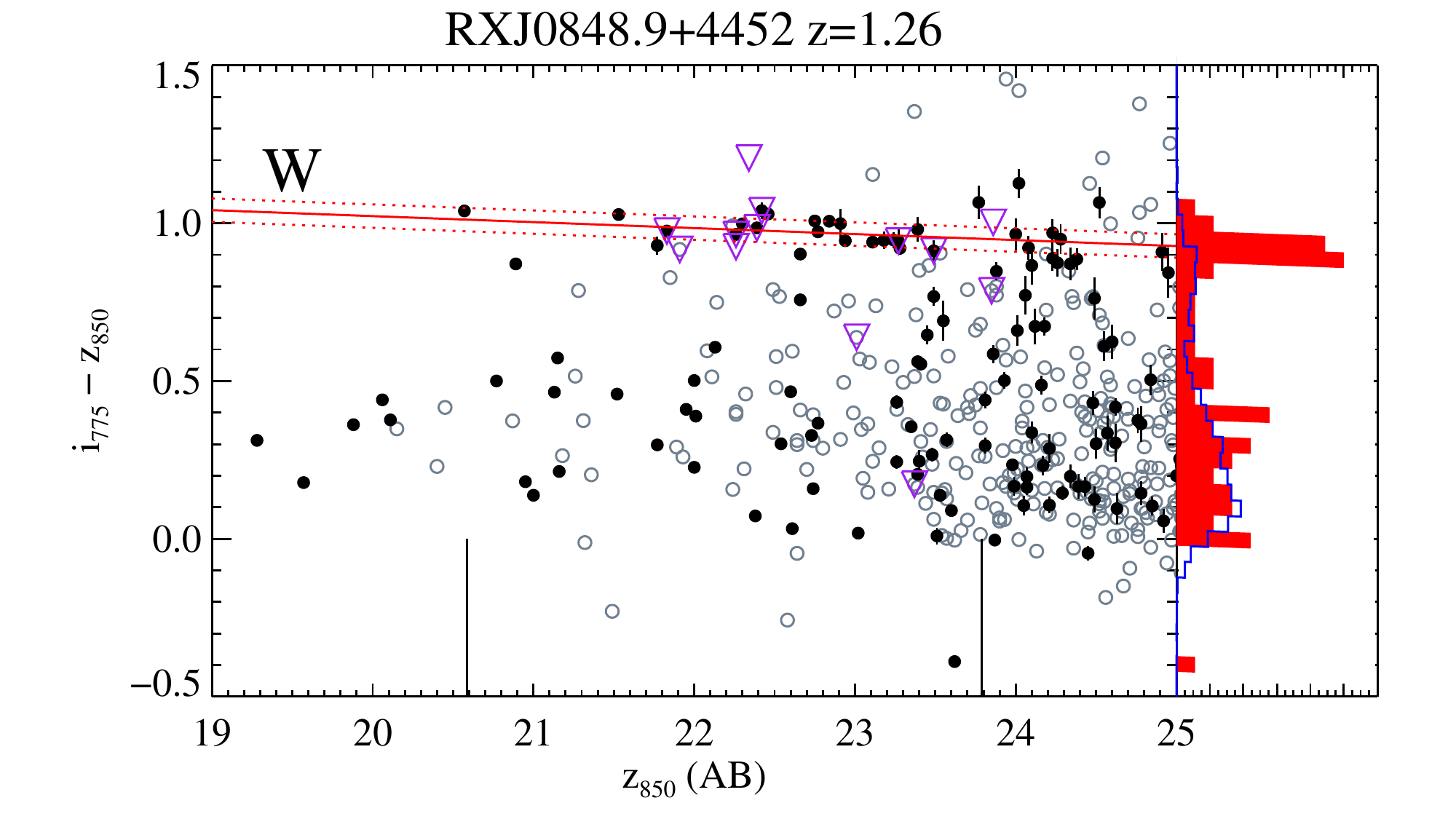}
  } \caption{Color magnitude diagrams (continued).}
\end{figure*}

\addtocounter{figure}{-1}
\begin{figure*}[H]
  \addtocounter{subfigure}{1}
  \centering \subfigure{ \label{fig:CMDpanel:h}
    \plotone{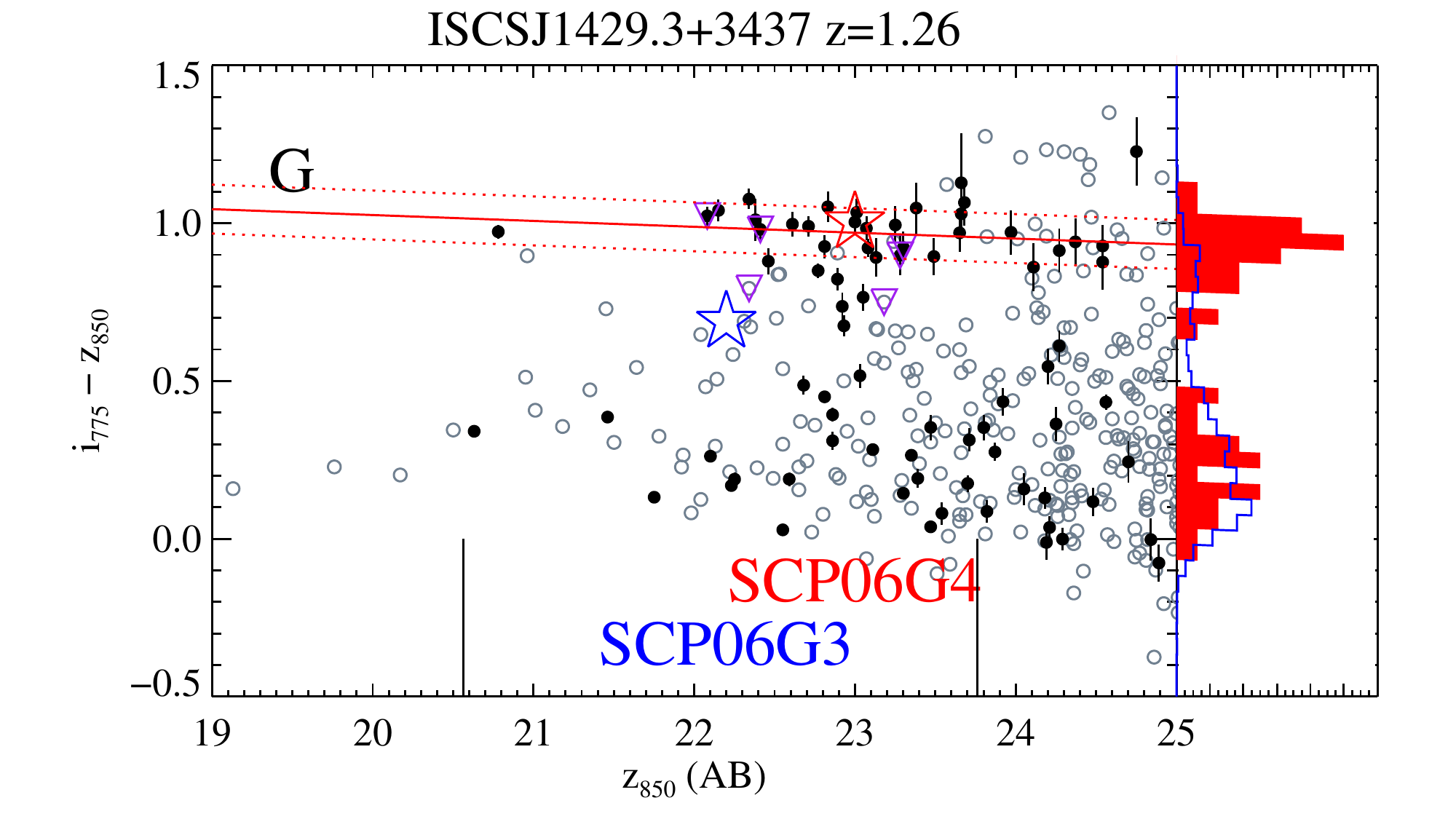}
 } \caption{Color magnitude diagrams (continued).}
\end{figure*}

\addtocounter{figure}{-1}
\begin{figure*}[H]
  \addtocounter{subfigure}{1}
  \centering \subfigure{ \label{fig:CMDpanel:i}
    \plotone{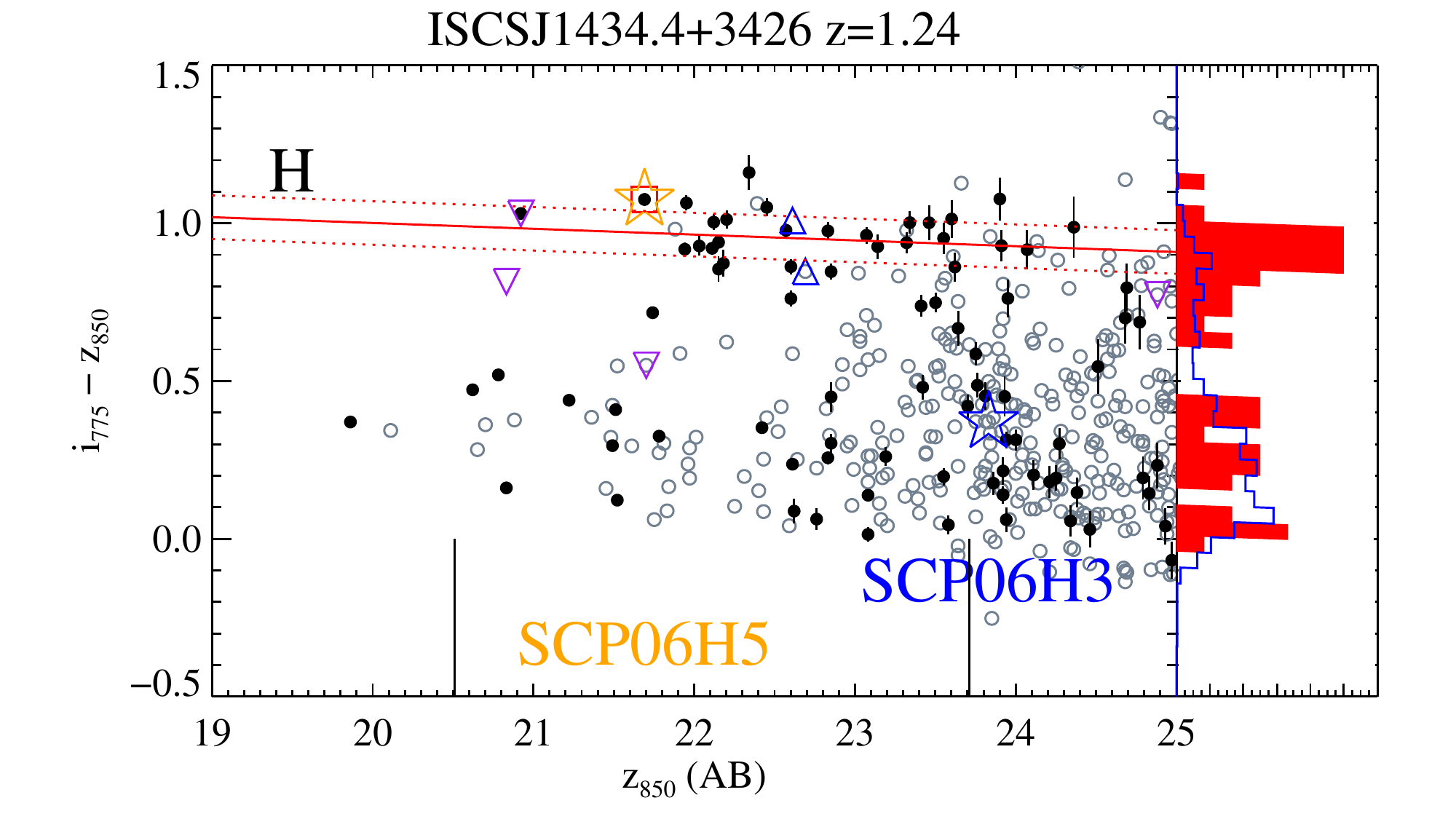}
  } \caption{Color magnitude diagrams (continued).}
\end{figure*}

\addtocounter{figure}{-1}
\begin{figure*}[H]
  \addtocounter{subfigure}{1}
  \centering \subfigure{ \label{fig:CMDpanel:j}
    \plotone{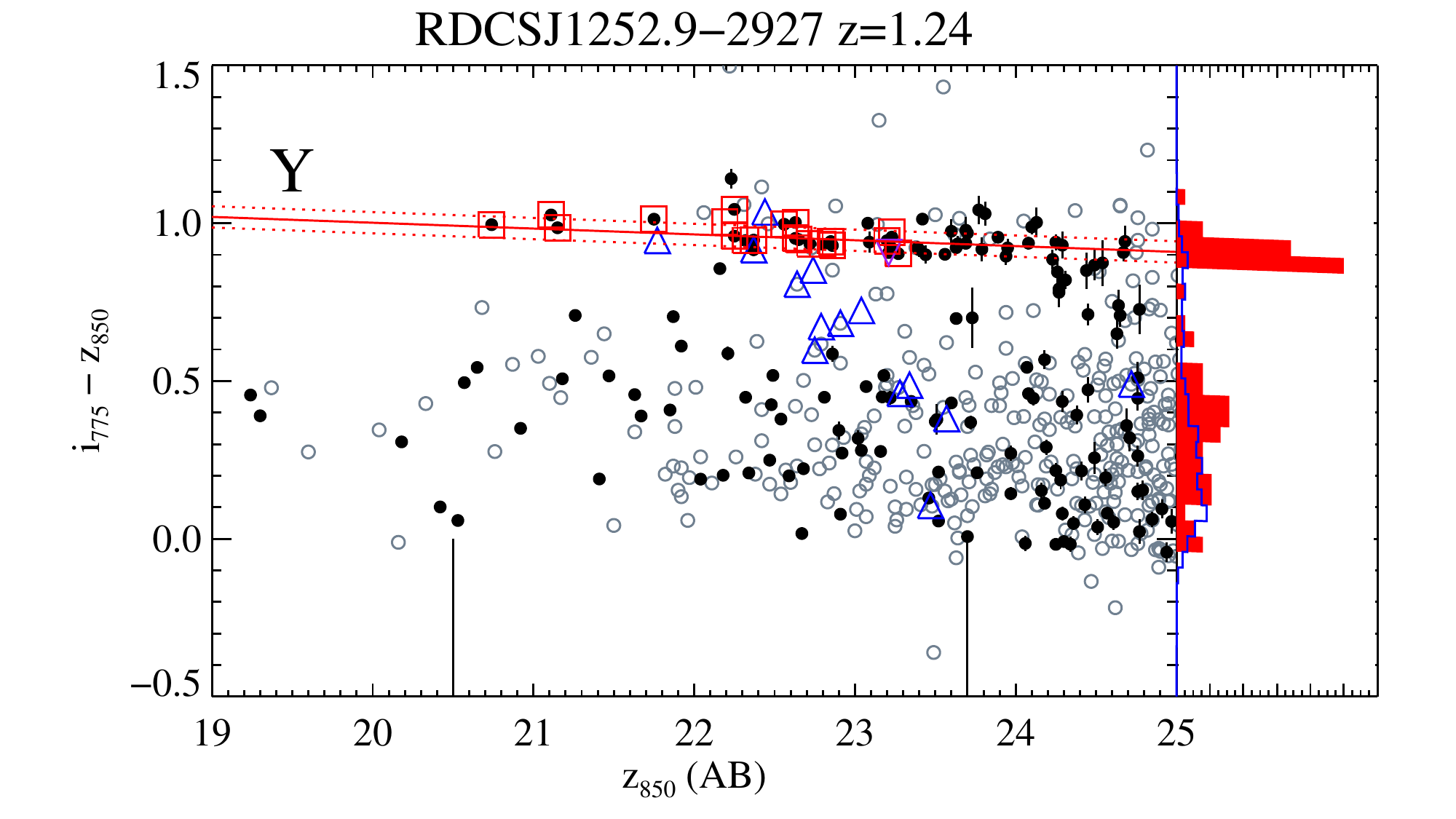}
  } \caption{Color magnitude diagrams (continued).}
\end{figure*}

\addtocounter{figure}{-1}
\begin{figure*}[H]
  \addtocounter{subfigure}{1}
  \centering \subfigure{ \label{fig:CMDpanel:k}
    \plotone{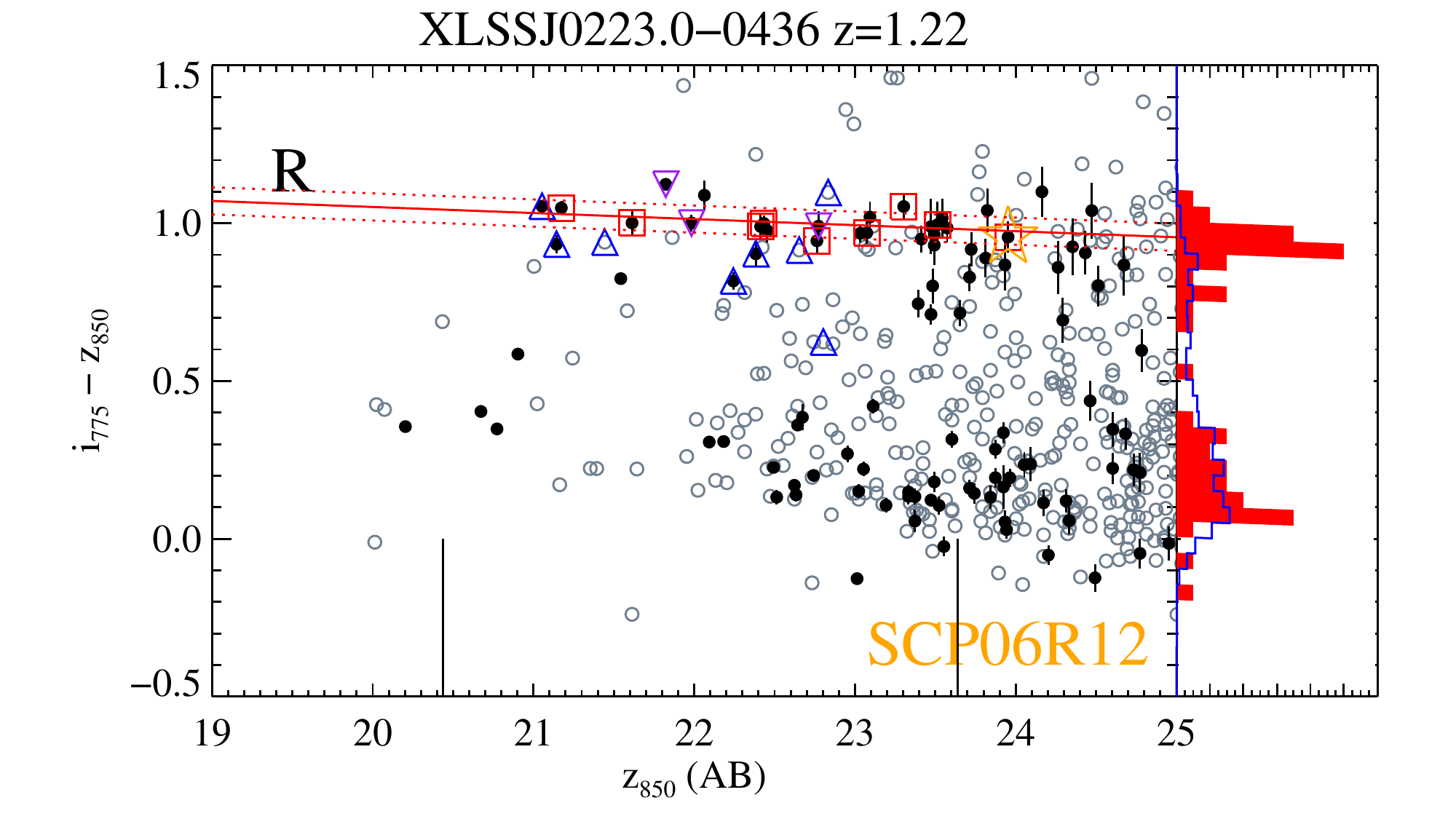}
  } \caption{Color magnitude diagrams (continued).}
\end{figure*}

\addtocounter{figure}{-1}
\begin{figure*}[H]
  \addtocounter{subfigure}{1}
  \centering \subfigure{ \label{fig:CMDpanel:l}
    \plotone{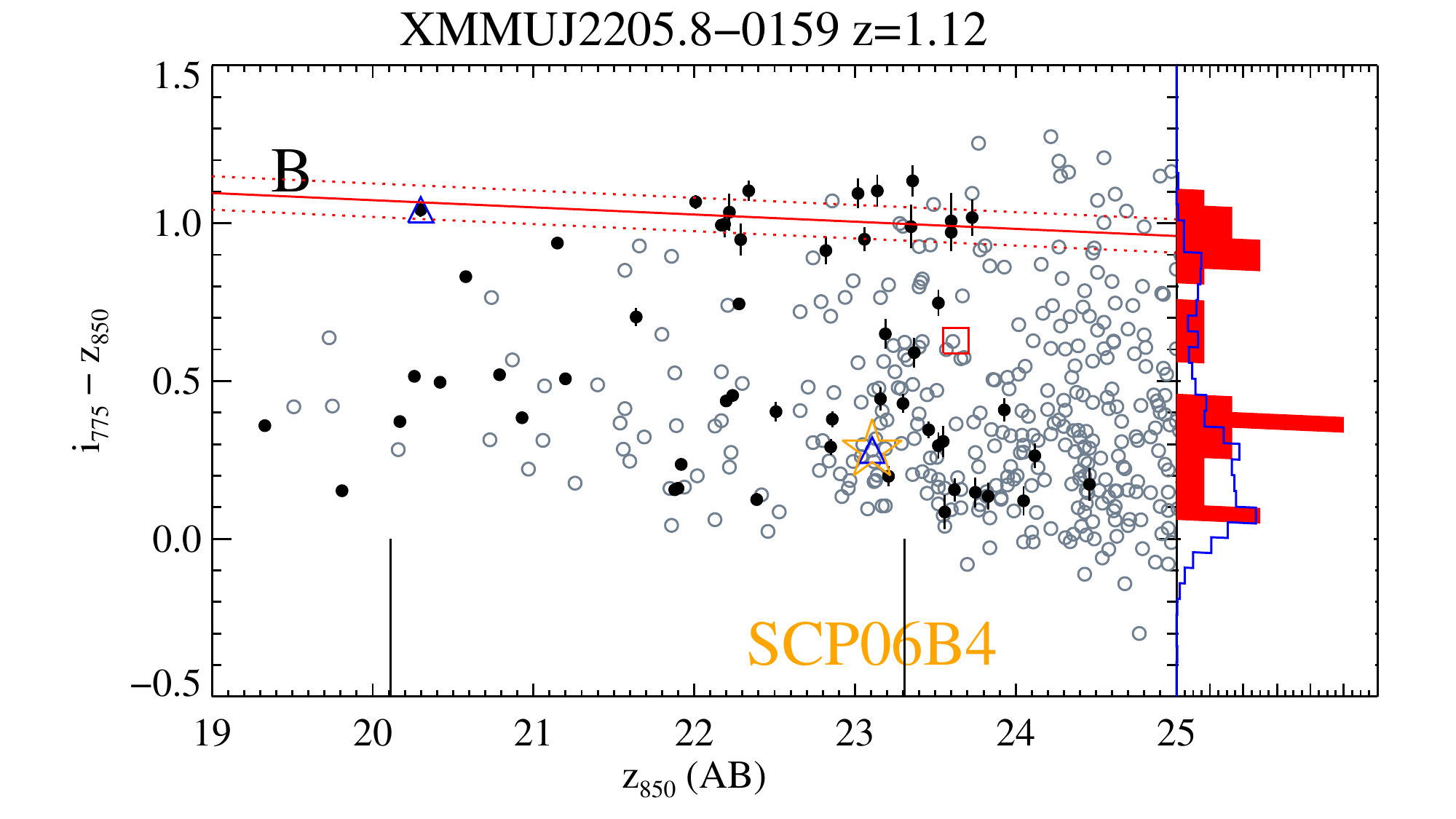}
  } \caption{Color magnitude diagrams (continued).}
\end{figure*}

\addtocounter{figure}{-1}
\begin{figure*}[H]
  \addtocounter{subfigure}{1}
  \centering \subfigure{ \label{fig:CMDpanel:m}
    \plotone{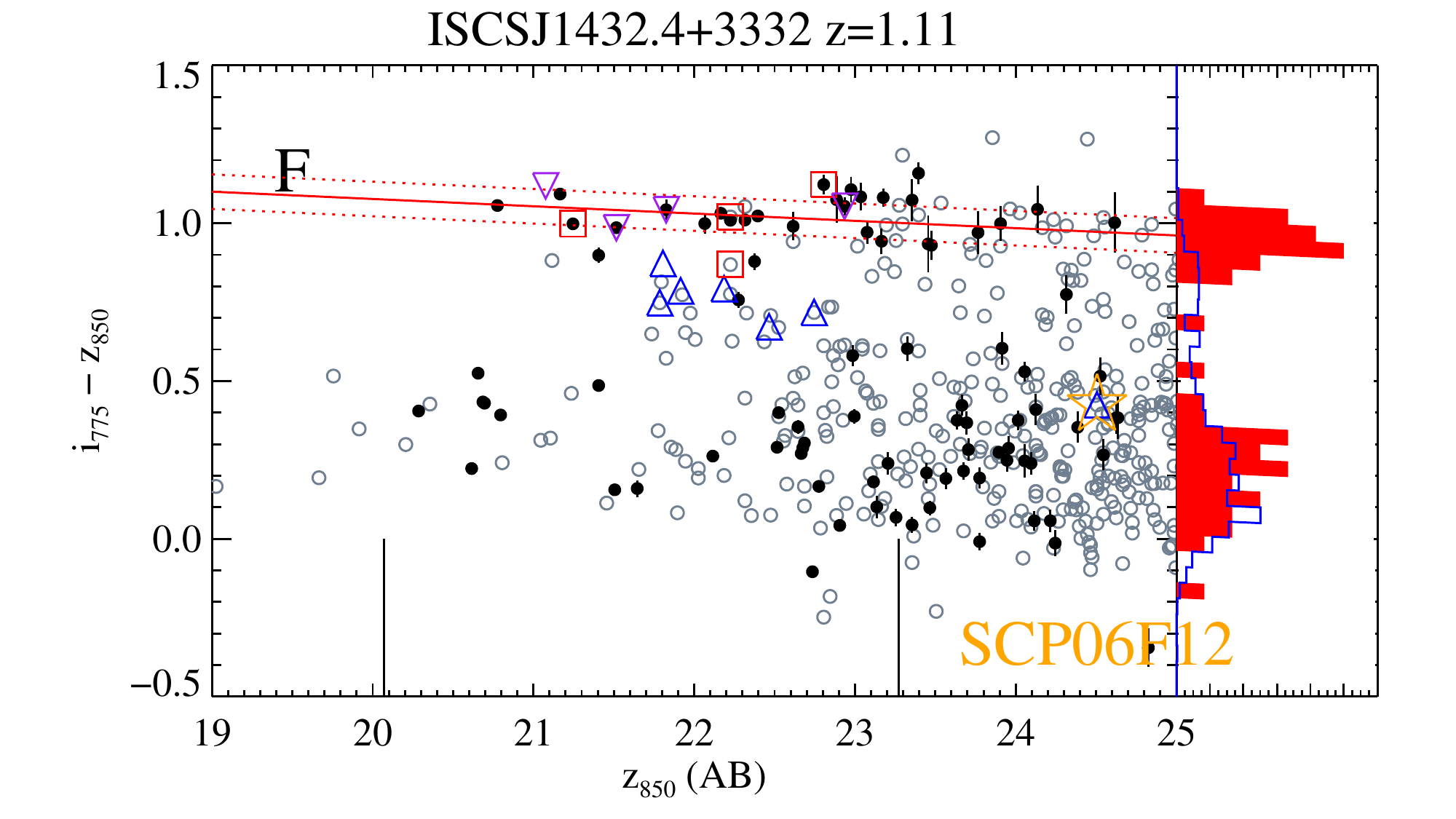}
  } \caption{Color magnitude diagrams (continued).}
\end{figure*}

\addtocounter{figure}{-1}
\begin{figure*}[H]
  \addtocounter{subfigure}{1}
  \centering \subfigure{ \label{fig:CMDpanel:n}
    \plotone{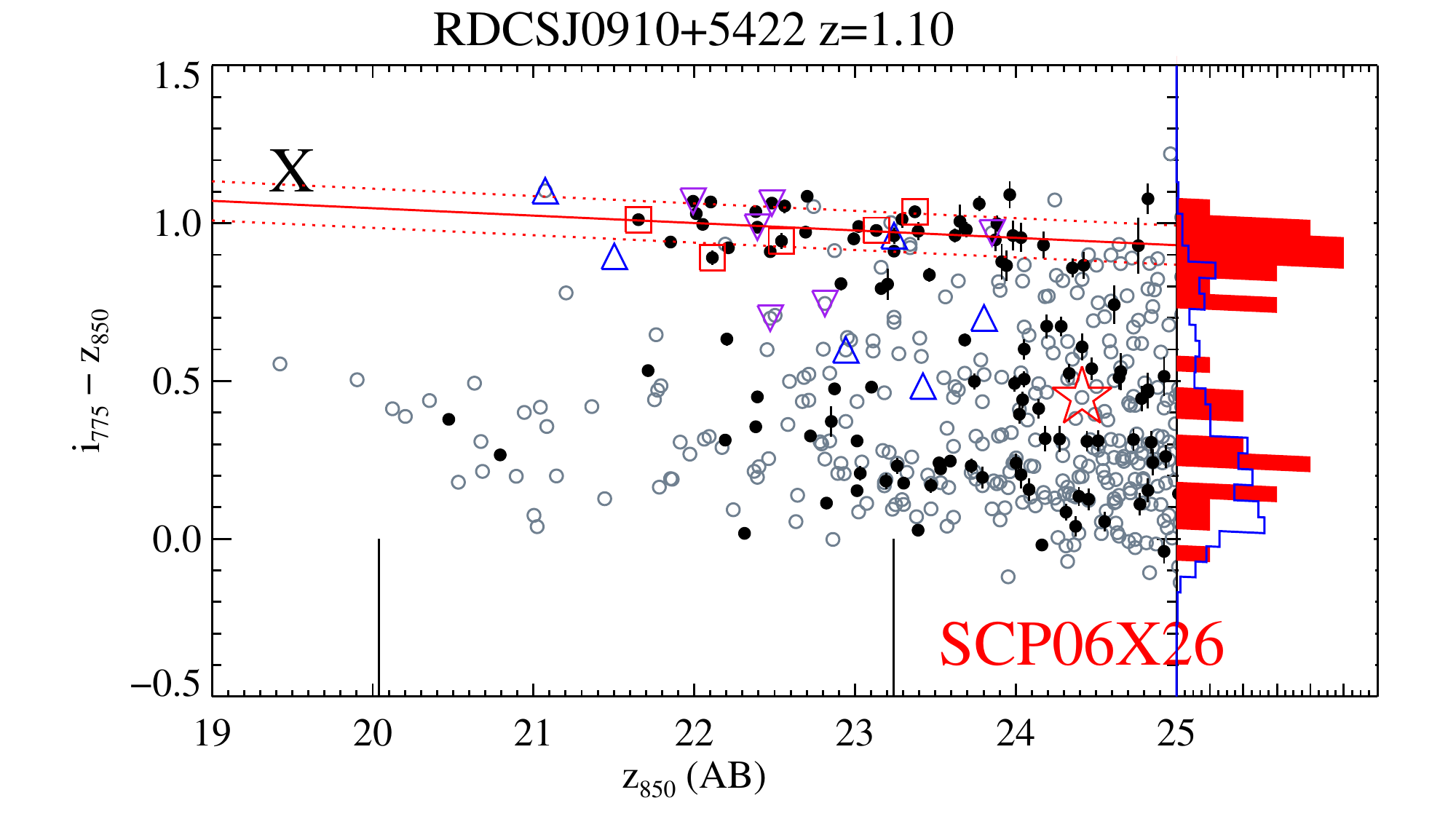}
  } \caption{Color magnitude diagrams (continued).}
\end{figure*}

\addtocounter{figure}{-1}
\begin{figure*}[H]
  \addtocounter{subfigure}{1}
  \centering \subfigure{ \label{fig:CMDpanel:o}
    \plotone{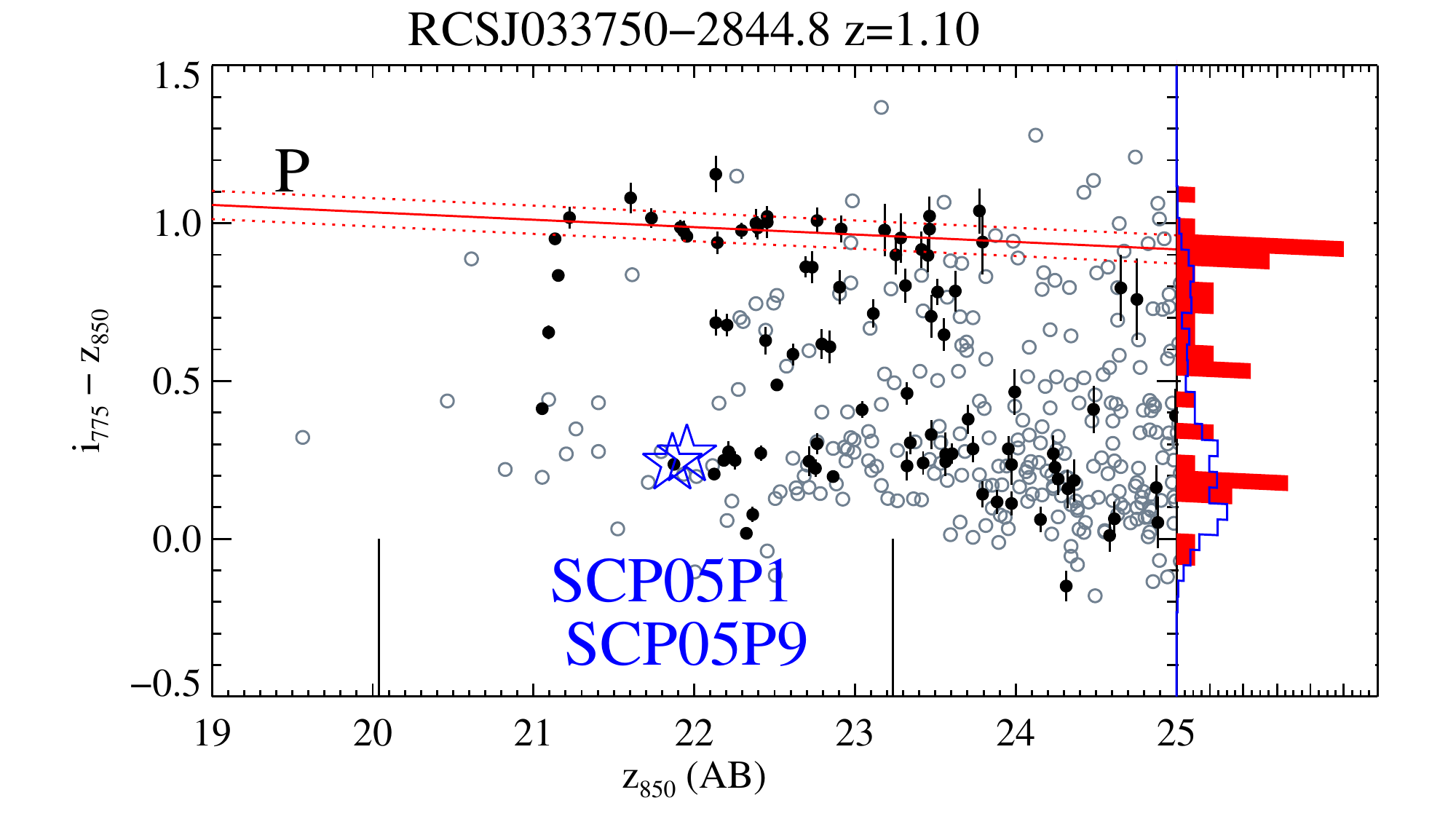}
  } \caption{Color magnitude diagrams (continued).}
\end{figure*}

\addtocounter{figure}{-1}
\begin{figure*}[H]
  \addtocounter{subfigure}{1}
  \centering \subfigure{ \label{fig:CMDpanel:p}
    \plotone{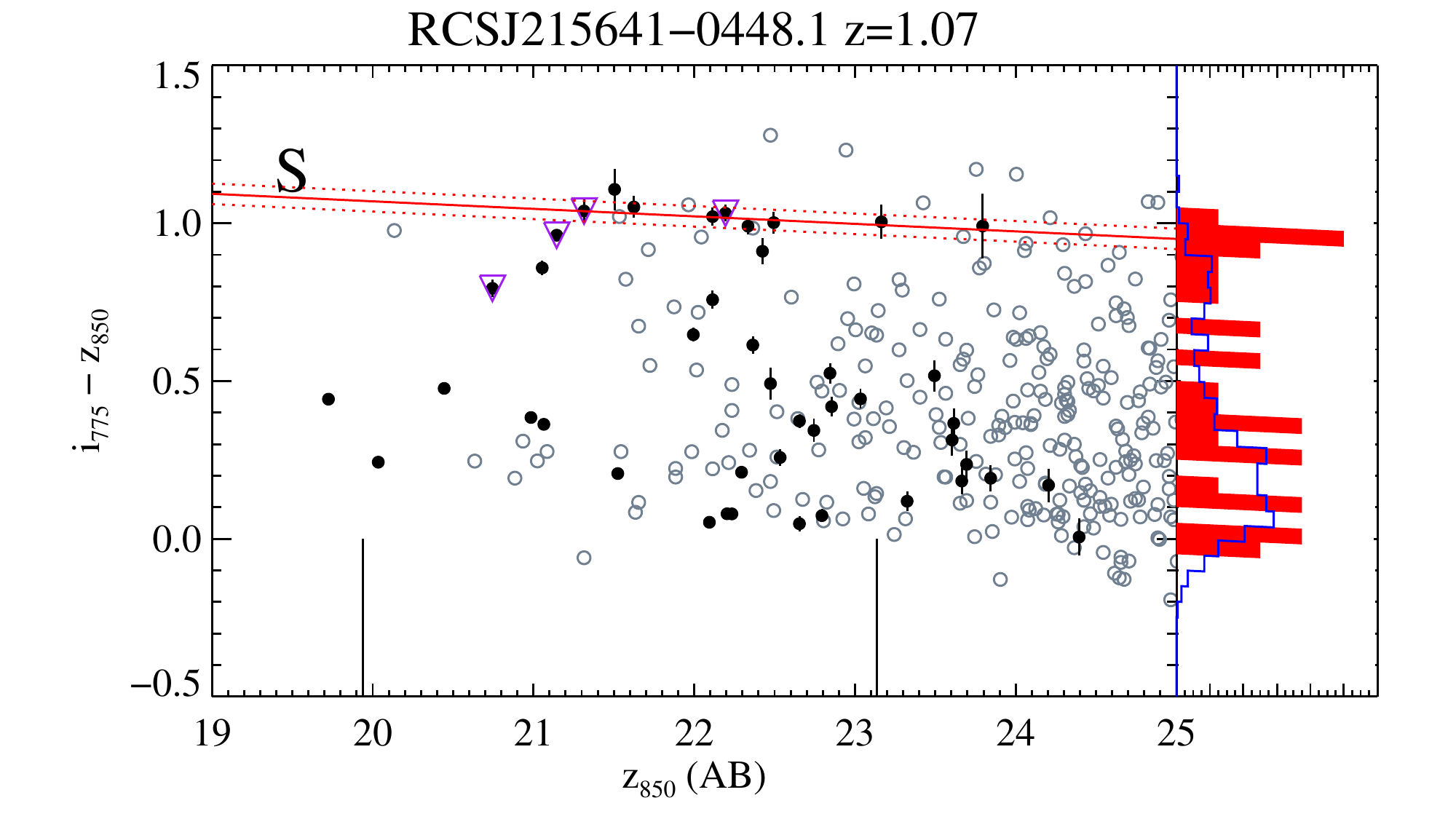}
  } \caption{Color magnitude diagrams (continued).}
\end{figure*}

\addtocounter{figure}{-1}
\begin{figure*}[H]
  \addtocounter{subfigure}{1}
  \centering \subfigure{ \label{fig:CMDpanel:q}
    \plotone{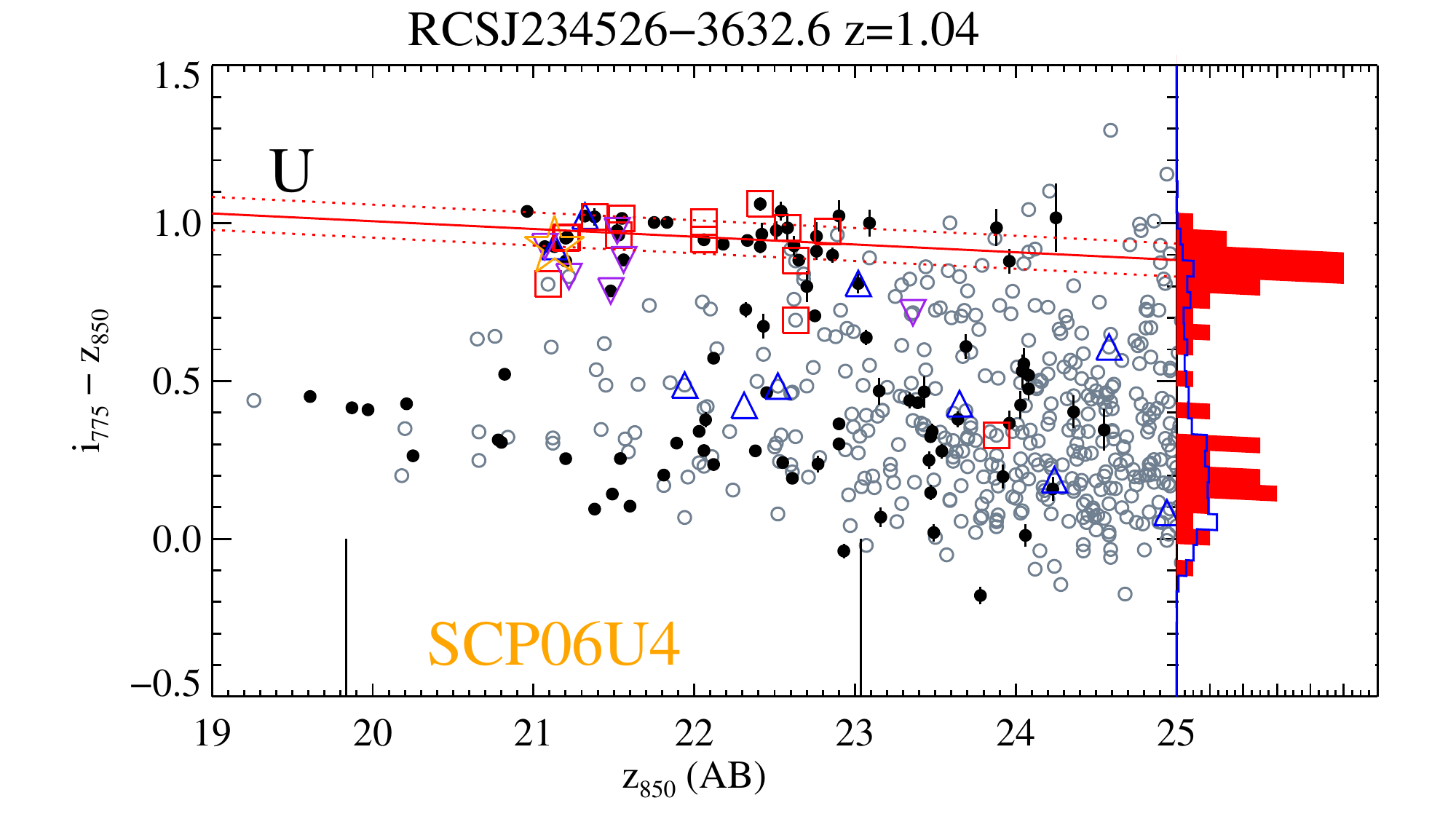}
  } \caption{Color magnitude diagrams (continued).}
\end{figure*}

\addtocounter{figure}{-1}
\begin{figure*}[H]
  \addtocounter{subfigure}{1}
  \centering \subfigure{ \label{fig:CMDpanel:r}
    \plotone{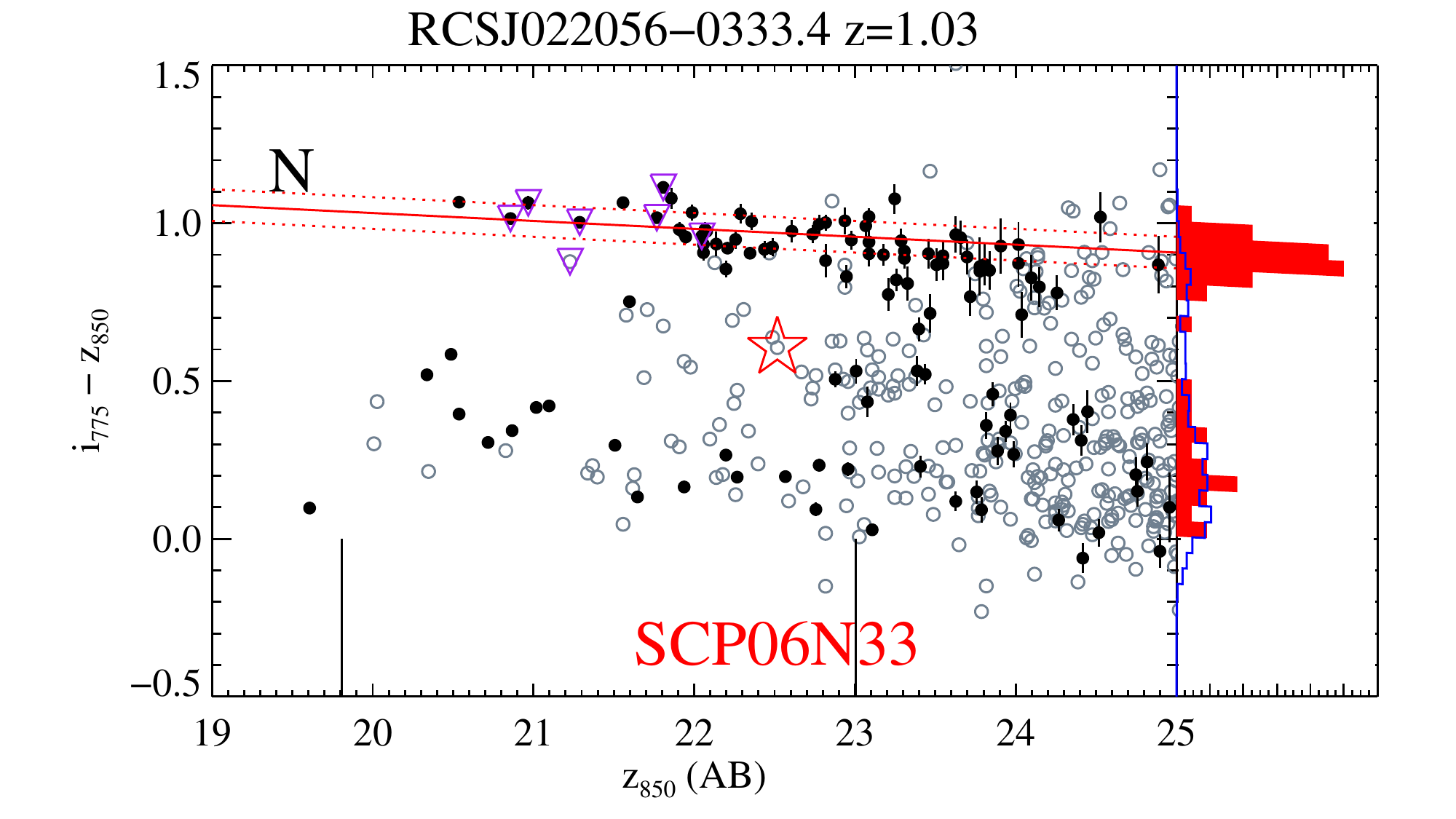}
  } \caption{Color magnitude diagrams (continued).}
\end{figure*}

\addtocounter{figure}{-1}
\begin{figure*}[H]
  \addtocounter{subfigure}{1}
  \centering \subfigure{ \label{fig:CMDpanel:s}
    \plotone{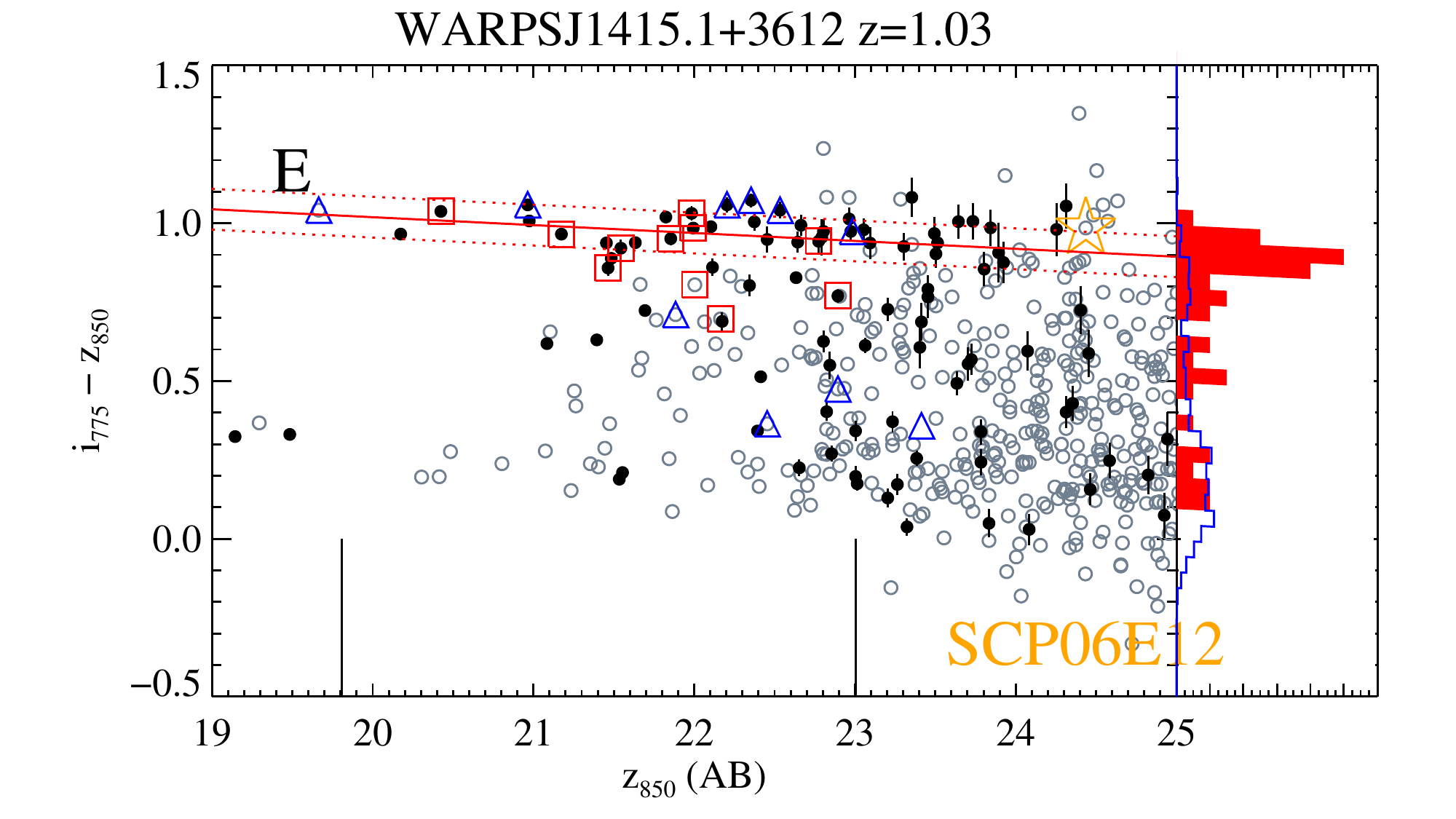}
  } \caption{Color magnitude diagrams (continued).}
\end{figure*}

\addtocounter{figure}{-1}
\begin{figure*}[H]
  \addtocounter{subfigure}{1}
  \centering \subfigure{ \label{fig:CMDpanel:t}
    \plotone{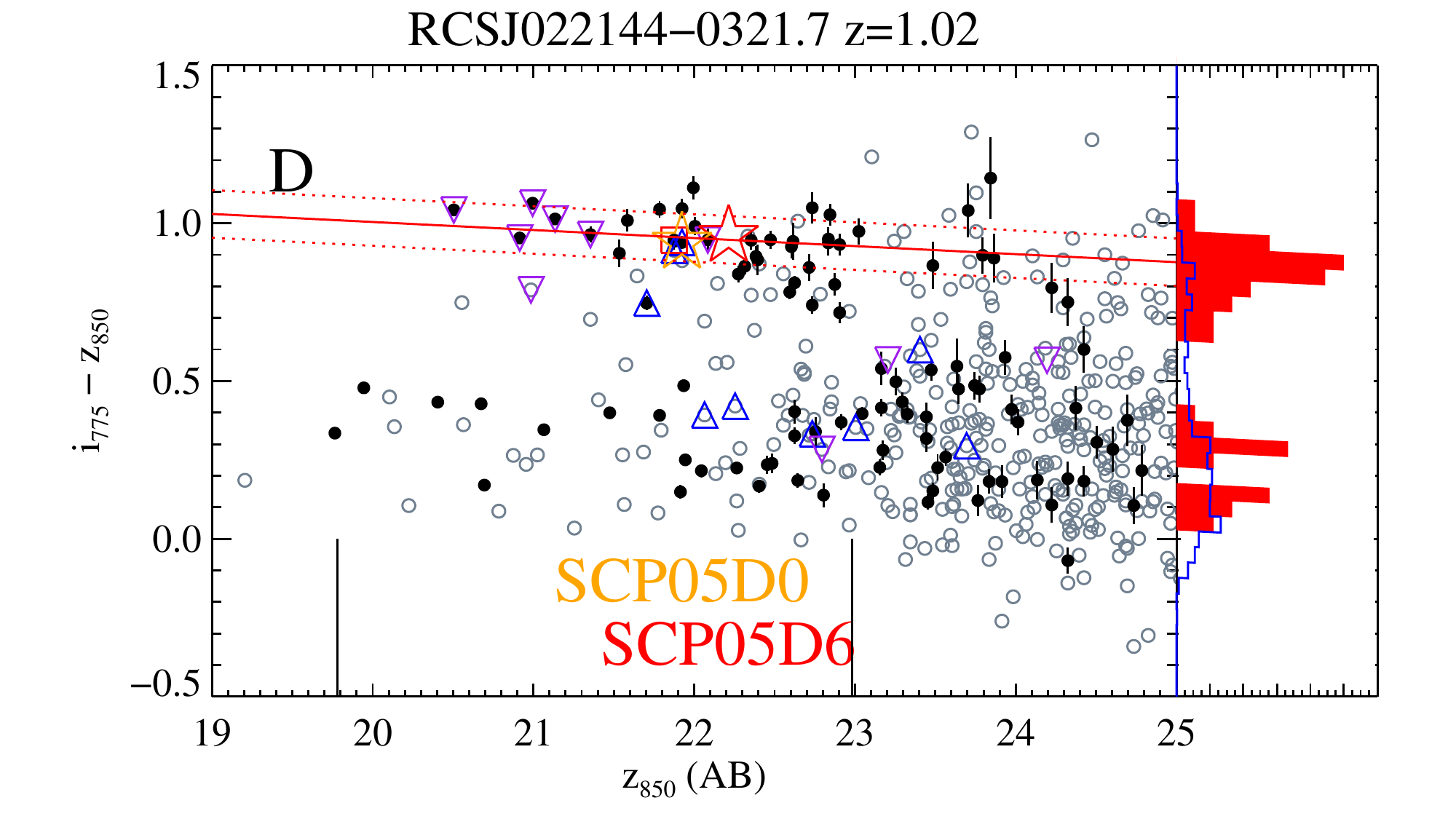}
  } \caption{Color magnitude diagrams (continued).}
\end{figure*}

\addtocounter{figure}{-1}
\begin{figure*}[H]
  \addtocounter{subfigure}{1}
  \centering \subfigure{ \label{fig:CMDpanel:u}
    \plotone{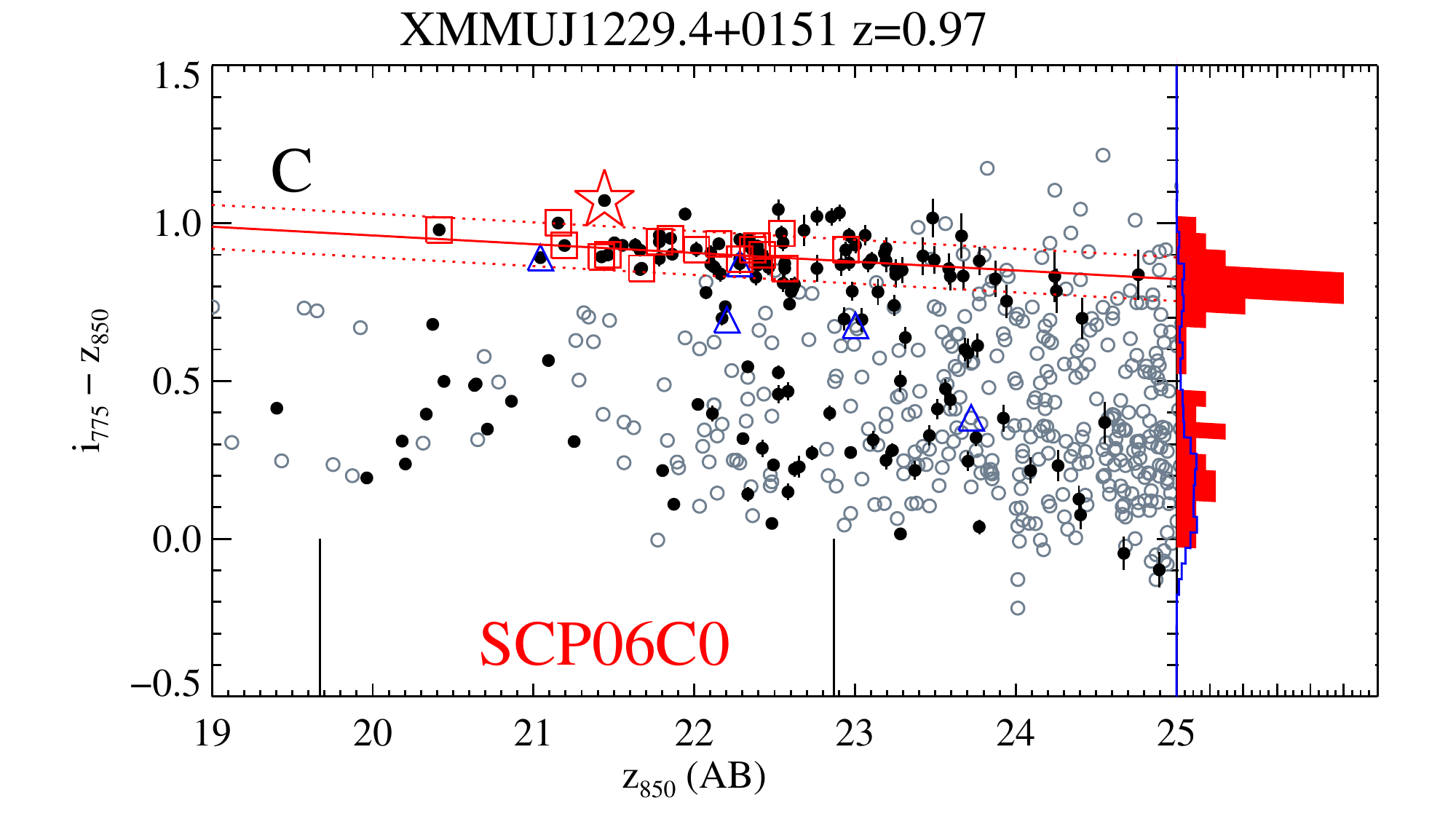}
  } \caption{Color magnitude diagrams (continued).}
\end{figure*}

\addtocounter{figure}{-1}
\begin{figure*}[H]
  \addtocounter{subfigure}{1}
  \centering \subfigure{ \label{fig:CMDpanel:v}
    \plotone{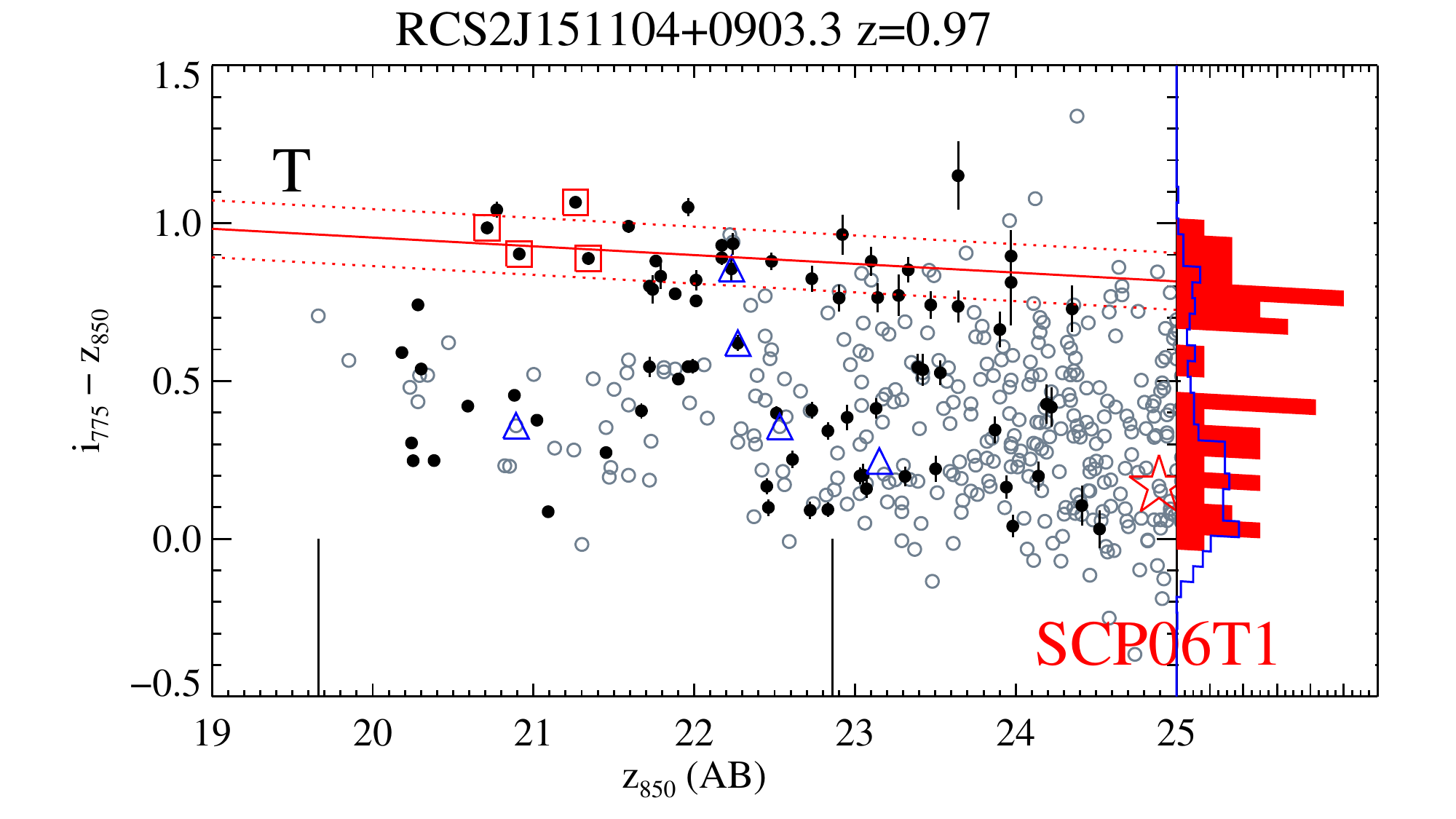}
  } \caption{Color magnitude diagrams (continued).}
\end{figure*}

\addtocounter{figure}{-1}
\begin{figure*}[H]
  \addtocounter{subfigure}{1}
  \centering \subfigure{ \label{fig:CMDpanel:w}
    \plotone{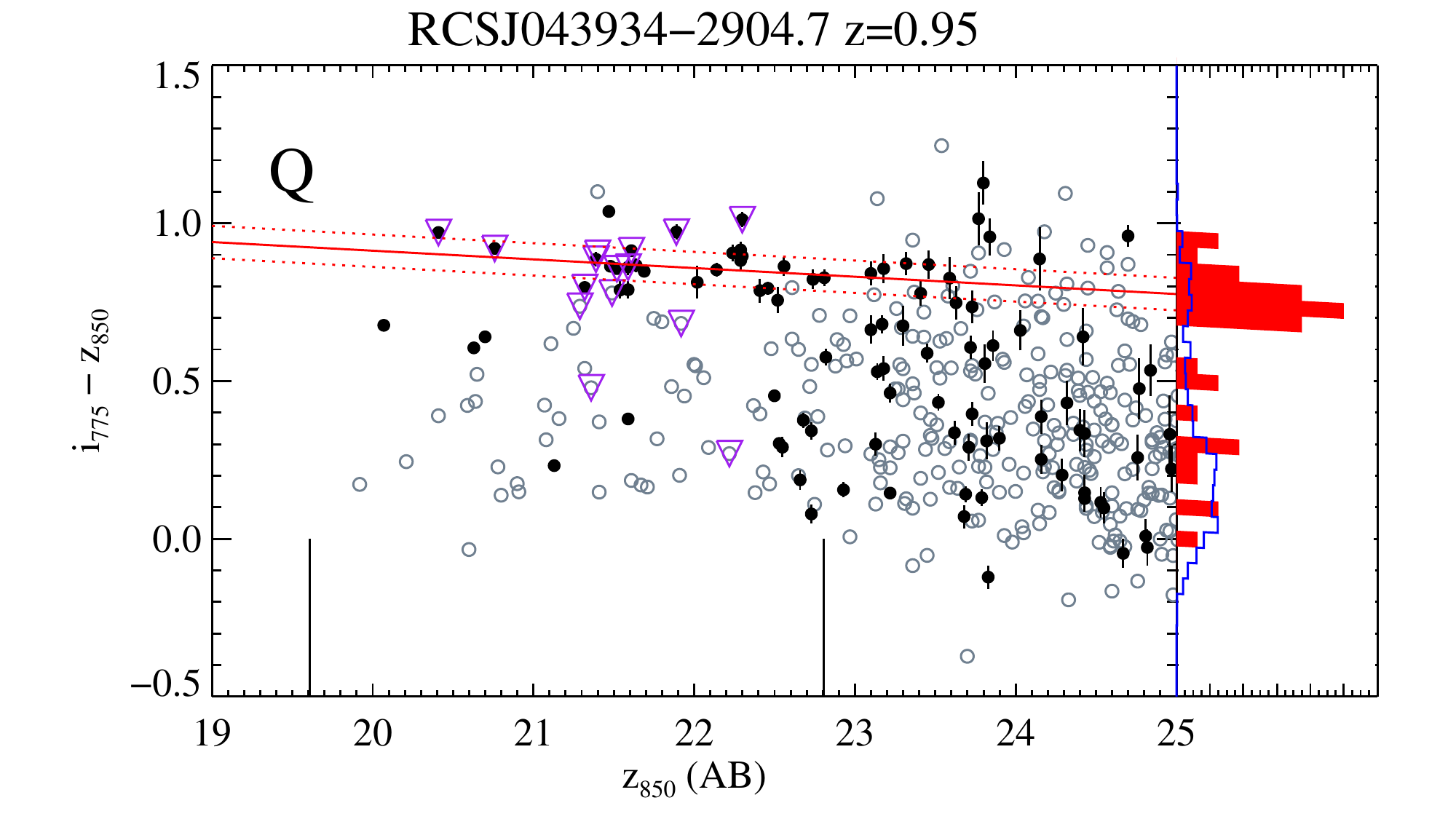}
  } \caption{Color magnitude diagrams (continued).}
\end{figure*}

\addtocounter{figure}{-1}
\begin{figure*}[H]
  \addtocounter{subfigure}{1}
  \centering \subfigure{ \label{fig:CMDpanel:x}
  \plotone{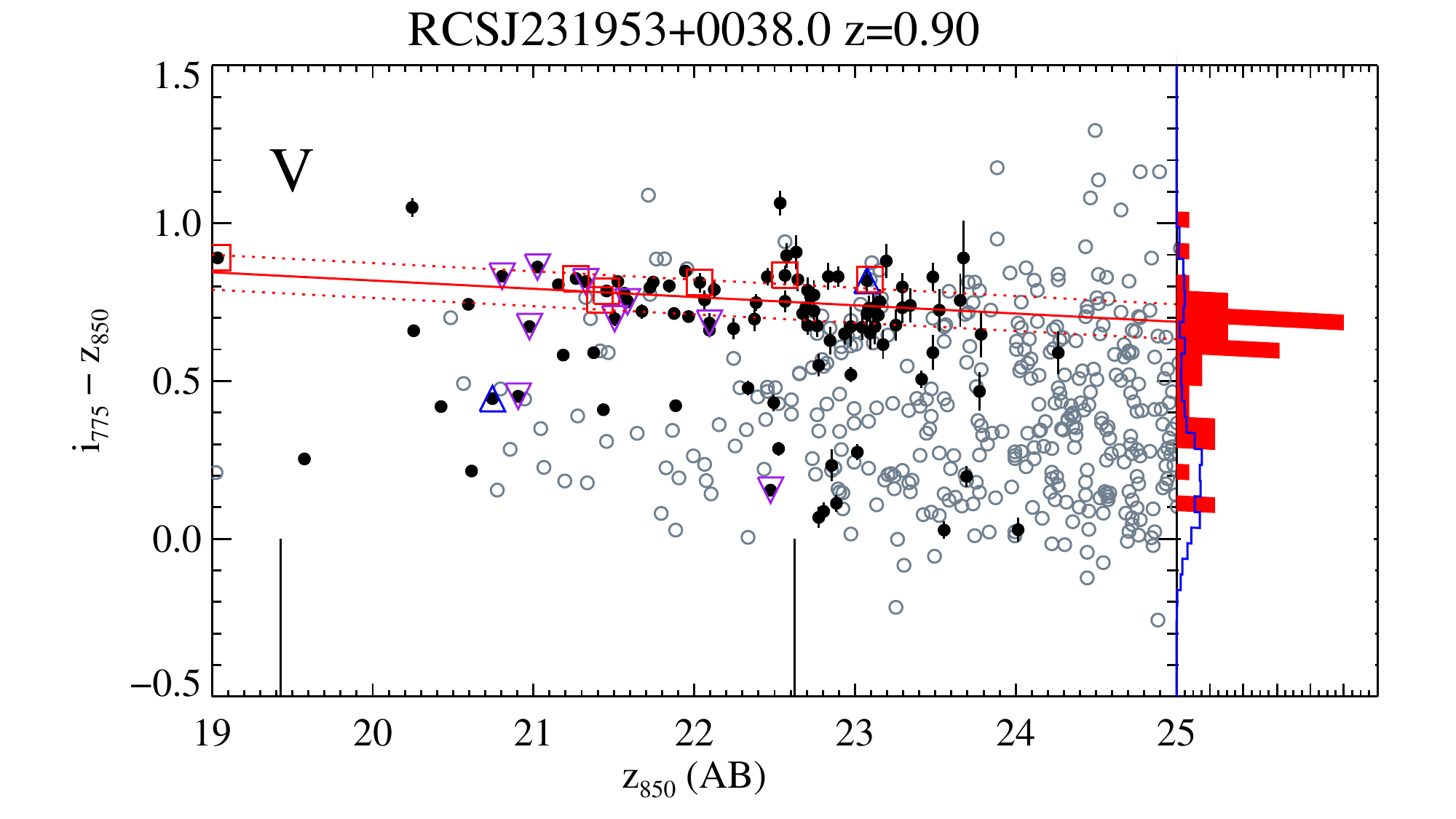}
  } \caption{Color magnitude diagrams (continued).}
\end{figure*}

\addtocounter{figure}{-1}
\begin{figure*}[H]
  \addtocounter{subfigure}{1}
  \centering \subfigure{ \label{fig:CMDpanel:y}
  \plotone{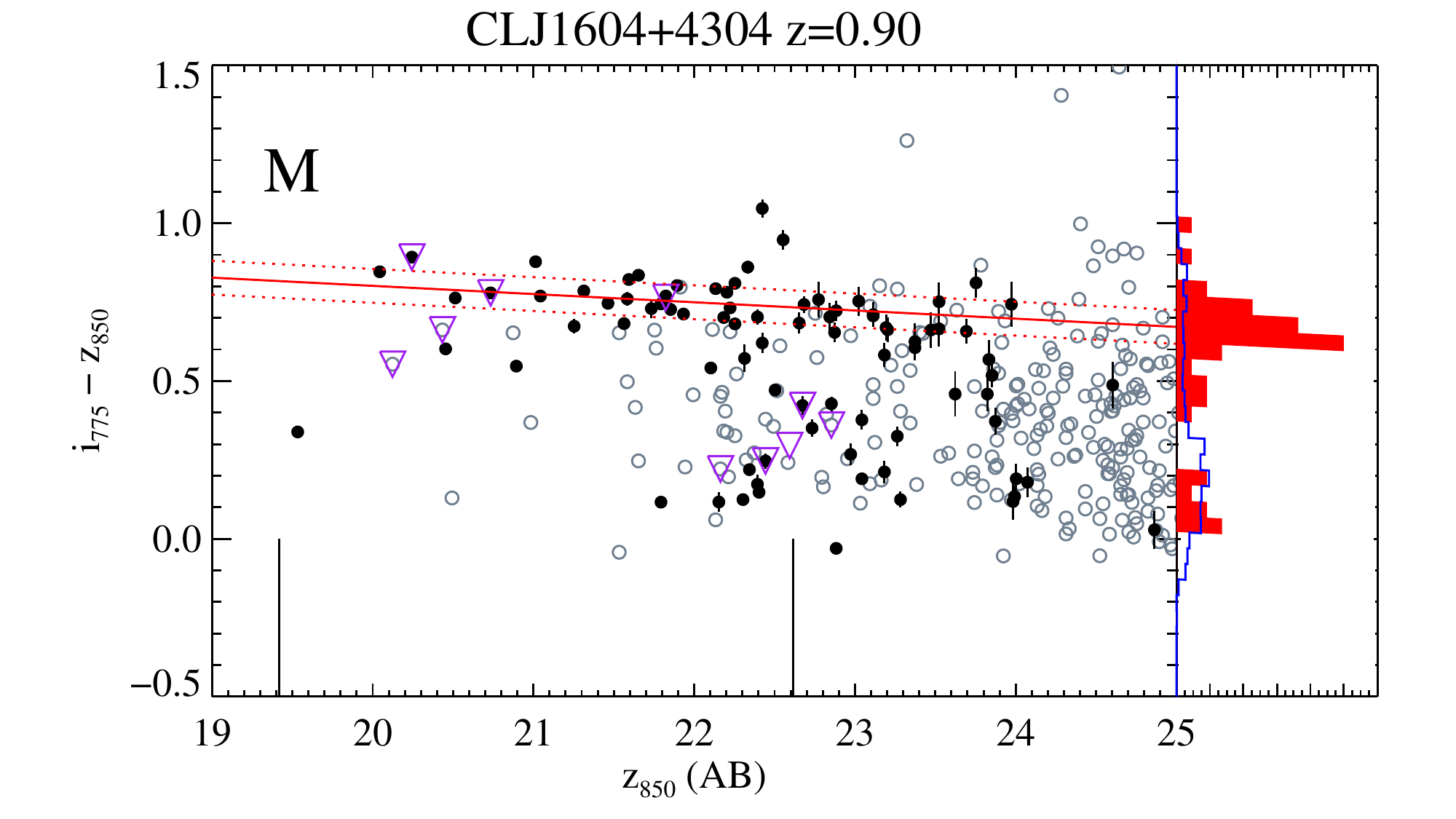}
  } \caption{Color magnitude diagrams (continued).}
\end{figure*}

\end{document}